\documentclass[11pt]{article}
\pdfoutput=1
\usepackage{jheppub}

\input{epsf}
\usepackage{epsfig}
\usepackage{amssymb}
\usepackage{amsfonts}
\usepackage{amsbsy}
\usepackage[all]{xy}
\usepackage{amsmath}
\usepackage{dsfont}
\usepackage{bbm}
\usepackage{upgreek}
\usepackage{amssymb,amscd}
\usepackage{mathrsfs}
\usepackage{amsmath,amsthm}
\usepackage{slashed}
\usepackage{mathtools}
\usepackage{dsfont}

\newcommand\blfootnote[1]{%
  \begingroup
  \renewcommand\thefootnote{}\footnote{#1}%
  \addtocounter{footnote}{-1}%
  \endgroup
}

\makeatletter
\DeclareFontFamily{OMX}{MnSymbolE}{}
\DeclareSymbolFont{MnLargeSymbols}{OMX}{MnSymbolE}{m}{n}
\SetSymbolFont{MnLargeSymbols}{bold}{OMX}{MnSymbolE}{b}{n}
\DeclareFontShape{OMX}{MnSymbolE}{m}{n}{
    <-6>  MnSymbolE5
   <6-7>  MnSymbolE6
   <7-8>  MnSymbolE7
   <8-9>  MnSymbolE8
   <9-10> MnSymbolE9
  <10-12> MnSymbolE10
  <12->   MnSymbolE12
}{}
\DeclareFontShape{OMX}{MnSymbolE}{b}{n}{
    <-6>  MnSymbolE-Bold5
   <6-7>  MnSymbolE-Bold6
   <7-8>  MnSymbolE-Bold7
   <8-9>  MnSymbolE-Bold8
   <9-10> MnSymbolE-Bold9
  <10-12> MnSymbolE-Bold10
  <12->   MnSymbolE-Bold12
}{}

\let\llangle\@undefined
\let\rrangle\@undefined
\DeclareMathDelimiter{\llangle}{\mathopen}%
                     {MnLargeSymbols}{'164}{MnLargeSymbols}{'164}
\DeclareMathDelimiter{\rrangle}{\mathclose}%
                     {MnLargeSymbols}{'171}{MnLargeSymbols}{'171}
\makeatother

\def\be{ \begin{equation} }
\def\ee{ \end{equation}}

\def\det{{\rm det}}

\def\exp{{\rm exp}}

\def\I{{\rm i}}

\renewcommand{\Im}{{\rm Im }}
\def\ker{{\rm ker}}
\def\Lie{{\rm Lie}}

\def\sgn{{\rm sgn\,}}

\def\half{\frac{1}{2}}

\def\one{{\hbox{ 1\kern-.8mm l}}}

\def\vx{{\vec{x}}}
\def\vy{{\vec{y}}}

\def\ti{\tilde{i}}
\def\tj{\tilde{j}}
\def\tk{\tilde{k}}

\def\CB{{\cal B}}

\def\CM {{\cal M}}
\def\CN {{\cal N}}

\def\CW {{\cal W}}
\def\CX {{\cal X}}

\def\CB {{\cal B}}

\def\CX {{\cal X}}
\def\CY {{\cal Y}}

\def\IR{{\mathbb{R}}}

\def\ft{\mathfrak{t}}

\def\ft{\mathfrak{t}}

\def\rmk#1{\bigskip\noindent{\bf Remark:} }

\newcommand\fro{{\overline{\underline{\Omega}}}}
\def\hk{hyperk\"ahler}

\def\fMM{\overline{\underline{\mathcal{M}}}}
\def\Dsl{\slashed{D}}

\DeclareMathAlphabet{\mathpzc}{OT1}{pzc}{m}{it}

\def\cp{\mathpzc{p}}

\def\cn{\mathpzc{n}}

\def\fMM{\underline{\overline{\mathcal{M}}}}

\def\Lie{\pounds}

\def\ie{{\it i.e.}}
\def\eg{{\it e.g.}}

\def\etc{{\it etc}}

\def\sgn{ \, \textrm{sgn} }

\def\rnk{ \, \textrm{rnk} }
\def\diag{ \, \textrm{diag}}

\def\ui{\underline{i}}
\def\uj{\underline{j}}
\def\uk{\underline{k}}

\def\gm{\gamma_{\rm m}}

\def\gmone{\gamma_{{\rm m},1}}
\def\gmtwo{\gamma_{{\rm m},2}}
\def\rM{\mathbf{M}}

\def\rG{\mathrm{G}}

\def\hi{\hat{i}}
\def\hj{\hat{j}}
\def\hk{\hat{k}}

\def\ha{\hat{a}}
\def\hb{\hat{b}}
\def\hc{\hat{c}}
\def\hp{\hat{p}}
\def\hq{\hat{q}}
\def\hr{\hat{r}}

\def\tC{\widetilde{C}}
\def\ti{\tilde{i}}
\def\tj{\tilde{j}}
\def\tk{\tilde{k}}
\def\tl{\tilde{l}}

\def\ualpha{\underline{\alpha}}
\def\ubeta{\underline{\beta}}
\def\ugamma{\underline{\gamma}}

\def\wtTheta{\widetilde{\Theta}}
\def\wtbC{\widetilde{{\bf C}}}

\def\Im{ \, \mathrm{Im} \, }

\def\NN{\mathcal{N}}

\def\half{\frac{1}{2}}

\def\RR{\mathcal{R}}

\def\WW{\mathcal{W}}

\def\MM{\mathcal{M}}

\def\SS{\mathcal{S}}

\def\XX{\mathcal{X}}

\def\YY{\mathcal{Y}}

\def\pd{\partial}
\def\uk{\underline{k}}

\def\vx{\vec{x}}
\def\vy{\vec{y}}

\title{Wall Crossing from Dirac Zeromodes }
\author[a]{T.~Daniel Brennan,}
\author[b]{Gregory W.~Moore,}
\author[c,\ast]{and Andrew B.~Royston\blfootnote{${}^\ast$Now at \emph{Penn State Fayette, The Eberly Campus, 2201 University Drive, Lemont Furnace, PA 15456, USA}}}

\affiliation[a,b]{NHETC and
Department of Physics and Astronomy, Rutgers University \\
126 Frelinghuysen Rd., Piscataway NJ 08855, USA}
\affiliation[c]{George P.\ \& Cynthia Woods Mitchell Institute for Fundamental Physics and Astronomy, \\
Texas A\&M University, College Station, TX 77843, USA}

\emailAdd{tdanielbrennan@physics.rutgers.edu}
\emailAdd{gmoore@physics.rutgers.edu}
\emailAdd{aroyston@physics.tamu.edu}

\abstract{We explore the physics of two-body decay of BPS states using semiclassical analysis to construct explicit solutions that illustrate the main features of wall crossing, for both ordinary and framed BPS states.  In particular we recover the primitive wall-crossing formula from an asymptotic analysis of certain Dirac-type operators on monopole moduli spaces.  Along the way we give an asymptotic metric for the moduli space of singular monopoles, analogous to the Gibbons--Manton and Lee--Weinberg--Yi metrics for the moduli space of smooth monopoles, and we find evidence for the existence of stable non-BPS boundstates.  Our discussion applies to four-dimensional $\CN$ = 2 super-Yang--Mills theories with general gauge group and general 't Hooft defects.}

\begin{document}
\begin{flushright} MI-TH-1884 \end{flushright}
\maketitle

\section{Introduction and Conclusion}\label{sec:Intro}

\parskip 7pt

 This paper continues an investigation into the semiclassical formulation of BPS states and their wall crossing initiated in \cite{Moore:2015szp,Moore:2015qyu,Brennan:2016znk}. In these papers, it was shown that semiclassical BPS states are described by the kernels of twisted Dirac operators on monopole moduli spaces. The goal of the present paper is to give a very explicit presentation of how these Dirac operators fail to be Fredholm on walls of marginal stability and how the change of the index exactly reproduces known wall-crossing formulae.
 
In the case of $\mathrm{SU}(3)$ gauge theory with a very special choice of magnetic charge the strongly centered moduli space is exactly the Taub-NUT space and, following a classic computation by Pope \cite{Pope:1978zx}, one can solve for the zeromodes of the twisted Dirac operator explicitly \cite{Gauntlett:1995fu,Lee:1996kz,Gauntlett:1996cw,Lee:1996if,Gauntlett:1999vc,Moore:2015szp,Bais:1979gv,Gibbons:1986df}. (See also \cite{Jante:2013kha,Boulton:2017jls}.)  These computations illustrate clearly how the wall crossing is related to the failure of this twisted Dirac operator to be Fredholm on a codimension one locus in parameter space.  In this paper we will generalize the discussion to two-body decays of BPS states carrying arbitrary magnetic charge in a theory with any simple gauge group $G$, and in the presence of an arbitrary set of 't Hooft defects.

The key idea will be to study the Dirac equation in a certain subregion of the standard asymptotic region of monopole moduli space. We will call the subregion the \emph{two-galaxy region}.\footnote{The `standard' asymptotic region is the one investigated by Gibbons and Manton \cite{Gibbons:1995yw} and Lee, Weinberg, and Yi \cite{Lee:1996kz}, corresponding to configurations in which all fundamental constituents are well separated relative to the inverse mass of the lightest $W$-boson.  Hence the two-galaxy region corresponds to a double scaling limit, or a hierarchy of scales.  In the language of the compactification of the moduli space as a manifold with corners, \cite{Kottke:2015rwx,FKS}, it is a neighborhood of the co-dimension two corner where the boundary face corresponding to the maximal partition of constituents meets the boundary face corresponding to a given two-partition of constituents.} Roughly speaking, in this limit the fundamental constituents making up the BPS state split into two widely separated clusters, or `galaxies', such that the moduli space splits as a product of the overall center of mass, two strongly centered moduli spaces for the two galaxies, and a relative moduli space modeled on the standard four-dimensional Taub-NUT space. Solving for the spectrum of the Hamiltonian is reduced to solving for the eigenspinors of the supercharge, a Dirac-type operator, on this Taub-NUT space.  As every multi-monopole moduli space has such two-galaxy regions, the wall crossing for a rank two group with monopole charge $\{1,1\}$ gives a universal mechanism of two-body decays of BPS boundstates.

Our analysis will give a semiclassical derivation of the \emph{primitive} wall-crossing formula \cite{Denef:2007vg}. As we know from the full wall-crossing formula \cite{Kontsevich:2008fj, Joyce:2008pc,Gaiotto:2008cd,Pioline:2011gf,MR3330788}, the change of the BPS Hilbert space across a wall of marginal stability is in general much more complicated. In the formulae for general wall crossing, the primitive wall-crossing contribution is only the first term in a series of multi-body decays which can be described using methods such as the attractor tree formalism \cite{Denef:2007vg,Andriyash:2010yf,Denef:2000nb,Manschot:2010xp}.  It is possible that an extension of the work in this paper may be used to derive these terms as well by allowing for arbitrary fractionalization of the two BPS galaxies. But we will not attempt that here. In ongoing work \cite{Kottke:2015rwx,FKS} a compactification of monopole moduli space as a manifold
with corners is being investigated. It would be quite fascinating if such a compactification could be used to understand the intricate combinatorics of the full Kontsevich--Soibelman wall-crossing formula.

It is also interesting to compare our discussion with other discussions of wall-crossing formulae based on Dirac operators. In \cite{Manschot:2010qz} Dirac operators on the moduli space of Denef's multicenter black holes were used to derive a formulation of the wall-crossing formula. In \cite{Stern:2000ie} a localization formula was used to derive the BPS index for the Dirac operators on the moduli space of $\{1,1,\ldots,1\}$-monopoles, where the Gibbons--Manton/Lee--Weinberg--Yi (GM/LWY) asymptotic metric is the exact metric \cite{Gibbons:1995yw,Lee:1996kz,Murray:1996hi}.  In the latter approach one analyzes a modification of the Dirac operator at a fixed point of the symmetry that lies deep in moduli space. Our approach will focus on the behavior of the Dirac operator in asymptotic regions of moduli space. The two approaches complement one another very nicely.

An interesting by-product of our analysis is the prediction of an infinite tower of non-BPS metastable states in the spectrum
which also disappear on the usual walls of marginal stability. We arrive at this prediction by considering the
full eigenspectrum of the Dirac-type operator on the relative Taub-NUT space. At the moment this prediction is conjectural
because one would need to establish that these boundstates extend to eigenstates of the full Dirac operator on the entire
moduli space. Indeed, the analysis of Appendix \ref{app:AppD} with $\ell_{\rm TN}<0$ provides a cautionary example where boundstate wavefunctions at large relative distance might or might not extend to BPS states of the full moduli space. 
Nevertheless, the analysis strongly suggests that if the BPS states exist, then so should the non-BPS hydrogen-like 
spectrum.  Furthermore in the $\gm = \{1,1\}$ example, Taub-NUT \emph{is} the full strongly centered space, so we know they exist in this special case.

Unlike their BPS counterparts, the same spectrum of non-BPS boundstates re-emerges on the other side of the wall.  The corresponding subspace of the Hilbert space is the same on either side of the wall, so they do not undergo wall crossing in the usual sense.  We find this observation intriguing as it suggests that, on the side of the wall where a given BPS state no longer exists, the lowest energy non-BPS boundstates carrying the same conserved charges will be completely stable.  Non-BPS boundstates in $\NN = 2$ theories have been studied before \cite{Ritz:2001jk,Ritz:2008jf}.  Our results, when restricted to the special case of zero Dirac--Schwinger--Zwanziger pairing between the constituent electromagnetic charges, are completely consistent with \cite{Ritz:2008jf}.

The outline of this paper is as follows. In Section \ref{sec:asymptoticM} we will review the asymptotic region of monopole moduli space and construct the metric on the two-galaxy region by taking a limit of the GM/LWY metric.  In Section \ref{sec:AsymptoticD} we will then calculate the twisted Dirac operator corresponding to one of the supercharge operators in the two-galaxy region and construct explicit solutions for the zeromode wavefunctions representing BPS states.  We will also comment on the non-BPS boundstates and their spectrum, based on the analysis of Appendix \ref{app:AppD}.  Then we will use these solutions to show that they decay at walls of marginal stability in such a way as to satisfy the primitive wall-crossing formula. In Section \ref{sec:Framed} we repeat this analysis for the case of framed BPS states, which generically take the form of a core-halo system.  This includes an analog of the GM/LWY result for the asymptotic region of singular monopole moduli space.  Again the expected results for the location of the walls of marginal stability and the primitive wall-crossing formula for framed BPS states are obtained.

The spectrum and eigenspinors of the Dirac operator we consider in Appendix \ref{app:AppD} were thoroughly analyzed in \cite{Jante:2015xra}, and our results are consistent.  We have chosen to reproduce some aspects of the analysis in our notation and conventions in order to keep the paper self contained.  Reference \cite{Jante:2015xra} also includes discussions of classical particle orbits, scattering states, their associated cross sections, and an algebraic construction of the boundstate spectrum.  Additionally, an investigation of the same operator appears in \cite{MurthyPioline}, where detailed results on the density of states are utilized.

\section{Asymptotic Regions of Moduli Space}\label{sec:asymptoticM}

It has been known for quite some time that the dynamics of semiclassical monopoles in the adiabatic limit can be interpreted as geodesic movement on monopole moduli space \cite{Manton:1981mp}. This construction has been applied to the study of semiclassical BPS states by the work of \cite{Gauntlett:1993sh,Sethi:1995zm,Gauntlett:1995fu,Lee:1996kz,Gauntlett:1999vc,Gauntlett:2000ks,Moore:2015szp,Moore:2015qyu,Brennan:2016znk} and many others. In this analysis, it was shown that the description of BPS states is captured by an $\CN=4$ supersymmetric quantum mechanics (SQM) on certain bundles over monopole moduli space.  BPS states sit in the kernel of one, and hence all, of the supercharge operators.  One of the supercharges takes the form of a Dirac-type operator.  In this section we summarize key properties of the moduli space itself and describe the asymptotic two-galaxy region.

\subsection{Moduli space facts}

Monopole moduli space, $\MM(\gm,\XX)$, is a hyperk\"ahler manifold depending on two pieces of data.  The vacuum expectation value (vev) $\XX$ is the limiting value of the adjoint Higgs field that participates in the Bogomolny equation, as we go to infinity in $\mathbb{R}^3$.  In a suitable gauge it can be taken to be constant on the two-sphere at infinity.\footnote{The relationship between this vev and the physical vacuum data of the $\NN = 2$ theory will be given below, after describing some further details.  See equation \eqref{SWscmap}.}  We assume $\XX \in \mathfrak{g}$ is regular and hence defines a unique Cartan subalgebra $\mathfrak{t} \in \mathfrak{g}$, basis of simple roots $\{ \alpha_I\}$ for $\mathfrak{t}^\ast$, and simple co-roots $\{ H_I \}$ for $\mathfrak{t}$, where $I = 1, \ldots, r := \rnk{\mathfrak{g}}$.  Physically this is the maximal symmetry-breaking case, such that the gauge group is broken to the Cartan torus $T \subset G$.  The magnetic charge takes the form $\gm = \sum_I n_{\rm m}^I H_I$, and $\MM$ is nonempty iff all of the $n_{\rm m}^I$ are non-negative integers and at least one is positive \cite{Taubes:1981gw}.  The quaternionic dimension of $\MM$ is $\sum_I n_{\rm m}^I$.  The physical interpretation \cite{Weinberg:1979ma,Weinberg:1979zt} is that there are $n_{\rm m}^I$ fundamental monopoles of type $I$ for each $I$, and each fundamental monopole carries four degrees of freedom:  three for its position and one for an internal phase conjugate to electric charge.

$\MM$ has isometries generated by Killing vectors forming an algebra $\mathbb{R}^3 \oplus \mathfrak{so}(3) \oplus \mathfrak{t}$, corresponding to the action of translations, rotations, and asymptotically nontrivial gauge transformations preserving the vev.\footnote{The action is effective if all $n_{\rm m}^I > 0$.  We will assume this in the following since the other cases can be reduced to this by embedding a smaller gauge group into $G$.  This is discussed in detail in \cite{Moore:2015szp}.}  The $\mathfrak{t}$-action is hyperholomorphic.  We denote by 
\begin{align}\label{Gmap}\begin{split}
{\rm G} : \mathfrak{t} &  \to \mathfrak{isom}_{\mathbb{H}}(\MM)\\
h &\mapsto {\rm G}(h)~,
\end{split}\end{align}
the Lie algebra homomorphism sending elements of the Cartan to the corresponding triholomorphic Killing vectors.  Gauge transformations act through the adjoint representation, so it is the $\rG(h^I)$, with $\{h^I\}$ the fundamental magnetic weights, that generate $2\pi$-periodic isometries of $\MM$.  The fundamental magnetic weights are the integral duals of the simple roots: $\langle\alpha_I, h^J \rangle = {\delta_I}^J$, where $\langle~,~\rangle :\mathfrak{t}^\ast \times \mathfrak{t} \to \mathbb{R}$ is our notation for the natural pairing between a vector space and its dual.   

The translational Killing vectors, along with ${\rm G}(\XX)$, are covariantly constant and generate a flat $\mathbb{R}^4$, such that the universal cover of the moduli space is metrically a product, $\widetilde{\MM}(\gm,\XX) = \mathbb{R}^4 \times \MM_0(\gm,\XX)$.  $\MM_0$ is an irreducible, smooth, and complete hyperk\"ahler manifold known as the strongly centered moduli space.  It encodes the relative positions and phases of the constituents while the $\mathbb{R}^4$ is associated with the overall center-of-mass degrees of freedom.  The fundamental group acts on the universal cover as the group of Deck transformations, $\mathbb{D}$, and one has
\begin{equation}\label{eq:topofactor}
\MM(\gm,\XX) = \mathbb{R}^3 \times \frac{ \mathbb{R} \times \MM_0(\gm,\XX)}{\mathbb{D}}~,
\end{equation}
where the distinguished $\mathbb{R}$ is the one generated by $\rG(\XX)$.  It is known from the rational map construction that $\mathbb{D} \cong \mathbb{Z}$.  However only an $\ell \cdot \mathbb{Z}$ subgroup is associated with the periodicity conditions of the asymptotically nontrivial gauge transformations.  Here $\ell = \gcd_I\{ n_{\rm m}^I \cp^I\}$ where $\cp^I = \alpha_{\rm long}^2/\alpha_{I}^2$ is the ratio of the length-squared of the long root to that of the $I^{\rm th}$ root \cite{Moore:2015szp}.  We will work with a Killing form $(~,~)$ on $\mathfrak{g}$ and $\mathfrak{g}^\ast$ such that the length-squared of long roots is two.  If $\phi$ is the isometry generating $\mathbb{D}$,  then for any $h \in \Lambda_{\rm mw}$ we have
\begin{equation}\label{muhomo}
\exp(2\pi {\rm G}(h)) = \phi^{\mu(h)}~,
\end{equation}
where $\mu(h) = (\gm,h)$.

Let $ds^2$ denote the hyperk\"ahler metric on $\MM$.  Using integration by parts and the explicit construction of the $\rG$-map, one can show that \cite{Moore:2015szp}
\begin{equation}\label{killing2metric}
ds^2(\rG(\XX),\rG(h)) = (\gm,h)~.
\end{equation}
Hence, despite the fact that $\XX$ will generically generate a non-closed curve in $T$, there is nevertheless a torus action by hyperk\"ahler isometries on $\MM_0$, generated by\footnote{This subspace is denoted $\mathfrak{t}_{\gm}^\perp$ in \cite{Moore:2015szp}.}
\begin{equation}\label{t0subspace}
\mathfrak{t}_0 := \{ h \in \mathfrak{t} ~|~ (\gm,h) = 0 \} \subset \mathfrak{t}~.
\end{equation}
This is the case if $\rnk{\mathfrak{g}} > 1$; if $\rnk{\mathfrak{g}} = 1$ then $\MM_0$ does not have any continuous family of hyperk\"ahler isometries.

One can introduce globally defined coordinates $\{ \vec{X},\chi \} \in \mathbb{R}^4$ on the center-of-mass factor of $\widetilde{\MM}$.  $\vec{X}$ is the position of the center of mass of the monopole configuration in $\mathbb{R}^3$ and $\chi$ is conveniently defined by the condition that $ds^2(\rG(\XX), \pd_{\chi}) = 1$.  In terms of these the metric takes the form
\begin{equation}\label{metproduct}
ds^2 = (\gm, \XX) d\vec{X}^2 + \frac{1}{(\gm,\XX)} d\chi^2 + ds_{0}^2~,
\end{equation}
where $ds_{0}^2$ is the hyperk\"ahler metric on $\MM_0$.  Explicit metrics on $\MM_0$ are known in special cases, but in order to make progress in general we turn to the asymptotic approximation of Gibbons--Manton \cite{Gibbons:1995yw} and Lee--Weinberg--Yi \cite{Lee:1996kz}.

\subsection{The GM/LWY asymptotic metric}

Let $\hi,\hj,\ldots = 1,\ldots, N$, with $N$ the total number of fundamental constituents.  (Indices $i,j$ will be reserved for a different range below.)  The GM/LWY metric on $\MM$ takes the form
\begin{equation}\label{eq:LWYMet}
ds^2 = M_{\hi\hj} d\vec{x}^{\,\hi} \cdot d\vec{x}^{\,\hj} + (M^{-1})^{\hi\hj} \Theta_{\hi} \Theta_{\hj}~,
\end{equation}
where
\begin{equation}
\Theta_{\hi} = d\xi_{\hi} + \sum_{\{\hj | \hj\neq \hi\}}\vec{W}_{\hi\hj} \cdot d\vec{x}^{\,\hj} ~,
\end{equation}
with
\begin{equation}\label{MW}
M_{\hi\hj} = \left\{ \begin{array}{l l} m_{\hi} - \sum_{\hk\neq \hi} \frac{D_{\hi\hk}}{r_{\hi\hk}} ~, & \hi = \hj \\ \frac{D_{\hi\hj}}{r_{\hi\hj}}~, & \hi \neq \hj~, \end{array} \right.  \quad \& \quad \vec{W}_{\hi\hj} = \left\{ \begin{array}{l l} -\sum_{\hk \neq \hi} D_{\hi\hk} \vec{w}_{\hi\hk}~, & \hi = \hj ~, \\ D_{\hi\hj} \vec{w}_{\hi\hj}~, & \hi \neq \hj \end{array} \right. ~.
\end{equation}
Here $\{ \vec{x}^{\,\hi},\xi_{\hi} \}$ are the spatial location and phase of the $\hi^{th}$ monopole and $\{ r_{\hi\hj},\theta_{\hi\hj},\phi_{\hi\hj} \}$ are standard spherical coordinates on $\mathbb{R}^3$ with radial vector $\vec{r}^{\,\hi\hj} := \vec{x}^{\,\hi} - \vec{x}^{\,\hj}$. $\vec{w}_{\hi\hj}$ is the Dirac potential in terms of the relative coordinates $\vec{r}^{\,\hi\hj}$ which is of the form
\begin{equation}\label{Diracmono}
\vec{w}_{\hi\hj} \cdot d\vec{r}^{\,\hi\hj} = \half (\pm 1 - \cos{\theta_{\hi\hj}}) d\phi_{\hi\hj}~.
\end{equation}
The $\xi_{\hi}$ are angular coordinates of periodicity $2\pi \cp^{\hi}$, where $\cp^{\hi}= 2/\alpha_{I(\hi)}^2$ is the ratio of the length-squared of the long root to that of the root associated with monopole $\hi$.  $I(\hi)$ is the type of monopole $\hi$. Note that the term $\sum_{\hj}\vec{W}_{\hi\hj}\cdot d\vec{x}^{\,\hj}$ can be rewritten in the form
\be
\sum_{\{\hj|\hj \neq \hi\}} \vec{W}_{\hi\hj}\cdot d\vec{x}^{\,\hj}=\sum_{\{\hi,\hj |\hj\neq \hi\}}\frac{D_{\hi\hj}}{2}(\pm 1-\cos(\theta_{\hi\hj}))d\phi_{\hi\hj}~.
\ee
The mass and coupling parameters in the above formulae are
\begin{equation}\label{constituentmasses}
m_{\hi} = (H_{I(\hi)}, \XX)~, \qquad D_{\hi\hj} = (H_{I(\hi)}, H_{I(\hj)})~,
\end{equation}
and the total magnetic charge is $\gm = \sum_{\hi} H_{I(\hi)}$.  The $m_{\hi}$, or more precisely $4\pi m_{\hi}/g_{0}^2$ with $g_0$ the bare YM coupling, are the classical masses of fundamental monopoles with magnetic charges $H_{I(\hi)}$.  By assumption $\XX$ sits in the fundamental Weyl chamber and thus the $m_{\hi}$ are positive.

The GM/LWY metric is hyperk\"ahler, with the triplet of K\"ahler forms given by \cite{MR953820,Gibbons:1995yw}
\be\label{hkstructure}
\omega^\alpha = \Theta_{\hi} \wedge d x^{\alpha \hi} - \frac{1}{2} M_{\hi\hj} \epsilon^{\alpha}_{\phantom{\alpha}\beta\gamma} dx^{\beta \hi} \wedge dx^{\gamma \hj} ~,\qquad \alpha,\beta,\gamma=1,2,3~.
\ee

This metric is a good approximation to the exact metric in the asymptotic region where all constituents are well separated relative the the scale of the lightest $W$-boson: $r_{\hi\hj} \gg \max_{\hk} \{ m_{\hk}^{-1}\}$, $\forall \hi\neq \hj$.  In fact for $\mathrm{SU}$ gauge groups it is known to be exponentially close to the exact metric with corrections of order $e^{-m_{\hi} r_{\hi\hj}}$ for those $\hi,\hj$ such that $I(\hi) = I(\hj)$, \cite{Bielawski:1998hj,Bielawski:1998hk}.  In other words there are corrections for pairs of constituents of the same type, but not for pairs of constituents of different types.  In particular the GM/LWY metric is exact when all $n_{\rm m}^I = 1$, a result conjectured in \cite{Lee:1996kz} and proven (for $\mathrm{SU}$ gauge groups) in \cite{Murray:1996hi}.

The space we have described above is the total space of an $N$-torus bundle over $\mathbb{R}^{3N} \setminus \Delta$, where $\Delta$ is the union of all hyperplanes defined by $r_{\hat{i}\hat{j}} = 0$.  The collection of $\Theta_{\hat{i}}$ provides a connection one-form on this bundle.  In order to obtain a space that is in 1:1 correspondence with the asymptotic region of monopole moduli space we must identify points under the action of the discrete group \cite{Gibbons:1995yw,Bielawski:1998hj,Bielawski:1998hk}
\begin{equation}\label{exchangesym}
S := S_{n_{\rm m}^1} \times \cdots \times S_{n_{\rm m}^r}~,
\end{equation}
where the $I^{\rm th}$ factor is a symmetric group corresponding to $n_{\rm m}^I$ identical particles of type $I$.  This group acts in the obvious way, by exchanging position and phase coordinates of monopoles of the same type.  These are hyperk\"ahler isometries of the GM/LWY metric.

The GM/LWY torus bundle is thus a discrete cover of the asymptotic region of monopole moduli space.  We note that this cover is not directly related to the cover constructed in terms of the strongly centered space described around \eqref{eq:topofactor}.  If one is working on the GM/LWY torus bundle, then one only needs to ensure that quantities (1) are invariant under the group action by $S$ and (2) respect the periodicities of the $\xi_{\hi}$, in order to have them well defined on monopole moduli space.  BPS states, however, are defined in terms of $L^2$ spinors on the strongly centered space, so it is of course important to understand how $\MM_0$ emerges in the GM/LWY picture.  We will come to this in due course.    

\subsection{The two-galaxy region}

Following our program, we will be focusing on a subregion of this asymptotic region which we call the \emph{two-galaxy region}. This limit can be taken as follows.  Partition the $\vec{x}^{\,\hi}$ into two sets, ${\rm S}_1$ and ${\rm S}_2$, representing the two distinct galaxies of size $N_1$ and $N_2$:
\begin{equation}\label{2gsplit}
\{ \vec{x}^{\,\hi} \}_{\hi=1}^N = {\rm S}_1 \cup {\rm S}_2~, \qquad {\rm S}_1 \cap {\rm S}_2 = \varnothing ~.
\end{equation}
Without loss of generality we label the $\vec{x}^{\,\hi}$ so that ${\rm S}_1$ corresponds to the first $N_1$ values of $\hi$:
\begin{equation}
{\rm S}_1 = \{\vec{x}^{\,\ha}\}_{\ha=1}^{N_1}~, \qquad {\rm S}_2 = \{ \vec{x}^{\,\hp} \}_{\hp=N_1 + 1}^{N}~,
\end{equation}
with $N_1 + N_2 = N$, such that $\min_{\ha,\hp} \{r_{\ha\hp} \} \gg \max\{ \max_{\ha,\hb} \{r_{\ha\hb} \}, \max_{\hp,\hq} \{r_{\hp\hq} \} \}$.  We will additionally use
\begin{equation}
\gamma_{1,{\rm m}} = \sum_{\ha=1}^{N_1} H_{I(\ha)}~, \qquad \gamma_{2,{\rm m}} = \sum_{\hp = N_1 + 1}^N H_{I(\hp)}~,
\end{equation}
to denote the total magnetic charge of each galaxy.  Indices $\ha,\hb,\ldots$ will always take values starting at 1 while indices $\hp,\hq,\ldots$ will take values starting at $N_1 + 1$.

Since we are examining the metric in the limit of large separation, we expand in $1/R$, where $R$ is the distance between the center of masses of the two galaxies.  To leading order, $r_{\ha\hp} \sim R$, $\forall \ha,\hp$. In order to determine both the spectrum of boundstates and the associated walls of marginal stability, we will (minimally) need to work to the lowest nontrivial order,  $O(1/R)$, so that the two galaxies are bound by an effective force.

We introduce center-of-mass and relative positions within each galaxy,
\begin{align}\label{cmrelcovg1}
& \vec{X}_1 = \frac{\sum_{\ha} m_{\ha} \vec{x}^{\,\ha} }{m_{\rm gal1}}~, \qquad \vec{y}^{\, a} = \vec{x}^{\,a} - \vec{x}^{\,a+1}~,~ a = 1,\ldots, N_1-1 ~, \cr
& \vec{X}_2 = \frac{\sum_{\hp} m_{\hp} \vec{x}^{\,\hp}}{m_{\rm gal2}}~,\qquad \vec{y}^{\,p} = \vec{x}^{\,p+1} - \vec{x}^{\,p+2}~, ~ p = N_1,\ldots, N -2~,
\end{align}
where $m_{\rm gal1} := \sum_{\ha} m_{\ha} = (\gamma_{1,{\rm m}},\XX)$ is the mass associated with galaxy 1, \etc.  The indices $a,b$ and $p,q$ run over the relative coordinates within galaxies 1 and 2 respectively, and we've built in a shift in the numerical values that $p,q$ run over so that these coordinates can be grouped together, 
\begin{equation}
\vec{y}^{\,i} = (\vec{y}^{\,a},\vec{y}^{\,p})~, \qquad i,j = 1,\ldots,N-2~,
\end{equation}
as will be convenient below.  The inverse transformations to \eqref{cmrelcovg1} are denoted
\begin{equation}\label{cmrelcovin}
(\vec{x}^{\,\hat{a}}) = {\bf J}_1 \left( \begin{array}{c} \vec{y}^{\,a} \\ \vec{X}_1 \end{array} \right)~, \qquad (\vec{x}^{\,\hat{s}}) = {\bf J}_2 \left( \begin{array}{c} \vec{y}^{\,s} \\ \vec{X}_2 \end{array} \right)~.
\end{equation}
There is a corresponding change of phase variables given by
\begin{equation}\label{fibercov}
(\xi_{\ha}) = ({\bf J}_{1}^T)^{-1} \left( \begin{array}{c} \psi_a \\ \chi_1 \end{array} \right)~, \qquad (\xi_{\hp}) = ({\bf J}_{2}^T)^{-1} \left( \begin{array}{c} \psi_{p} \\ \chi_2 \end{array} \right)~,
\end{equation}
and we will denote by $\psi_i = \{\psi_a,\psi_p\}$ the collection of relative phases.  See Appendix \ref{app:A} for further details including the explicit form of the matrices ${\bf J}_{1,2}$.

Let us consider, for the moment, galaxy one in isolation.  There is an associated moduli space $\MM_1 := \MM(\gmone,\XX)$.  The coordinates $\{\vec{y}^{\, a},\psi_a\}$ parameterize the asymptotic region of the strongly centered space $\MM_{1,0} := \MM_0(\gmone,\XX)$, while $\{\vec{X}_1,\chi_1\}$ parameterize the $\mathbb{R}^4$ of the universal cover $\widetilde{\MM}_1 = \mathbb{R}_{(1)}^4 \times \MM_{1,0}$.  There is an exchange symmetry, $S_1$, which is a product of symmetric groups for each type of monopole.  As we describe in Appendix \ref{app:deck}, the exchange symmetry acts on $\MM_{1,0}$ only, while $\{\vec{X}_1,\chi_1\}$ are $S_1$ invariants.  We also describe there how the quotient by the group of deck transformations, $\MM_1 = \widetilde{\MM}_1/\mathbb{D}_1$ follows from the periodicities of the individual $\xi_{\ha}$.

Similarly, we can associate a moduli space $\MM_2 := \MM(\gmtwo,\XX)$ to galaxy two.  The coordinates $\{\vec{y}^{\, p}, \psi_p \}$ parameterize the strongly centered space $\MM_{2,0} := \MM_0(\gmtwo,\XX)$, and $\{\vec{X}_2,\chi_2\}$ parameterize an $\mathbb{R}_{(2)}^4$.

Now we return to the full picture where these two galaxies are interacting with each other.  Using the center-of-mass coordinates for each galaxy we can construct the overall center-of-mass coordinates $\{\vec{X},\chi \}$ introduced earlier and the relative-galaxy coordinates $\{\vec{R},\uppsi \}$ as follows: 
\begin{align}\label{globalcom}\begin{split}
& \vec{X}=\frac{m_{\rm gal1}\vec{X}_1+m_{\rm gal2}\vec{X}_2}{m_{\rm gal1}+m_{\rm gal2}} ~,\qquad \vec{R}=\vec{X}_1-\vec{X}_2~,\\
& \chi=\chi_1+\chi_2 ~, \qquad  \uppsi= \frac{ m_{\rm gal2} \chi_1 - m_{\rm gal1} \chi_2}{m_{\rm gal1} + m_{\rm gal 2}} ~.
\end{split}\end{align}
Then it is the collection of position coordinates $\{\vec{y}^{\,i},\vec{R}\}$ and phase coordinates $\{\psi_i,\uppsi\}$ that parameterize the strongly centered space $\MM_0(\gm,\XX)$ for the whole system.  

We implement these transformations on \eqref{eq:LWYMet} and expand the result in $1/R$, with $R = |\vec{R}|$.  After some computation we find a metric of the form \eqref{metproduct} with the strongly centered piece given by
\begin{align}\label{hatmetric}
ds_{0}^2 =&~ \left( d\vec{{\bf y}}^T, d\vec{R} \right) \left( \begin{array}{c c} \wtbC + \frac{1}{R} \delta {\bf C} & \frac{1}{R} {\bf L} \\  \frac{1}{R} {\bf L}^T & \mu H(R) \end{array} \right) \left( \begin{array}{c} \cdot d\vec{{\bf y}} \\ \cdot d\vec{R} \end{array} \right) + \cr
&~ +  \left( \Theta_0, \Theta_\uppsi \right) \left( \begin{array}{c c} \wtbC + \frac{1}{R} \delta\mathbf{ C} & \frac{1}{R} {\bf L} \\  \frac{1}{R} {\bf L}^T & \mu H(R) \end{array} \right)^{-1} \left( \begin{array}{c} \Theta_0 \\ \Theta_{\uppsi} \end{array} \right) + \cdots~, \cr
=: &~ d\hat{s}_{0}^2 + \cdots~,
\end{align}
where
\be\label{reducedmass}
\mu=\frac{m_{\rm gal1} m_{\rm gal2}}{m_{\rm gal1}+m_{\rm gal2}}~,
\ee
is the reduced mass of the two-galaxy system.  Note that $m_{\rm gal1} + m_{\rm gal2} = (\gm,\XX)$ is the total mass appearing in the center-of-mass factor of the metric, \eqref{metproduct}.

Here we have introduced a length $N-2$ column vector $\vec{{\bf y}}$ whose components are the $\vec{y}^{\, i}$.  We have also introduced a corresponding collection of connection one-forms $\Theta_0 = (\Theta_a, \Theta_p)^T$ given, along with $\Theta_{\uppsi}$, by
\begin{equation}\label{Thetarels1}
\left(\begin{array}{c} \Theta_0 \\ \Theta_\uppsi \end{array} \right) = \left( \begin{array}{c} \wtTheta_0 \\ d\uppsi \end{array} \right) + \left( \begin{array}{c c} \delta {\bf C} & {\bf L} \\ {\bf L}^T & - (\gamma_{1,{\rm m}}, \gamma_{2,{\rm m}}) \end{array} \right)\otimes \vec{w}(\vec{R}) \cdot \left( \begin{array}{c} d \vec{{\bf y}} \\ d \vec{R} \end{array} \right)~,
\end{equation}
where
$\wtTheta_0 = (\wtTheta_a, \wtTheta_p)^T$ with
\begin{equation}\label{Thetarels2}
\wtTheta_a = d\psi_a + (\vec{\WW}_1)_{ab} \cdot d\vec{y}^{\,b} ~, \qquad \wtTheta_p = d\psi_p + (\vec{\WW}_2)_{pq} \cdot d\vec{y}^{\,q} ~.
\end{equation}
The matrix $\wtbC$ is block-diagonal with respect to the two-galaxy structure and $H(R)$ is a harmonic function the $\mathbb{R}^3$ parameterized by $\vec{R}$:
\be
H(R)=\left(1-\frac{(\gamma_{1,m},\gamma_{2,m})}{\mu R}\right)\quad,\qquad \wtbC = \left(\begin{array}{cc}(C_1)_{ab}&0\\0&(C_2)_{pq}\end{array}\right)~.
\ee
Finally the matrix $\delta{\bf C}$ and column vector ${\bf L}$ are constants, depending on the Higgs data (\ie\ masses) and magnetic charges.  Detailed expressions can be found in the appendix.

The quantities $(C_1)_{ab}$ and $(\vec{\WW}_1)_{ab}$ depend on the $\vec{y}^{\,a}$ only and are precisely the data that one would use to construct the GM/LWY asymptotic metric on the strongly centered moduli space of galaxy one in isolation.  This metric is $(C_1)_{ab} d\vec{y}^{\,a} \cdot d\vec{y}^{\,b} + (C_{1}^{-1})^{ab} \wtTheta_a \wtTheta_b$.  Likewise, $C_2,\vec{\WW}_2$ depend on the $\vec{y}^{\,p}$ only and give the metric on the strongly centered moduli space for galaxy two in isolation in the same way.  At leading order in $1/R$ the metric \eqref{hatmetric} on the full strongly centered moduli space reduces to a direct sum of these two metrics with a flat metric on $\mathbb{R}^4$ parameterized by $(\vec{R},\uppsi)$:
\begin{equation}\label{leadsplitting}
\MM_0 ~ \xrightarrow[R \to \infty]{} ~\MM_{1,0} \times \MM_{2,0} \times \mathbb{R}_{\rm rel}^4~.
\end{equation}

The terms proportional to $\delta{\bf C}, {\bf L}$, and $(\gamma_{1,{\rm m}}, \gamma_{2,{\rm m}})$ encode the first $O(1/R)$ corrections.  Specifically $\delta{\bf C}$ takes into account $1/R$ corrections to the strongly centered metrics within each galaxy and also the leading couplings between the two strongly centered metrics.  ${\bf L}$ encodes the leading couplings of the strongly centered metrics to the galaxy-relative coordinates $(\vec{R},\uppsi)$.  It will play a crucial role in the following and has components ${\bf L} = (L_a, L_p)^T$, with
\begin{equation}\label{Lcoefs}
L_a = - \langle \beta_a, \gmtwo \rangle~, \qquad L_p = \langle \beta_p, \gmone \rangle~,
\end{equation}
where the $\beta$ are certain linear combinations of the duals of the constituent charges $H_{I(\hat{a})}, H_{I(\hat{p})}$ determined by the same linear transformations appearing in \eqref{cmrelcovin}.  (See \eqref{beta1def}, \eqref{beta2def}.)  Finally the $O(1/R)$ terms proportional to $(\gamma_{1,{\rm m}}, \gamma_{2,{\rm m}})$, coming from both $H(R)$ and $\vec{w}(\vec{R})$ --- which is itself $O(1/R)$ --- give the first corrections to the flat metric on the relative $\mathbb{R}_{\rm rel}^4$.  The corrected metric is consistent with the Taub--NUT metric at $O(1/R)$.

The ellipses in \eqref{hatmetric} denote higher order terms in the expansion.  Strictly speaking, the second line should be  expanded and only terms through $O(1/R)$ kept, but we find it more convenient to work directly with the metric $d\hat{s}_{0}^2$ in \eqref{hatmetric}.  This is acceptable provided we remember that computations with it should only be trusted through $O(1/R)$.

A well known feature of the spaces we are approximating is that they are hyperk\"ahler. We should then expect that the asymptotic metric we are dealing with is hyperk\"ahler to the appropriate order in $1/R$.  This is confirmed in Appendix \ref{app:A}.

Once we truncate the metric on the strongly centered space to $d \hat{s}_{0}^2$, we break the exchange symmetry \eqref{exchangesym} to a subgroup,
\begin{equation}
S \longrightarrow S_1 \times S_2~,
\end{equation}
where $S_{1}(S_2)$ is the exchange symmetry corresponding to galaxy $1(2)$ in isolation and acts only on $\MM_{1,0}(\MM_{2,0})$.  If the two galaxies have equal numbers of constituents of each type then an additional $\mathbb{Z}_2$ corresponding to exchanging the galaxies will be preserved.

\rmk~  Based on the techniques developed in \cite{Kottke:2015rwx}, it is expected\footnote{ABR thanks M.~Singer for communication on this point.} that there should exist a coordinate system on the asymptotic region of moduli space in which the off-diagonal blocks of the metric can be eliminated at $O(1/R)$.  We can indeed remove these off-diagonal terms by making the further coordinate transformation $\{\psi_i,\vec{R}\} \mapsto \{\sigma_i,\vec{\RR}\}$ given by
\be\label{eq:coordtrans}
\psi_{i}=\sigma_{i} +\frac{L_{i}}{\mu \RR}\uppsi ~,\qquad \vec{R}= \vec{\RR}-\frac{{\bf L}^T \vec{{\bf y}}}{\RR}~.
\ee
%

\subsection{Triholomorphic killing vectors and the electric charge operator}

In order to study the BPS spectrum we will need to understand the the relationship between triholomorphic Killing vectors of the GM/LWY metric and the $r=\rnk{\mathfrak{g}}$ triholomorphic Killing vectors of the exact metric, $\rG(h^I)$, that generate $2\pi$-periodic isometries.

It is clear from \eqref{eq:LWYMet} and \eqref{hkstructure} that
\be
\pounds_{\partial_{\xi_{\hi}}} \omega^\alpha=0 = \pounds_{\partial_{\xi_{\hi}}} ds^2~,
\ee
and hence the $\pd_{\xi_{\hi}}$ are triholomorphic Killing fields.  This means that in the asymptotic region the hyperk\"ahler action of the Cartan torus $T\subset G$ is enhanced to $U(1)^N$.    Since the $\{ \pd_{\psi_a},\pd_{\psi_p},\pd_{\uppsi},\pd_{\chi} \}$ are just linear combinations of the $\pd_{\xi_{\hi}}$, they are triholomorphic Killing fields as well.  However, only $r$ linear combinations of the $\pd_{\xi_{\hi}}$ will extend to triholomorphic Killing vectors of the exact metric.  This is described in Appendix \ref{app:deck} with the result that
\begin{equation}\label{exactTKVlimit}
\rG(h^I) \longrightarrow \cp^I \sum_{\hi_I=1}^{n_{\rm m}^I} \frac{\pd}{\pd \xi_{{\hi}_I}}~,
\end{equation}
exponentially fast in the asymptotic region, where $\xi_{\hi_I}$ are the phases associated with constituents of type $I$.

An important application of these hyperholomorphic isometries is the construction of the semiclassical electric charge operator $\hat{\gamma}^{\rm e}$.  The electric charge operator takes the form
\begin{equation}
\hat{\gamma}^{\rm e} = \I \sum_{I=1}^{\rnk{\mathfrak{g}}} \alpha_I \Lie_{\rG(h^I)} ~,
\end{equation}
in the collective coordinate quantization.  This operator commutes with the supercharge operator and hence states $\Psi$ can be labeled by eigenvalues $\gamma^{\rm e}$ of $\hat{\gamma}^{\rm e}$, which sit in the root lattice $\gamma^{\rm e} \in \Lambda_{\rm rt} \subset \mathfrak{t}^\ast$ thanks to the periodicity of the isometries generated by the $\rG(h^I)$.

We can decompose $\gamma^{\rm e}$ into a component parallel to $\gamma_{\rm m}^\ast$ and a component $\gamma^{\rm e}_0$ that has zero pairing with $\XX$.  The component parallel to $\gamma_{\rm m}^\ast$ is denoted $q$:
\begin{equation}
\gamma^{\rm e} = q \gamma_{\rm m}^\ast + \gamma_{0}^{\rm e} ~, \qquad \textrm{such that} \quad q = \frac{\langle \gamma^{\rm e},\XX\rangle}{(\gm, \XX)} ~, \quad \langle \gamma_{0}^{\rm e}, \XX\rangle = 0~.
\end{equation}
The projection operator onto the subspace $\mathfrak{t}_{0}^\ast = \{ \beta ~|~ \langle \beta, \XX \rangle =0 \} \subset \mathfrak{t}^\ast$ is in fact dual to the projection used in \eqref{t0subspace}, \cite{Moore:2015qyu}.

All of these quantities can be expressed in terms of the triholomorphic Killing vectors of the GM/LWY metric.  First using \eqref{exactTKVlimit} we have
\begin{equation}
\hat{\gamma}^{\rm e} \to  \I \sum_{I} \alpha_I \cp^I \sum_{\hi_I = 1}^{n_{\rm m}^I} \Lie_{\pd_{\xi_{\hi_I}^I}} = \I \sum_{\hi = 1}^N H_{I(\hi)}^\ast \Lie_{\pd_{\xi_{\hi}}} ~.
\end{equation}
From here it is easy to construct operators measuring the total electric charge in each galaxy:
\begin{equation}
\hat{\gamma}_{1}^{\rm e} := \I \sum_{\ha = 1}^{N_1} H_{I(\ha)}^\ast \Lie_{\pd_{\xi_{\ha}}} ~, \qquad \hat{\gamma}_{2}^{\rm e} := \I \sum_{\mathclap{\hp = N_1 + 1}}^{N} H_{I(\hp)}^\ast \Lie_{\pd_{\xi_{\hp}}} ~.
\end{equation}
Each of these can be broken into center of mass and relative pieces by changing the basis of Killing fields and utilizing the definition of the $\beta_a, \beta_s$ that appeared previously in \eqref{Lcoefs}\footnote{The steps to deriving this are analogous to the ones in \eqref{Gdecomp1} and \eqref{Gtwistpsis}.}:
\begin{equation}
\hat{\gamma}_{1}^{\rm e} = \I \gmone^\ast \Lie_{\pd_{\chi_1}} +\I \sum_{a=1}^{N_1 -1} \beta_{a} \Lie_{\pd_{\psi_a}} ~, \qquad \hat{\gamma}_{2}^{\rm e} = \I \gmtwo^\ast \Lie_{\pd_{\chi_2}} +\I \sum_{p=N_1}^{N -2} \beta_{p} \Lie_{\pd_{\psi_p}} ~.
\end{equation}
Finally, changing variables $\{\chi_1,\chi_2\}\mapsto\{\chi,\uppsi\}$ with \eqref{globalcom} we also express these as
\begin{align}\label{echargeconstituents}
\hat{\gamma}_{1}^{\rm e} =&~ \I \gamma_{{\rm m},1}^\ast \left(\pd_\chi + \frac{m_{\rm gal2}}{m_{\rm gal1} +m_{\rm gal2}} \pd_{\uppsi} \right) + \I \sum_{a=1}^{N_1 -1} \beta_{a} \Lie_{\pd_{\psi_a}} ~, \cr
\hat{\gamma}_{2}^{\rm e} =&~  \I \gamma_{{\rm m},2}^\ast \left(\pd_\chi - \frac{m_{\rm gal1}}{m_{\rm gal1} +m_{\rm gal2}} \pd_{\uppsi} \right) + \I \sum_{p=N_1}^{N -2} \beta_{p} \Lie_{\pd_{\psi_p}}~, 
\end{align}
and recombining them gives $\hat{\gamma}^{\rm e} = \I \gm^\ast \pd_{\chi} + \hat{\gamma}_{0}^{\rm e}$, with the relative charge operator
\begin{equation}\label{relechargeop}
\hat{\gamma}_{0}^{\rm e} :=  \I \frac{ m_{\rm gal2} \gamma_{{\rm m},1}^\ast - m_{\rm gal1} \gamma_{{\rm m},2}^\ast }{m_{\rm gal1} + m_{\rm gal 2}} \pd_{\uppsi} + \I \sum_{i = 1}^{N-2} \beta_{i} \Lie_{\pd_{\psi_i}} ~,
\end{equation}
whose eigenvalues are $\gamma_{0}^{\rm e}$.

\section{The Asymptotic Dirac Operator}\label{sec:AsymptoticD}

Having described the metric in the two-galaxy region, our next goal is to construct the relevant Dirac operators.  Specifically, in the case we focus on initially --- pure $\NN = 2$ SYM theory without defects ---  collective coordinate quantization leads to a SQM on the Dirac spinor bundle over the moduli space $\MM(\gm,\XX)$.  The bundle factorizes into a $\mathbb{C}^4$ factor tensored with the Dirac spinor bundle over the strongly centered moduli space, $\MM_0(\gm, \CX)$.

Supersymmetry then dictates that BPS states are represented by sections of the Dirac spinor bundle that are in the kernel of any (and hence all) of the supercharge operators. Let $\YY_0 \in \mathfrak{t}_0 \subset \mathfrak{t}$, \eqref{t0subspace}, such that $(\gm, \YY_0) = 0$.  One finds that BPS spinors $\Psi$ should be of the form $e^{-i q \chi} s \otimes \Psi_0$, where $s \in \mathbb{C}^4$ is a constant, and $\Psi_0$ is an $L^2$ section of the Dirac spinor bundle over $\MM_0$ which is annihilated by the Dirac-type operator
\be\label{eq:D}
\slashed{D}^{\CY_0}= \slashed{D}-\I \slashed{\rG}(\CY_0)~.
\ee
$\slashed{D}$ is the ordinary Dirac operator on $\MM_0$, and $\slashed{\rG}(\CY_0)$ is the Clifford contraction of the triholomorphic Killing field associated with a nontrivial gauge transformation that asymptotes to $\YY_0$.  By \eqref{killing2metric}, $(\gm,\YY_0) = 0$ ensures that $\rG(\YY_0)$ is metric-orthogonal to $\rG(\XX)$ and hence restricts to $\MM_0$.

The $L^2$ kernel of \eqref{eq:D} can be graded by the eigenvalues of $\hat{\gamma}_{0}^{\rm e}$.  Let us denote a wavefunction in the $\gamma_{0}^{\rm e}$-subspace of the $L^2$ kernel of \eqref{eq:D} by $\Psi_0^{(\gamma_{0}^{\rm e})}$.  If the dual of the magnetic charge is a primitive element of the root lattice, then $e^{-i q \chi} s \otimes \Psi_0^{(\gamma_{0}^{\rm e})}$ represents a BPS state of electric charge $\gamma^{\rm e} = q \gamma_{\rm m}^\ast + \gamma_{0}^{\rm e}$.  If, however, $\gm^\ast$ is not primitive then to get a physical state we must in general impose a $\mathbb{Z}_{\ell}$ equivariance condition on $\Psi_0$ with respect to the (lift to the spinor bundle of) the action of the isometry $\phi$ generating the group of deck transformations \cite{Moore:2015qyu}.  This ensures that the spinor we've constructed on $\mathbb{R}^4 \times \MM_0$ descends to one on $\MM$.  However in the GM/LWY asymptotic description, the quotient by deck transformations is associated with the periodicities of the constituent phases.  See the discussion in Appendix \ref{app:deck}.  Any wavefunction that respects those periodicities will automatically satisfy the required equivariance condition.  Hence we will not need to worry about this point in the following. 

The physical data of the $\NN = 2$ theory that labels BPS states involves a point $u$ on the Coulomb branch and an element $\gamma$ of the electromagnetic charge lattice over $u$.  The following relationship between the $\NN = 2$ data and the Dirac operator/moduli space data was conjectured in \cite{Moore:2015szp}, based on a detailed collective coordinate analysis of the $\NN = 2$ theory including some one-loop effects.  First, we define $\YY \in \mathfrak{t}$ by
\begin{equation}
\YY := \YY_0 - \frac{\langle \gamma^{\rm e}, \XX \rangle}{(\gm,\XX)} \XX = \YY_0 - q \XX~,
\end{equation}
and then we have
\begin{align}\label{SWscmap}
& \XX = \Im\left[ \zeta^{-1}_{\rm van} a(u) \right]~, \qquad \YY = \Im\left[ \zeta_{\rm van}^{-1} a_{\rm D}(u) \right] ~, \qquad \gm \oplus \gamma^{\rm e} = \gamma ~,
\end{align}
where $\zeta_{\rm van} := - Z_\gamma(u)/|Z_\gamma(u)|$ is the phase of negative of the central charge.  Note that the choice of electromagnetic duality frame, which induces the splitting of the charge lattice $\gamma \in \Gamma = \Gamma_{\rm m} \oplus \Gamma_{\rm e} \subset \Lambda_{\rm mw} \oplus \Lambda_{\rm wt}$, is specified by the condition that $\XX$ be in the fundamental Weyl chamber of $\mathfrak{t}$.  Then the symplectic pairing on $\Gamma$ is given in terms of the canonical pairing on $\mathfrak{t} \times \mathfrak{t}^\ast$ by
\begin{equation}
\llangle \gamma_1, \gamma_2 \rrangle = \llangle \gamma_{1,{\rm m}} \oplus \gamma_{1}^{\rm e}, \gamma_{2,{\rm m}} \oplus \gamma_{2}^{\rm e} \rrangle = \langle \gamma_{1}^{\rm e}, \gamma_{2,{\rm m}}\rangle - \langle \gamma_{2}^{\rm e}, \gamma_{1,{\rm m}} \rangle~.
\end{equation}
Meanwhile $a(u)$ and $a_{\rm D}(u)$ are the special and dual special coordinates on the Coulomb branch, which are given by period integrals of $\lambda_{\rm SW}$ on the Seiberg--Witten curve \cite{Seiberg:1994rs,Seiberg:1994aj}.

For generic values of the Higgs data, $\{\XX,\YY_0\}$, the operator \eqref{eq:D} acting on the space of $L^2$-spinors on $\MM_0$ is expected to be Fredholm.  From the mathematical viewpoint, this statement and the ones in the remainder of the paragraph are conjectures, but they are physically well-grounded and we will adopt them.  In physical terms, there will be a finite-dimensional space of $L^2$-normalizable zeromodes for a given electric charge, and a positive gap to the continuum of scattering states. The wall-crossing phenomena we are interested in occurs when the parameters are such that the operator actually fails to be Fredholm. For special Higgs data lying on a real co-dimension one wall in the space of $\{\XX,\YY_0\}$, the gap to the continuous spectrum vanishes, while the $L^2$ boundstates fail to be normalizable and mix with the scattering states. 

Since the monopole moduli space is smooth, the only source of a failure of the Fredholm property is expected to come from the behavior of the differential operator in the asymptotic regions of moduli space -- if the moduli space were compact, the spectrum of the Dirac-type operator would be discrete and there would be no wall crossing.  Since the metric simplifies considerably in the GM/LWY asymptotic region, we can -- in principle --  study the eigenvalue problem just on the spinors in this region, assuming some smooth, but unspecified extension into the interior. Then we can, in principle, derive the gap to the continuous spectrum and also see whether there are exponentially decaying eigenfunctions. In practice, this program is still too difficult to carry out analytically, but below we show how it can be carried out in the two-galaxy region.  We conjecture that this subregion, which corresponds to a partition of the total number of constituents into two clusters, is the one responsible for the leading contribution to the wall-crossing formula -- namely the primitive wall-crossing formula.

\subsection{The Dirac operator in the two-galaxy region}

We are now tasked with computing the asymptotic form of the twisted Dirac operator, \eqref{eq:D}, in the two-galaxy region of $\MM_0$.  The details are outlined in Appendix \ref{app:B} and here we summarize the result, which can be cast in the form
\be\label{eq:SimpleDiracOperator}
\Dsl^{\CY_0} = \Lambda\left( \Dsl_{12}^{\CY_0} + \Dsl_{\rm rel}^{\CY_0} + O(1/R^2) \right)\Lambda^{-1}~.
\ee
Here $\Lambda$ is the lift of a local frame rotation to the Dirac spinor bundle; it approaches the identity as $R \to \infty$ but deviates in a $\vec{{\bf y}}$-dependent way at $O(1/R)$.  The virtue of this frame rotation is that it effectively absorbs all of the $O(1/R)$ off-diagonal mixing terms of $\slashed{D}^{\YY_0}$ such that the action of the rotated operator in the parentheses is block-diagonal with respect to the asymptotic factorization $\SS(\MM_0) \to \SS(\MM_{1,0} \times \MM_{2,0}) \otimes \SS(\mathbb{R}_{\rm rel}^4)$ of the spin bundle over $\MM_0$ as $R \to \infty$.  We believe that the existence of this frame rotation is a consequence of the existence of the coordinate transformation \eqref{eq:coordtrans} that block-diagonalizes the metric at $O(1/R)$.

In order to characterize the operators $\slashed{D}_{\rm 12}^{\YY_0}$ and $\slashed{D}_{\rm rel}^{\YY_0}$, we first collect position and phase coordinates by introducing indices $\mu,\nu = 1,\ldots,4$ and writing $y^{\mu i} = \{\vec{y}^{\,i}, \psi_i\}$, $R^\mu = \{\vec{R},\uppsi\}$.  Next we let $\Gamma^{\underline{\mu i}},\Gamma^{\underline{\mu R}}$ denote the corresponding gamma matrices satisfying the Clifford algebra
\begin{equation}\label{gammamatrices}
[ \Gamma^{\underline{\mu i}}, \Gamma^{\underline{\nu j}} ]_+ = 2\delta^{\underline{\mu\nu}} \delta^{\underline{ij}}~, \qquad [ \Gamma^{\underline{\mu R}}, \Gamma^{\underline{\nu R}} ]_+ = 2 \delta^{\underline{\mu\nu}} ~, \qquad [ \Gamma^{\underline{\mu i}}, \Gamma^{\underline{\nu R}} ]_+ = 0~,
\end{equation}
where underlined indices refer to an orthonormal frame.  Then $\slashed{D}_{12}^{\YY_0}$ involves only the $\Gamma^{\underline{\mu i}}$, while $\slashed{D}_{\rm rel}^{\YY_0}$ involves only the $\Gamma^{\underline{\mu R}}$.  
All terms in $\slashed{D}^{\YY_0}$ at order $O(1/R)$ involving both types of gamma matrices are captured by the frame rotation.

As we mentioned above, all of the vector fields $\{ \pd_{\psi_a}, \pd_{\psi_p}, \pd_{\uppsi}\} = \{ \pd_{\psi_i}, \pd_{\uppsi} \}$ are triholomorphic with respect to the GM/LWY hyperk\"ahler structure on $\MM_0$.  Furthermore the corresponding Lie derivatives commute with the Dirac operator $\slashed{D}^{\YY_0}$ constructed from the GM/LWY metric, so in particular they commute with the operators appearing in the two-galaxy expansion, \eqref{eq:SimpleDiracOperator}.  Since the operators $\Lie_{\psi_i},\Lie_{\uppsi}$ comprise the electric charge operator $\hat{\gamma}_{0}^{\rm e}$, \eqref{relechargeop}, we have that our Hilbert space is graded by the electric charge. An eigenspinor of the $\{ \Lie_{\pd_{\psi_i}},\Lie_{\pd_{\uppsi}} \}$ operators is a state of definite relative electric charge,
\begin{equation}\label{LWYspinor}
\Psi_{0}^{(\gamma_{0}^{\rm e})} = \Lambda  e^{ \I \nu^i \psi_i + \I \upnu \uppsi} \Psi_{\nu^i,\upnu}(\vec{y}^{\,i}, \vec{R})~,
\end{equation}
with $\gamma_{0}^{\rm e}$ given by
\begin{equation}\label{LWYspinorcharge}
\gamma_{\rm 0}^{\rm e} = - \frac{ m_{\rm gal2} \gamma_{{\rm m},1}^\ast - m_{\rm gal1} \gamma_{{\rm m},2}^\ast }{m_{\rm gal1} + m_{\rm gal 2}} \upnu - \sum_{i = 1}^{N-2} \beta_{i}^\ast \nu^i ~.
\end{equation}
Then we can consider the restriction of our Dirac operator to a given $\{ \nu^i,\upnu\}$ eigenspace.  

Now introduce a decomposition of the gamma matrices with respect to the factorization $\SS(\MM_{1,0} \times \MM_{2,0}) \otimes \SS(\mathbb{R}_{\rm rel}^4)$:
\begin{equation}
\Gamma^{\underline{\mu i}} = \gamma^{\underline{\mu i}} \otimes \mathbbm{1}_4 ~, \qquad \Gamma^{\underline{\mu R}} = \bar{\gamma}_{12} \otimes \gamma^{\underline{\mu}}~,
\end{equation}
where $\gamma^{\underline{\mu i}}$ are gamma matrices for the $\MM_{1,0} \times \MM_{2,0}$ factor, $\gamma^{\underline{\mu}}$ are gamma matrices for the $\mathbb{R}_{\rm rel}^4$ factor, and $\bar{\gamma}_{12}$ is the chirality operator on the $\MM_{1,0} \times \MM_{2,0}$ factor.  Then
\begin{align}
\left. \slashed{D}_{12}^{\YY_0} \right|_{\{\nu^i,\upnu\}} =&~ \breve{\slashed{D}}_{12}^{\YY_0} \otimes \mathbbm{1}_4 := \left( \slashed{D}_{\MM_{1,0} \times \MM_{2,0}}^{\YY_0} + (\breve{\slashed{D}}_{12}^{\YY_0})^{(1)} \right) \otimes \mathbbm{1}_4~, \cr
\left. \slashed{D}_{\rm rel}^{\YY_0} \right|_{\{\nu^i,\upnu\}} =&~ \bar{\gamma}_{12} \otimes \breve{\slashed{D}}_{\rm rel}^{\YY_0} ~,
\end{align}
where $\slashed{D}_{\MM_{1,0} \times \MM_{2,0}}^{\YY_0}$ is the leading order term and $(\breve{\slashed{D}}_{12}^{\YY_0})^{(1)}$ is the $O(1/R)$ correction in $\breve{\slashed{D}}_{12}^{\YY_0}$.  Here $\slashed{D}_{\MM_{1,0} \times \MM_{2,0}}^{\YY_0}$ is precisely the $\rG(\YY_0)$-twisted Dirac operator on the direct product $\MM_{1,0} \times \MM_{2,0}$.  Explicit expressions can be found in Appendix \ref{app:B3}.  Meanwhile,
\begin{align}\label{DrelY0}
\breve{\slashed{D}}_{\textrm{{\rm rel}}}^{\CY_0} :=&~ \frac{1}{\sqrt{\mu}} \left( 1 + \frac{ (\gamma_{{\rm m},1}, \gamma_{{\rm m},2})}{2\mu R} \right) \bigg\{ \gamma^{\ualpha} \delta_{\ualpha}^{\phantom{\alpha}\alpha} \left[ \pd_{R^\alpha} - i p w_{\alpha} \right] - \I\gamma^{\underline{4}} \left( x - \frac{p}{2 R} \right) \bigg\} ~,
\end{align}
where $w_\alpha$ is the Dirac potential and with
\begin{equation} \label{pxparameters}
p := L_i \nu^i - (\gmone,\gmtwo) \upnu ~, \qquad x := (\gmone,\YY_0) - \mu \upnu~.
\end{equation}
%

\subsection{Zeromode asymptotics}

An important consequence of the above is that $\slashed{D}_{12}^{\YY_0}$ , $\slashed{D}_{\rm rel}^{\YY_0}$ are asymptotically anticommuting,
\begin{equation}
\left[ \slashed{D}_{12}^{\CY_0} \,, \, \slashed{D}_{\textrm{{\rm rel}}}^{\CY_0} \right]_+ = O(1/R^2)~,
\end{equation}
and hence $\slashed{D}^{\YY_0} \Psi_0 = 0$ implies
\begin{equation}\label{Diracpair}
\left( \breve{\slashed{D}}_{12}^{\YY_0} \otimes \mathbbm{1}_4 + O(1/R^2) \right) \Psi_{\nu^i,\upnu} = 0 \quad \& \quad \left( \bar{\gamma}_{12} \otimes \breve{\slashed{D}}_{\rm rel}^{\YY_0} + O(1/R^2) \right)   \Psi_{\nu^i,\upnu} = 0 ~.
\end{equation}
We look for a solution of the form\footnote{The reason it is necessary to allow $\uptheta,\upphi$ dependence in $\Psi_{12}^{(1)}$ is that $(\breve{\slashed{D}}_{12}^{\YY_0})^{(1)}$ has $\uptheta,\upphi$ dependence. Derivatives with respect to $R,\uptheta,\upphi$ in $\breve{\slashed{D}}_{\rm rel}^{\YY_0}$ will act on the $\Psi_{12}^{(1)}$ term, but these contributions will be suppressed by $O(1/R^2)$ relative to the leading terms in the equation.}
\begin{equation}\label{Psiansatz}
\Psi_{\nu^i,\upnu}(\vec{y}^{\,i},\vec{R}) = \left( \Psi_{12,\nu^i}^{(0)}(\vec{y}^{\,i}) + \frac{1}{R} \Psi_{12,\nu^i}^{(1)}(\vec{y}^{\,i};\uptheta,\upphi) + O(1/R^2) \right) \otimes \Psi_{\rm rel}^{(p)}(\vec{R})~,
\end{equation}
where $\{R,\uptheta,\upphi\}$ are spherical coordinates centered on $\vec{R} = 0$.  Note that the $\upnu$ dependence on the right-hand side of \eqref{Psiansatz} occurs through $p$, via \eqref{pxparameters}.  

The latter equation of \eqref{Diracpair} will be satisfied provided
\begin{equation}\label{Drelzeromodes}
\breve{\slashed{D}}_{\rm rel}^{\YY_0} \Psi_{\rm rel}^{(p)} = 0~,
\end{equation}
which is exactly the equation studied in Appendix C of \cite{Moore:2014jfa}.\footnote{The quantities $p,x$ were denoted $p_\mu,x_\mu$ in \cite{Moore:2014jfa}, but this would be confusing notation here.}  We review and expand on this computation in Appendix \ref{app:AppD} below.  There is an angular momentum $j = \half (|p| -1)$ multiplet of exponentially decaying solutions iff $x$ and $p$ have the same sign.  The wavefunctions take the form
\begin{equation}\label{Drelzeromodesres}
\Psi_{\rm rel}^{(p)} \propto R^{j-1/2} e^{-\sgn(p) x R} \Psi_{j,m}(\uptheta,\upphi)~.
\end{equation}
The explicit form of the angular part is not needed here, but see \eqref{BPSangular} below.

Determining the asymptotic behavior of solutions to \eqref{Drelzeromodes}, if they exist, does not require knowledge of the interior region of moduli space where our large $R$ expansion breaks down.  However, determining the above stated degeneracy of such solutions certainly does require an assumption about the interior.  Specifically, in Appendix \ref{app:AppD} we show that a certain Dirac operator on Taub-NUT space agrees with \eqref{Drelzeromodes} to the order we work in $1/R$, and then we solve the Dirac equation on Taub-NUT.  Taking \eqref{eq:SimpleDiracOperator}, dropping the $O(1/R^2)$ corrections, and replacing \eqref{DrelY0} with this Taub-NUT Dirac operator defines a model Dirac operator that agrees with the exact one, \eqref{eq:D}, asymptotically.    

One can then ask whether or not this model Dirac operator is a Fredholm deformation of the exact one.  This is the pertinent question.  On the one hand, the indices will agree if and only if this is the case.  On the other hand, the index of the exact Dirac operator is identified with the index of BPS states in \cite{Moore:2015szp}.

However the behavior of the Taub-NUT-like metric in the interior, and the boundary conditions one should impose there, depend on the sign of the parameter
\begin{equation}\label{TNell}
\ell_{\rm TN} = - \frac{(\gmone,\gmtwo)}{\mu} ~,
\end{equation}
appearing in the harmonic function $H = 1 + \ell_{\rm TN}/R$.  Our analysis leading to the $\Psi_{\rm rel}$ wavefunctions, \eqref{Drelzeromodesres}, assumes that $\ell_{\rm TN}$ is positive and imposes a simple regularity condition at $R = 0$.  If $\ell_{\rm TN} <0$, the metric is singular at $R = -\ell_{\rm TN}$ and we would instead have to impose boundary conditions there.  Without a more detailed analysis it is not clear what the appropriate boundary conditions should be.  This issue requires further investigation and henceforth we will restrict to the $\ell_{\rm TN} > 0$ (\ie\ $(\gmone,\gmtwo) < 0$) case.  

The results below will show that the wall-crossing properties of the kernel of the model Dirac operator agree with those of BPS states -- and hence the model Dirac operator is a Fredholm deformation of the exact one -- but only if the constituent electromagnetic charges, $\gamma_1,\gamma_2$, are primitive.  

Returning to our analysis, then, consider the first equation of \eqref{Diracpair} acting on our ansatz \eqref{Psiansatz}.  At leading order in $R$, we find that
\begin{equation}
\slashed{D}_{\MM_{1,0} \times \MM_{2,0}}^{\YY_0} \Psi_{12,\nu^i}^{(0)} = 0~.
\end{equation}
$L^2$ solutions, when they exist, will be of the form
\begin{equation}
\Psi_{12,\nu^i}^{(0)} = \Psi_{1,0}^{(\nu^a)}(\vec{y}^{\,a}) \otimes \Psi_{2,0}^{(\nu^p)}(\vec{y}^{\,p})~,
\end{equation}
where each factor is in the $L^2$ kernel of the $\rG(\YY_0)$-twisted Dirac operator on the respective strongly centered moduli spaces (and restricted to the appropriate $\pd_{\psi}$-eigenspaces).  Note we must demand these wavefunctions are invariant under the exchange symmetries $S_1$ and $S_2$ in order that they be well-defined on $\MM_{1,0}$ and $\MM_{2,0}$ respectively.  Assuming this has been done, then at $O(1/R)$ we get
\begin{equation}\label{firstperturbation}
\slashed{D}_{\MM_{1,0} \times \MM_{2,0}}^{\YY_0} \Psi_{12,\nu^i}^{(1)}  = - \lim_{R \to \infty} \left\{ R\left( \left. (\breve{\slashed{D}}_{12}^{\YY_0})^{(1)} \right|_{\pd_R \to -\sgn(p) x} \right)  \Psi_{12,\nu^i}^{(0)} \right\}  ~.
\end{equation}
Here some further explanation about factoring off the $\Psi_{\rm rel}$ is required, since the perturbing operator $(\breve{\slashed{D}}_{12}^{\YY_0})^{(1)}$ has a term involving $\frac{1}{R} \pd_{R^\alpha}$, that could act on $\Psi_{\rm rel}$.  Working in spherical coordinates $\{R,\uptheta,\upphi\}$, the derivatives along the angular directions are suppressed by an extra power of $R$ and do not contribute at the order we are working.  The derivative with respect to $R$ only contributes when acting on the exponential factor in $\Psi_{\rm rel}$.  This is what is meant by the replacement rule in \eqref{firstperturbation}.  Then, with this understood, both terms in the $O(1/R)$ part of the equation are proportional to $\Psi_{\rm rel}$, which can then be factored out.  Setting the $O(1/R)$ terms to zero results in \eqref{firstperturbation}.  We are free to assume that the perturbation of the wavefunction $\Psi_{\nu^i,12}^{(1)}$ is orthogonal to the kernel of the leading order operator, and hence \eqref{firstperturbation} can be solved to give a unique $\Psi_{12,\nu^i}^{(1)}$ for a give $\Psi_{12,\nu^i}^{(0)}$.

Therefore the leading order behavior of $L^2$ solutions is
\begin{equation}\label{L2zmleading}
\Psi_{\nu^i,\upnu} \to  R^{j-1/2} e^{-\sgn(p)x R} \, \Psi_{1,0}^{(\nu^a)}(\vec{y}^{\, a}) \otimes \Psi_{2,0}^{(\nu^p)}(\vec{y}^{\,p}) \otimes \Psi_{j,m}(\uptheta,\upphi)~,
\end{equation}
with the angular momentum quantum number $j = \half (|p|-1)$, and we have shown how the first order correction is computed.  Let us assume that the perturbation theory converges to give a zeromode of the exact Dirac operator.  Next we consider the implications.  

\subsection{Recovering the primitive wall-crossing formula}

The wavefunction \eqref{LWYspinor} corresponds to a state $\Psi^{(\gamma^{\rm e})} = e^{-i q \chi} s \otimes \Psi_{0}^{(\gamma_{0}^{\rm e})}$ of electric charge $\gamma^{\rm e} = \gamma_{1}^{\rm e} + \gamma_{2}^{\rm e}$, with the constituent charges given by
\begin{equation}
\gamma_{1}^{\rm e} =  \left( q - \frac{ (\gamma_{{\rm m},2},\XX)\upnu }{(\gm,\XX)} \right) \gamma_{{\rm m},1}^\ast - \sum_{a=1}^{N_1-1} \nu^a \beta_{a} ~, \qquad \gamma_{2}^{\rm e} = \left( q + \frac{ (\gamma_{{\rm m},1},\XX)\upnu }{(\gm,\XX)} \right) \gamma_{{\rm m},2}^\ast - \sum_{\mathclap{p=N_1}}^{N-2} \nu^p \beta_{p} ~,
\end{equation}
where we used \eqref{echargeconstituents} and the expressions for the galaxy masses in terms of the vev $\XX$.  With these and \eqref{Lcoefs} one quickly discovers
\begin{equation}\label{presult}
\llangle \gamma_1,\gamma_2 \rrangle = \langle \gamma_{1}^{\rm e}, \gamma_{{\rm m},2} \rangle - \langle \gamma_{2}^{\rm e},\gamma_{{\rm m},1} \rangle = - (\gmone,\gmtwo) \upnu + \nu^i L_i = p~.
\end{equation}
Thus the parameter $p$, \eqref{pxparameters}, is none other than the DSZ pairing of the constituent electromagnetic charges!  

Similarly, recalling that the $\beta_i$ satisfy $\langle \beta_i, \XX \rangle = 0$, one can observe that
\begin{align}
\left\langle \gamma_{1}^{\rm e} - \frac{\langle \gamma^{\rm e}, \XX\rangle}{(\gm, \XX)} \gamma_{{\rm m},1}^\ast, ~ \XX \right\rangle =&~ - \frac{ (\gamma_{{\rm m},1}, \XX) (\gamma_{{\rm m},2}, \XX) }{(\gm,\XX)} \upnu = - \mu \upnu~.
\end{align}
Hence the parameter $x$, \eqref{pxparameters}, is precisely the quantity appearing in the prediction from \cite{Moore:2015szp}:
\begin{equation}\label{xresult}
x = (\gamma_{{\rm m},1} ,\YY_0) + \left\langle \gamma_{1}^{\rm e} - \frac{\langle \gamma^{\rm e}, \XX\rangle}{(\gm, \XX)} \gamma_{{\rm m},1}^\ast, ~ \XX \right\rangle ~.
\end{equation}
Using the map \eqref{SWscmap}, one has 
\begin{equation}\label{xresult2}
x = \Im \left[ \zeta_{\rm van}^{-1} Z_{\gamma_1}(u) \right]~,
\end{equation}
and the condition $x = 0$ is equivalent to the usual condition $\Im [Z_{\gamma_1}(u) \overline{Z_{\gamma_2}(u)} ] = 0$ for the walls of marginal stability.  Furthermore the condition $\sgn(p) x > 0$ for the existence of the $L^2$ zeromode $\Psi_{\nu^i,\upnu}$ is equivalent to the usual stability condition for a BPS state: $\llangle \gamma_1, \gamma_2 \rrangle \Im[Z_{\gamma_1}(u) \overline{Z_{\gamma_2}(u)} ]  > 0$.

Let us note here that $x$ and $p$ are related to Denef's boundstate radius \cite{Denef:2000nb,Gaiotto:2010be} in a simple way:
\begin{equation}\label{xtorD}
r_{\rm D} := \half  \llangle \gamma_1, \gamma_2 \rrangle \frac{ |Z_{\gamma_1+\gamma_2}(u) |}{2 \Im [ \overline{Z_{\gamma_1}(u)} Z_{\gamma_2(u)} ]} = \frac{\llangle \gamma_1, \gamma_2 \rrangle}{2 \Im[ \zeta_{\rm van}^{-1} Z_{\gamma_1}(u) ]} = \frac{p}{2x}~.
\end{equation}

Now let us consider the degeneracy of $L^2$ zeromodes \eqref{L2zmleading} gained or lost when we cross the locus $x =0$.  For each pair of zeromodes $\Psi_{1,0} \in \ker_{L^2}( \slashed{D}_{\MM_{1,0}}^{\YY_0} )$ and $\Psi_{2,0} \in \ker_{L^2} ( \slashed{D}_{\MM_{2,0}}^{\YY_0})$, we have a spin $j = \half (|p|-1)$ multiplet of states.  Hence the total degeneracy is a product of the dimensions of the $L^2$ kernels of the constituent strongly centered Dirac operators, times a factor of $|p| = |\llangle \gamma_1,\gamma_2\rrangle |$ from the relative moduli space.  This agrees with the primitive wall-crossing formula for the indices of vanilla BPS states, $\Omega(\gamma,u)$.  Furthermore the ${\rm SO}(3)$ isometry of monopole moduli space is the collective coordinate manifestation of the field theory $\mathfrak{su}(2)$ that is used in defining the protected spin characters \cite{Gaiotto:2010be,Moore:2015szp}.  Hence the Dirac operator analysis in the two-galaxy region reproduces the primitive wall-crossing formula of \cite{Denef:2007vg,Diaconescu:2007bf} for the protected spin characters:
\begin{equation}
\Delta\Omega(\gamma_1+\gamma_2,u;y)=\Omega(\gamma_1,u;y) \, \Omega(\gamma_2,u;y) \, \chi_{|\llangle \gamma_1,\gamma_2\rrangle|}(y)~,
\end{equation}
where $\chi_{n}(y)$ is the character of the ${\rm SU}(2)$ representation of dimension $n$ as a polynomial in $y$.

There are a number of ways to interpret this result, depending on one's viewpoint.  If we take the wall-crossing formulae for BPS states and the identification \eqref{SWscmap} between Seiberg--Witten and semiclassical data for granted, then this result demonstrates that our model asymptotic Dirac operator is a Fredholm deformation of the exact one, provided the constituent charges are primitive.  Alternatively, if one could prove the Fredholm property, then by combining that proof with the identification \eqref{SWscmap}, our analysis here yields a semiclassical derivation of primitive wall crossing for BPS states.  Finally, one could combine the Fredholm proof with the wall-crossing formula to verify the identification \eqref{SWscmap}.\footnote{Note that a first-principles derivation of \eqref{SWscmap} from quantum field theory would require the computation of instanton corrections to the collective coordinate metric for monopole moduli space!}

The fact that we do \emph{not} recover the full wall-crossing formula in non-primitive cases is also interesting.  In these cases our model Dirac operator is apparently not a Fredholm deformation of the exact Dirac operator.  This supports the idea that the physical picture of fractionalization of non-primitive constituent charges is related to considering different $n$-galaxy asymptotic subregions of monopole moduli space, and that the additional Dirac zeromodes contributing to wall crossing in these cases are only made visible by considering such limits.

\subsection{Non-BPS boundstates}\label{ssec:nonBPS}

Near the walls of marginal stability we can study how the spectrum of $\breve{\slashed{D}}_{{\rm rel}}^{\YY_0}$ degenerates. Specifically, in Appendix \ref{app:AppD} we review the solution for the spectrum,
\be\label{eq:BREVEDIRAC}
\breve{\slashed{D}}_{\rm rel}^{\CY_0}\Psi_{{\rm rel},\lambda}^{(p)}= -\I \lambda \, \Psi_{{\rm rel},\lambda}^{(p)}~,
\ee
taking $\breve{\slashed{D}}_{\rm rel}$ to be our toy-model Taub-NUT Dirac operator.  See also \cite{Jante:2015xra}.  We expect the resulting features described below to be qualitatively correct when $(\gmone,\gmtwo) <0$.  They are quantitatively correct in the special examples where the exact strongly centered moduli space \emph{is} Taub-NUT.  

The asymptotic behavior of solutions to \eqref{eq:BREVEDIRAC} is dictated by $\Psi_{{\rm rel},\lambda}^{(p)}\sim e^{-\kappa_\lambda r}$ where
\be
\kappa_\lambda = \sqrt{x^2- \mu\lambda^2}~.
\ee
Note that the Dirac operator represents a supercharge operator, so its eigenvalues have units of energy${}^{1/2}$.  The square of the eigenvalue is twice the energy above the BPS bound of the corresponding state, 
\begin{equation}
\lambda^2 = 2 \Delta E~,
\end{equation}
where the factor of two originates from the normalization of the collective coordinate supercharge in terms of the Dirac operator.\footnote{As we mentioned below \eqref{constituentmasses}, in this paper we have set a factor of $4\pi/g_{0}^2$ equal to one.  Here $g_0$ is the classical Yang--Mills coupling and this factor should appear as an overall factor relating the physical metric on moduli space to the one we have been using.  In particular, when it multiplies the constituent masses in the GM/LWY metric it gives them their proper physical value.  There are similar factors in the normalization of the collective coordinate Hamiltonian and supercharges.  Specifically, the supercharge corresponding to the Dirac operator, that squares to the collective coordinate Hamiltonian, is $Q = \frac{g_0}{2\sqrt{2\pi}} \I \slashed{D}^{\YY}$.  (See \cite{Moore:2015szp}.)  Hence, in our conventions, $(\I \slashed{D}^{\YY})^2 = 2 Q^2 = 2 H_{\rm c.c.}$.}   

The spectrum of eigenspinors is divided into two parts by a critical value $\lambda_{\rm gap}=|x|/\sqrt{\mu}$.  There is a continuous spectrum of plane-wave normalizable states for $|\lambda| > \lambda_{\rm gap}$, while for $|\lambda|<\lambda_{\rm gap}$ there is an infinite discrete spectrum, with the BPS states at $\lambda=0$ and accumulation points at $\lambda \to \pm \lambda_{\rm gap}$.  The exact discrete spectrum depends on the parameters $\mu,\ell_{\rm TN},p,x$ given in \eqref{reducedmass}, \eqref{TNell}, \eqref{presult}, and \eqref{xresult2}, and is given by
\begin{equation}\label{specx}
\lambda_{p,n}^2 = \frac{2}{\mu \ell_{\rm TN}^2} \left[ n \sqrt{ n^2 + p \ell_{\rm TN} x + \ell_{\rm TN}^2 x^2} - n^2 - \tfrac{p}{2} \ell_{\rm TN} x \right]~.
\end{equation}
BPS states correspond to $n = |p|/2$ with $p \neq 0$, and the allowed values of $n$ increase from this in integer steps.  If $p =0$ there are no BPS states and $n \in \mathbb{N}$.  Meanwhile the allowed values of the angular momentum quantum number run from $j = \half (|p|-1)$ in the BPS case up to $j = n- \half$ in integer steps.  More details on the eigenspinors themselves and the degeneracy of each eigenvalue are given in the appendix.  When $p$ is nonzero we can make use of \eqref{xtorD} to trade $x$ for Denef's boundstate radius, $r_{\rm D}$, and express the result as follows:
\be\label{specrD}
\lambda_{p,n}^2 = \frac{2}{\mu \ell_{\rm TN}^2} \left(  n \sqrt{n^2 - \tfrac{p^2}{4} + \tfrac{p^2}{4} H^2(r_{\rm D}) } - (n^2 - \tfrac{p^2}{4} + \tfrac{p^2}{4} H(r_{\rm D})  ) \right)~,
\ee
where $H(r_{\rm D}) = 1 + \frac{\ell_{\rm TN}}{r_{\rm D}}$.

The expression \eqref{specrD} is, however, a bit misleading.  We were able to determine the relation between the parameter $x$ and quantum-exact Seiberg--Witten data, \emph{viz.}\ \eqref{xresult2}, using the map \eqref{SWscmap}.  In turn, that map was conjectured in \cite{Moore:2015szp} by comparing results from semiclassical analysis and Seiberg--Witten theory in the weak-coupling regime where they should agree.  See especially the discussion in section 4.5 there.  These arguments utilized the fact that the kernel of the supercharge operator, represented by the Dirac operator, determines the BPS states, whose existence is unaffected by higher order D-terms in the SQM for the collective coordinates.  However the normalization of the supercharge operator certainly will be affected by such terms.  This does not affect the kernel but does modify the nonzero spectrum.  Therefore we have no right to use \eqref{specrD}, as an expression for the energy of non-BPS boundstates, beyond leading order in the semiclassical approximation.  

The semiclassical nature of \eqref{specrD} is indeed quite evident if we consider the overall prefactor involving the reduced mass $\mu$.  The expression for $\mu$, \eqref{reducedmass}, involves only the magnetic contribution to the masses of the constituents.  This is the dominant contribution in the semiclassical limit.  Since the central charge vectors of the constituents are both near the positive imaginary axis in the semiclassical limit, a convenient expansion parameter is provided by the angle $\theta_{\gamma_1\gamma_2}$ between them.  This angle satisfies $\Im[ \overline{Z_{\gamma_1}(u)} Z_{\gamma_2}(u) ] = |Z_{\gamma_1}| |Z_{\gamma_2}| \sin{\theta_{\gamma_1\gamma_2}}$.  Hence from \eqref{xtorD},
\begin{equation}
r_{\rm D} = \frac{p}{2} \frac{ \sqrt{ |Z_{\gamma_1}|^2 + |Z_{\gamma_2}|^2 + 2 |Z_{\gamma_1}| |Z_{\gamma_2}| \cos{\theta_{\gamma_1\gamma_2}} }}{| Z_{\gamma_1}| | Z_{\gamma_2}| \sin{\theta_{\gamma_1 \gamma_2}}} = \frac{p (|Z_{\gamma_1}| + |Z_{\gamma_2}|)}{2 |Z_{\gamma_1}| |Z_{\gamma_2}| \theta_{\gamma_1 \gamma_2}} \left( 1 + O(\theta_{\gamma_1\gamma_2}^2) \right)~.
\end{equation}
However the magnitudes of the individual central charges are dominated by the magnetic term, up to corrections $\sim \theta_{\gamma_1\gamma_2}^2$.  Therefore setting $|Z_{\gamma_{1,2}}| \to m_{{\rm gal}1,2}$, we recognize the prefactor in terms of the reduced mass to leading order:
\begin{equation}
r_{\rm D} = \frac{p}{2\mu \theta_{\gamma_1\gamma_2}} \left(1 + O(\theta_{\gamma_1\gamma_2}^2)\right) \qquad \textrm{or} \qquad x = \mu \theta_{\gamma_1 \gamma_2} \left( 1 + O(\theta_{\gamma_1\gamma_2}^2) \right)~.
\end{equation}
We also note that the binding energy of BPS states in this limit becomes\footnote{This quantity is finite when $p = 0$ since $r_{\rm D} = 0$ in that case as well.  The coefficient can alternatively be expressed as $E_{\rm binding} = - (\mu \theta_{\gamma_1\gamma_2}^2/2) (1 + O(\theta_{\gamma_1\gamma_2}^2))$.}
\begin{align}
E_{\rm binding} :=&~ | Z_{\gamma_1} + Z_{\gamma_2} | - |Z_{\gamma_1}| - |Z_{\gamma_2}| = - \frac{ |Z_{\gamma_1}| |Z_{\gamma_2}| \theta_{\gamma_1 \gamma_2}^2}{2 (|Z_{\gamma_1}| + |Z_{\gamma_2}|)} \left( 1 + O(\theta_{\gamma_1\gamma_2}^2)\right) \cr
=&~ - \frac{p^2}{8 r_{\rm D}^2 \mu} \left( 1 + O(\theta_{\gamma_1\gamma_2}^2) \right)~.
\end{align}
Using these results, the discrete energy spectrum above the BPS bound is given by
\begin{align}
\Delta E_{p,n} =&~ \half \lambda_{p,n}^2 = \frac{1}{\mu \ell_{\rm TN}^2}  \left[ n \sqrt{ n^2 + p \ell_{\rm TN} \mu \theta_{\gamma_1\gamma_2} + \ell_{\rm TN}^2 \mu^2 \theta_{\gamma_1\gamma_2}^2} - n^2 - \tfrac{p}{2} \ell_{\rm TN} \mu \theta_{\gamma_1\gamma_2} +O(\theta_{\gamma_1\gamma_2}^3) \right] \cr
=&~ - E_{\rm binding} \left( 1 - \frac{p^2}{4 n^2} + O(\theta_{\gamma_1\gamma_2}) \right) ~. \raisetag{16pt}
\end{align}
Now we clearly see how the discrete spectrum accumulates at the threshold for the continuum as $n \to \infty$, and that the gap is consistent with the binding energy of the BPS states.  Note if $p=0$ then the $O(\theta_{\gamma_1\gamma_2})$ terms determine the form of the non-BPS spectrum.  This case was analyzed in \cite{Ritz:2008jf} from the perspective of the Seiberg--Witten low-energy effective theory.  Our results are compatible with this analysis.  

As we approach the marginal stability wall at $x = 0$ (or equivalently $\theta_{\gamma_1\gamma_2} = 0$), the gap shrinks and the entire discrete spectrum is squeezed down.  Our analysis in Appendix \ref{app:AppD} shows, however, that the non-BPS boundstates exist on both sides of the wall.  This strongly suggests, especially in those cases where Taub-NUT is the exact strongly centered moduli space, that the lowest-energy non-BPS states are completely stable on the side of the wall where the BPS states do not exist.

We expect generic non-BPS boundstates to be unstable when one includes interactions with the massless abelian vectormultiplets on the Coulomb branch.  These interactions should lead to effective interactions in the collective coordinate Hamiltonian, and one could in principle use those to estimate the lifetime of these boundstates. We expect them to become long-lived near a wall of marginal stability.  It would be interesting to investigate these points further.

\section{Framed BPS States and Haloes}\label{sec:Framed}

We can now apply this analysis with some simple modifications to the study the wall-crossing behavior of framed BPS states. This follows the line of thought presented in \cite{Tong:2014yla,Moore:2015szp,Moore:2015qyu,Brennan:2016znk}. In these papers, the authors showed that the story for vanilla BPS states is modified by replacing the moduli space of smooth monopoles with the moduli space of singular monopoles for the case of magnetically charged defects and by coupling to a vector bundle with a natural hyperholomorphic connection for the case of electrically charged defects.  In this paper we will restrict ourselves to the case of purely magnetic (\ie\ 't Hooft) defects.

\subsection{'t Hooft defects and the moduli space of singular monopoles}

Singular monopole moduli space $\fMM(\{P_n\},\gamma_{\rm m},\CX)$ describes the moduli space of smooth monoples in the presence of a collection of singular monopoles with charges $\{P_n\}\in \Lambda_{G}^\vee$ inserted at locations $\{\vx_n\}\in\IR^3$.  This is a hyperk\"ahler manifold, possibly with singularities on co-dimension four and higher loci.  Here $\Lambda_{G}^{\vee} := \{ H \in \mathfrak{t} ~|~ \exp(2\pi H) = 1_{G} \} \cong \mathrm{Hom}(\mathrm{U}(1),T)$.  The data $\{(P_n,\vx_n)\}$ specifies the `t Hooft defects which impose the (singular) boundary conditions on the gauge and Higgs field,
\be
F\to\half P_n d\Omega_n +O(r_{n}^{-3/2})~,\qquad \Phi \to -\frac{P_n}{2r_n}+O(r_n^{-1/2})~,\qquad \text{as }r_n=|\vx-\vx_n|\to 0~,
\ee
and the data $\{\gamma_{\rm m},\CX\}$ fixes the asymptotic boundary conditions as in the case of smooth monopoles. 

There is a torus action of hyperholomorphic isometries associated with asymptotically nontrivial gauge transformations and generated by the image of the ${\rm G}$-map
\begin{equation}
{\rm G} : \mathfrak{t} \to \mathfrak{isom}_{\mathbb{H}}(\fMM)~.
\end{equation}
't Hooft defects break the translational isometries.  Correspondingly the metric on $\fMM$ does not factorize into center of mass and strongly centered pieces.  In the case of a single defect, $\fMM$ has an ${\rm SO}(3)$ isometry corresponding to spatial rotations about the defect.

\begin{figure}
\begin{center}
\includegraphics[scale=1.3,trim= 4cm 24cm 11.5cm 2cm]{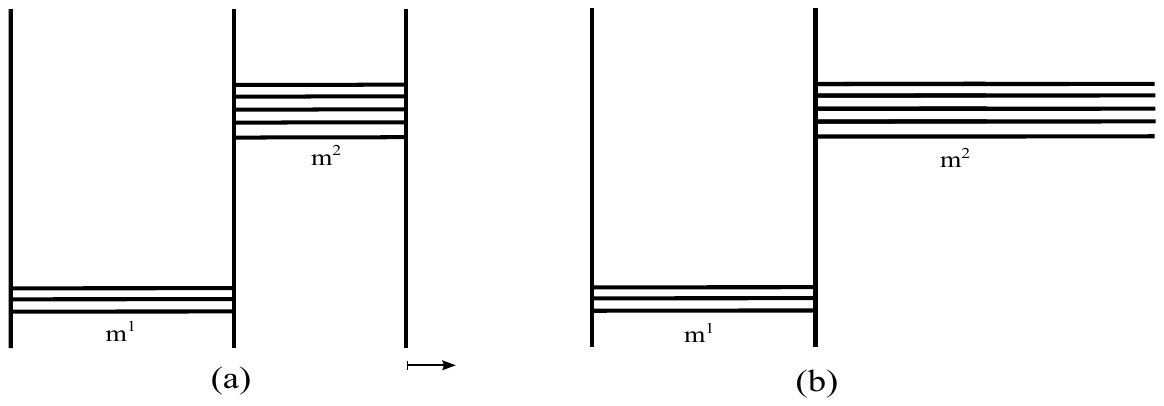}
\end{center}
\caption{This figure demonstrates the brane construction of line defects in ${\rm PSU}(2)$ gauge theory from ${\rm SU}(3)$ gauge theory. (a) shows the fundamental monopoles in the ${\rm SU}(3)$ theory. By taking the rightmost brane to infinity, we arrive at (b) with line defects in the ${\rm PSU}(2)$ theory.}\label{fig:PSU2}
\end{figure}

\subsection{The asymptotic metric on the moduli space of singular monopoles}

In order to describe framed BPS states in the presence of  't Hooft defects, we'll need a geometric description of the asymptotic region of singular monopole moduli space. In \cite{Moore:2014gua}, the authors gave a method for obtaining singular $\mathrm{PSU}(\mathfrak{N})$ monopoles as a limit of smooth ${\rm SU}(\mathfrak{N}+1)$-monopoles.  This was motivated by using the string theory interpretation of smooth ${\rm SU}(\mathfrak{N})$ monopoles as D1-strings stretched between ${\mathfrak{N}}$ D3-branes \cite{Diaconescu:1996rk} and singular monopoles as semi-infinite D1-strings \cite{Hanany:1996ie}.  The semi-infinite strings are obtained by starting with a set of $\mathfrak{N}+1$ D3-branes and sending the rightmost brane to infinity.  See Figure \ref{fig:PSU2}.

In field theory, this construction is described by viewing the 't Hooft defects of the ${\rm PSU}(\mathfrak{N})$ theory initially as smooth monopoles in the ${\rm SU}(\mathfrak{N}+1)$ theory.  Let us begin with a configuration of $N_{\rm def}$ smooth monopoles of magnetic charge $\hat{P}_n=H_{\mathfrak{N}}$, the $\mathfrak{N}^{\rm th}$ simple co-root, where $n=1,..., N_{\rm def}$.  We also allow for some number of additional smooth monopoles that do not have magnetic charges with components along $H_{\mathfrak{N}}$.  We denote the charges of the latter by $H_{I(\hat{i})}$, for $\hat{i} = 1,\ldots,N$, with $I(\hat{i}) < {\mathfrak{N}}$, $\forall \hat{i}$.  Then we make a projection $\Pi : \mathfrak{su}({\mathfrak{N}}+1) \to \mathfrak{su}({\mathfrak{N}})$ on the Lie algebra-valued field configurations followed by a limit on the Higgs vev, $\CX_{\mathfrak{N}}\to \infty$, where
\be
\CX=\sum_{I=1}^{\mathfrak{N}} \CX_{I}h^I ~,\quad \XX_I = \langle \alpha_I, \XX \rangle~. 
\ee
We recall that the $\{h^I\}$ are the fundamental magnetic weights satisfying $\langle \alpha_I, h^J\rangle = {\delta_I}^J$.  This procedure removes certain degrees of freedom by making them infinitely heavy and simultaneously converts the $N_{\rm def}$ smooth monopoles into singular ones with defect charges $P_n = \Pi(\hat{P}_n)$, where $\Pi$ is a projector.  

The projection is described in detail in \cite{Moore:2014gua}.  On the Cartan subalgebra is acts as follows:
\begin{equation}
\Pi(H) = H - \frac{(h^{{\mathfrak{N}}}, H)}{(h^{\mathfrak{N}}, h^{\mathfrak{N}})} h^{\mathfrak{N}} ~, \qquad \forall H \in \mathfrak{t}~.
\end{equation}
This gives $\Pi(H_I) = H_I$ for $I=1,\ldots,{\mathfrak{N}}-1$, and $\Pi(H_{\mathfrak{N}}) = - h^{{\mathfrak{N}}-1}$ for the simple co-roots of the $\mathfrak{su}({\mathfrak{N}}+1)$ theory.  In particular the charges of the 't Hooft defects in the configuration we are considering are $P_n =  -h^{{\mathfrak{N}}-1}$.  The fact that these charges are given by a fundamental magnetic weight show that the global form of the gauge group in the reduced theory must be taken as the adjoint group, ${\rm PSU}({\mathfrak{N}}) = {\rm SU}({\mathfrak{N}})/\mathbb{Z}_{\mathfrak{N}}$.  Note that the Killing pairing between the magnetic charges of the defects and those of the smooth monopoles is unaffected by the projection:
\begin{align}\begin{split}
(P_n,H_{I(\hat{i})})&=(\Pi(H_{\mathfrak{N}}),\Pi(H_{I(\hat{i})}))=(H_{\mathfrak{N}},H_{I(\hat{i})})-\frac{{\mathfrak{N}}+1}{\mathfrak{N}}(h^{\mathfrak{N}},H_{I(\hat{i})})\\
&= (H_{\mathfrak{N}},H_{I(\hat{i})})~.
\end{split}\end{align}

Since this procedure of producing singular monopoles by taking the limit of smooth monopoles is semiclassically well defined, we can construct the asymptotic metric on singular monopole moduli space by taking the same limit of the GM/LWY metric for smooth ${\rm SU}({\mathfrak{N}}+1)$ monopoles.

Now we determine the form of this metric.  We start with the GM/LWY metric for $N + N_{\rm def}$ fundamental $\mathfrak{su}({\mathfrak{N}}+1)$ monopoles with charges $H_{I(\hat{i})}$, $\hat{i} = 1,\ldots,N$, and $\hat{P}_n = H_{\mathfrak{N}}$, $n=1,\ldots,N_{\rm def}$, respectively.  \emph{First} we restrict to the subregion of fixed location and phase for the $N_{def}$ fundamental monopoles with charges $H_{\mathfrak{N}}$. \emph{Then} we take the limit as $\CX_{\mathfrak{N}}\to \infty$:
\be
ds^2_{\fMM}=\lim_{\XX_{\mathfrak{N}} \to \infty}\left[ds^2_{\CM}\Big{|}_{(\vec{x}^{\,n},\xi_n)=(\vec{x}_{\rm def}^{\,n},\xi_{n}^{\rm def})\,\textrm{fixed}}\right]~,
\ee
where the defects are located at $\{\vec{x}^{\,n}_{\rm def}\}_{n=1}^{N_{\rm def}}$. 

It is important to check that this procedure leads to a well defined metric.  In order to accomplish this we need to examine the behavior of $M_{\bar{i}\bar{j}}$ and $(M^{-1})^{\bar{i}\bar{j}}$ where now $\bar{i},\bar{j}=1,...,N+N_{\rm def}$.  For $M_{\bar{i}\bar{j}}$ we only need to consider the case where both indices correspond to smooth monopoles $\bar{i} \bar{j} \to \hat{i}\hat{j}$, since we are restricting to the subspace of fixed $\vec{x}^{
,n}_{\rm def}$.  Then, from \eqref{MW}, we see that $M_{\hat{i}\hat{j}}$ is clearly well defined in the limit.  For $(M^{-1})^{\bar{i}\bar{j}}$ we need to consider each possibility: $(M^{-1})^{\hat{i}\hat{j}}$, $(M^{-1})^{\hat{i} n}$, and $(M^{-1})^{mn}$.  The corresponding coefficients of the cofactor matrix diverge as $c_{\hat{i}\hat{j}} \sim O(\XX_{\mathfrak{N}}^{N_{\rm def}})$ and $c_{\hat{i}n},c_{mn} \sim O(\XX_{\mathfrak{N}}^{N_{\rm def}-1})$, while the determinant diverges as $\det{M} \sim O(\XX_{\mathfrak{N}}^{N_{\rm def}})$.  Hence only $(M^{-1})^{\hat{i}\hat{j}}$ survives the limit and is given by
\be \label{limitVrestrict}
\lim_{\CX_{\mathfrak{N}}\to \infty}\left[ (M^{-1})^{\hat{i}\hat{j}}\Big{|}_{\vec{x}^{\,n}=\vec{x}_{\rm def}^{\,n}} \right] =\left(\lim_{\CX_{\mathfrak{N}}\to \infty} \left[ M_{\hat{i}\hat{j}}\Big{|}_{\vec{x}^{\,n}=\vec{x}_{\rm def}^{\,n}} \right]\right)^{-1}~.
\ee

The resulting metric on $\fMM$ takes the form 
\begin{align}\begin{split}\label{eq:singmetric}
ds^2_{\fMM}=\overline{M}_{\hat{i}\hat{j}}d\vec{x}^{\,\hat{i}}\cdot d\vec{x}^{\,\hat{j}}+(\overline{M}^{-1})^{\hat{i}\hat{j}}\overline{\Theta}_{\hat{i}} \overline{\Theta}_{\hat{j}}~,
\end{split}\end{align}
where
\begin{align}\label{singmeta}\begin{split}
\overline{\Theta}_{\hat{i}} := d\xi_{\hat{i}}+\sum_{\{\hi,\hat{j}| \hat{j} \neq \hat{i}\}}\frac{D_{\hat{i}\hat{j}}}{2}(\pm 1-\cos(\theta_{\hat{i}\hat{j}}))d\phi_{\hat{i}\hat{j}}+\sum_{n=1}^{N_{\rm def}} \frac{(P_n,H_{I(\hat{i})})}{2}(\pm 1-\cos(\theta_{\hat{i}n}))d\phi_{\hat{i}n}~,
\end{split}\end{align}
and
\begin{align}\label{singmetb}\begin{split}
\overline{M}_{\hat{i}\hat{j}} := \begin{cases}
m_{\hat{i}}-\sum_{\hat{k}\neq \hat{i}}\frac{D_{\hat{i}\hat{k}}}{r_{\hat{i}\hat{k}}}-\sum_{n=1}^{N_{\rm def}}\frac{(P_n,H_{I(\hat{i})})}{r_{\hat{i}n}} &,\quad \hat{i}=\hat{j}\\
\frac{D_{\hat{i}\hat{j}}}{r_{\hat{i}\hat{j}}}&,\quad i\neq j
\end{cases}~.
\end{split}\end{align}
It is easy to check that this metric, which is of the general Pedersen-Poon form \cite{MR953820}, satisfies the necessary and sufficient equations for hyperk\"ahlerity.  (See Appendix \ref{app:hk} and especially \eqref{hkrelns} for the form of those equations.)  The triplet of K\"ahler forms is
\be\label{singhkstructure}
\overline{\omega}^\alpha = \overline{\Theta}_{\hi} \wedge d x^{\alpha \hi} - \frac{1}{2} \overline{M}_{\hi\hj} \epsilon^{\alpha}_{\phantom{\alpha}\beta\gamma} dx^{\beta \hi} \wedge dx^{\gamma \hj} ~,\qquad \alpha,\beta,\gamma=1,2,3~.
\ee
Although we have derived this metric for a special set of defects in a ${\rm PSU}({\mathfrak{N}})$ theory as motivated by a brane construction, the result clearly makes sense for a generic set of defects in a theory with any simple $G$.  We therefore  conjecture that \eqref{eq:singmetric} with \eqref{singmeta} and \eqref{singmetb} is the analog of the GM/LWY asymptotic metric for the moduli space of singular monopoles.  This should be confirmed by carrying out the analysis of point dyons interacting with the fixed defects, following \cite{Gibbons:1995yw,Lee:1996kz}.    

Furthermore, inspired by the results of Bielawski \cite{Bielawski:1998hj,Bielawski:1998hk} and Murray \cite{Murray:1996hi}, following the conjecture of Lee, Weinberg, and Yi in \cite{Lee:1996kz}, we conjecture that this metric is exponentially close to the exact metric with corrections of order $e^{-m_{\hat{i}} r_{\hat{i}\hat{j}}}$ for those $\hat{i},\hat{j}$ such that $I(\hat{i}) = I(\hat{j})$.  In particular, if we have no more than one smooth monopole of each type, we conjecture that this asymptotic metric is the exact metric on the moduli space of singular monopoles.

Note that the $\mathrm{PSU}({\mathfrak{N}})$ defects we obtained from the limiting procedure have charges $P_n = -h^{{\mathfrak{N}}-1}$, which are in the closure of the antifundamental Weyl chamber.  This is important since only in this case will the dimension of $\fMM$, as computed from the exact formula \cite{Moore:2014jfa}, be equal to $4N$, the dimension of the space on which \eqref{eq:singmetric} is defined.  Therefore when claiming that \eqref{eq:singmetric} provides an asymptotic metric on $\fMM$, it should be understood that all 't Hooft charges are to be taken in the closure of the antifundemantal Weyl chamber.\footnote{This is not a physical restriction on the set of 't Hooft defects we consider, since 't Hooft defects only depend on the Weyl orbit of the 't Hooft charge.}  If one considers 't Hooft charges that are not restricted in this way, then \eqref{eq:singmetric} will only describe the asymptotic metric on a subspace of $\fMM$, where some smooth monopole positions have been fixed and are coincident with those defect positions.  This issue is explored from the D-brane perspective in \cite{Moore:2014gua}.

Additionally, in order to get a 1:1 map between our model space and the asymptotic region of $\fMM$, we expect that one should quotient the model space by exchange symmetries analogous to \eqref{exchangesym}, but where the $n_{\rm m}^I$ are replaced by the numbers of smooth monopoles of type $I$.  It is easy to see that exchanging the coordinates of smooth monopoles of the same type is a hyperk\"ahler isometry of \eqref{eq:singmetric}.

Asymptotic metrics on the moduli space of singular monopoles are not entirely new.  Specific examples of such metrics have been discussed previously in, \eg\ \cite{Cherkis:1997aa,Cherkis:1998hi,Houghton:1997ei}.  Also, an asymptotic metric on the moduli space of ${\rm U}(n)$ instantons over Taub-NUT space was described in \cite{Cherkis:2010bn} and recovered from a three-dimensional gauge theory computation in \cite{Cherkis:2011ee}.  Certain $\mathrm{U}(1)$-invariant instantons over Taub-NUT, which correspond to a special class of \emph{Cheshire} bows \cite{Blair:2010vh} in the bow formalism of \cite{Cherkis:2010bn}, are equivalent to singular monopoles via the work of Kronheimer \cite{Kronheimer}.

\subsection{The two-galaxy region}

Due to \eqref{limitVrestrict}, the analysis used to write down the two-galaxy limit of the smooth monopole moduli space can be implemented, with minor substitutions, in the singular case as well.  Some details are presented in Appendix \ref{app:C}.  We separate the smooth constituents into two groups and take the limit where the distance between the two groups, $R$, is much larger than any other distance scale.  In particular, one of the two groups will remain relatively close to the full set of 't Hooft defects.

Hence the physical picture of this limit is very similar to before, but with a slightly different interpretation.  In the case of framed BPS states -- that is, BPS states bound to a collection of line defects -- there is a generic core-halo structure. This means that there is a core of ``vanilla" BPS states which is centered around the collection of defects and additionally a halo of vanilla BPS particles.  In the limit we are considering, the halo is composed of a single galaxy of fundamental constituents, and the moduli space splits asymptotically as $\fMM \to \fMM_{\rm c} \times \MM_{{\rm h},0} \times \mathbb{R}_{\rm rel}^4$.  The first factor is the moduli space of singular monopoles associated with the core galaxy which contains all of the defects.  The second factor is the strongly centered moduli space associated with the halo galaxy.  The final factor describes the relative positions of the two galaxies as before.  Associated with this decomposition, the symmetry group exchanging identical smooth monopoles reduces to $S \to S_{\rm h} \times S_{\rm c}$, so that we only exchange identical smooth monopoles within each galaxy.

We now summarize this limit at the level of the metric \eqref{eq:singmetric}.  We will take our origin of coordinates to be within the core region, for example at the location of one of the defects.  Let $N_{\rm core}$ and $N_{\rm halo}$ be the number of smooth monopoles in the core and halo respectively with $N_{\rm core}+N_{\rm halo}=N$.  We use indices $\hat{a},\hat{b} = 1,\ldots,N_{\rm core}$, to label the core constituents and $\hat{p},\hat{q} = N_{\rm core} + 1,\ldots,N$, to label the halo constituents.  The corresponding coordinates are $\{\vx^{\,\hat{a}},\xi_{\hat{a}}\}$ and $\{\vx^{\,\hat{p}},\xi_{\hat{p}}\}$.  In the halo galaxy we change coordinates to center of mass and relative variables:
\begin{align}
\vec{R}=\frac{\sum_{\hat{p}} m_{\hat{p}} \vx^{\,\hat{p}}}{m_{\rm halo}} ~, \qquad \vy^{\,p}=\vx^{\,p}-\vx^{\,p+1}~,
\end{align}
where $m_{\rm halo}=\sum_{\hat{p}} m_{\hat{p}}$ is the mass of the halo galaxy and $p,q = N_{\rm core}+1, \ldots, N - 1$.  There is an analogous map for the halo phases, $\{ \xi_{\hat{p}} \} \mapsto \{ \psi_p, \uppsi \}$.  See Appendix \ref{app:C} for further details.  We then combine the core constituent coordinates and the halo relative coordinates into a single set,
\begin{equation}
\vec{y}^{\,i} = (\vec{x}^{\,\hat{a}}, \vec{y}^{\,p})~, \qquad \psi_i = (\xi_{\hat{a}},\psi_p)~,
\end{equation}
where now $i,j = 1,\ldots, N-1$.  

We also introduce the total magnetic charges for the core and halo,
\begin{equation}
\gamma_{{\rm c,m}} = \sum_{\hat{a}=1}^{N_{\rm core}} H_{I(\hat{a})} + \sum_{n=1}^{N_{\rm def}} P_n ~, \qquad \gamma_{{\rm h,m}} = \sum_{\hat{p} = N_{\rm core}+1}^{N} H_{I(\hat{p})} ~,
\end{equation}
and note that $m_{\rm halo} = (\gamma_{{\rm h,m}}, \XX)$.

Then the metric \eqref{eq:singmetric} on $\fMM$ can be written as $ds^2 = d\hat{s}^2 + O(1/R^2)$ with
\begin{align}\label{2singmetric}
d\hat{s}^2 :=&~  \left( d\vec{\bf y}, d\vec{R}\right) \left( \begin{array}{c c} \widetilde{\overline{\mathbf{C}}} + \frac{1}{R} \delta\mathbf{ \overline{C}} & \frac{1}{R} {\bf \overline{L}} \\  \frac{1}{R} {\bf \overline{L}}^T & m_{\rm halo} \overline{H}(R) \end{array} \right) \left( \begin{array}{c} \cdot d\vec{\bf y} \\ \cdot d\vec{R} \end{array} \right) + \cr
&~ + \left( \overline\Theta_0, \overline\Theta_\uppsi \right) \left( \begin{array}{c c} \widetilde{\overline{\mathbf{C}}} + \frac{1}{R} \delta\mathbf{ \overline{C}} & \frac{1}{R} {\bf \overline{L}} \\  \frac{1}{R} {\bf \overline{L}}^T & m_{\rm halo} \overline{H}(R) \end{array} \right)^{-1} \left( \begin{array}{c} \overline\Theta_0 \\ \overline\Theta_{\uppsi} \end{array} \right)~.
\end{align}
Here $\overline{H}(R)=1-\frac{(\gamma_{\rm h,m},\gamma_{\rm c,m})}{M_{\rm halo} R}$ and 
\begin{align}\begin{split}
\left(\begin{array}{c}
\overline{\Theta}_0\\
\overline{\Theta}_{\uppsi}
\end{array}\right)=
\left(\begin{array}{c}
\widetilde{\overline{\Theta}}_0\\d\uppsi
\end{array}\right)
+\left(\begin{array}{cc}
\delta \overline{\bf C}&\overline{\bf{L}}\\
\overline{\bf L}^T& -(\gamma_{\rm h,m}, \gamma_{\rm c,m})
\end{array}\right)\otimes \vec{w}(\vec{R}) \cdot \left(\begin{array}{c}
d\vec{\bf y}\\  d\vec{R}
\end{array}\right)~,
\end{split}\end{align}
where $\widetilde{\overline{\Theta}}_0=(\widetilde{\overline{\Theta}}_{\hat{a}}, \widetilde{\Theta}_p)^T$ with
\begin{align}\begin{split}
&\widetilde{\overline{\Theta}}_{\hat{a}}=d\xi_{\hat{a}}+\sum_{\{\ha,\hat{b} | \hat{b} \neq \hat{a}\}}\frac{D_{\hat{a}\hat{b}}}{2}(\pm 1-\cos(\theta_{\hat{a}\hat{b}}))d\phi_{\hat{a}\hat{b}}+\sum_{n=1}^{N_{\rm def}} \frac{(P_n,H_{I(\hat{a})})}{2}(\pm 1-\cos(\theta_{\hat{a}n}))d\phi_{\hat{a}n}~,\\
&\widetilde{\Theta}_{p}=d\psi_{p}+(\vec\CW_{\rm halo})_{pq}\cdot d\vec{y}^{\,q}~.
\end{split}\end{align}
In the last line, $(\vec\CW_{\rm halo})_{pq}$ is defined like $(\vec\CW_{2})_{pq}$ in the smooth case, \eqref{curlyW}, with the halo galaxy playing the role of galaxy two.  Additionally $\mathbf{\widetilde{\overline{C}}}$ is the direct sum metric on $\fMM_{\rm c} \times \MM_{{\rm h},0}$, where the first factor corresponds to the singular monopole metric that would be associated with the core galaxy in isolation, and the second factor corresponds to the (smooth) strongly centered monopole metric that would be associated with the halo galaxy in isolation.  Finally, $\overline{{\bf L}}$ and $\delta\overline{{\bf C}}$ are constant vector and matrix.  In particular, $\overline{{\bf L}} = (\overline{L}_{\hat{a}},\overline{L}_p)^T$ with
\begin{equation}
\overline{L}_{\hat{a}} = - (H_{I(\hat{a})}, \gamma_{{\rm h,m}})~, \qquad \overline{L}_{p} = \langle \beta_p, \gamma_{{\rm c,m}} \rangle~.
\end{equation}
%

\subsection{Framed BPS states for pure 't Hooft defects}

According to \cite{Moore:2015szp}, framed BPS states are identified with sections of the $L^2$ kernel of the twisted Dirac operator,
\begin{equation}\label{DiracfMM}
\slashed{D}^{\YY} = \slashed{D} - \I \slashed{\rG}(\YY)~,
\end{equation}
on the moduli space of singular monopoles.\footnote{Framed BPS states are boundstates of the field theory Hamiltonian rather than asymptotic one-particle states.  Correspondingly, the moduli space of singular monopoles does not have a center of mass factor that decouples, and framed BPS states are represented by $L^2$ zeromodes of the Dirac operator on the full moduli space.  Likewise, there is no decomposition of the Higgs data $\YY$ into center of mass and strongly centered pieces.}  The semiclassical data $\XX,\YY \in \mathfrak{t}$ defining the metric and Dirac operator is related to the physical data of the $\CN=2$  theory by picking a point $u\in \CB$ on the Coulomb branch and a phase $\zeta\in U(1)$  for the line defects and then defining the Higgs vevs $\CX$ and $\CY$ as
\be
\CX={\rm Im}~[\zeta^{-1} a(u)]~, \qquad \CY={\rm Im}~[\zeta^{-1}a_{\rm D}(u)]~.
\ee
As before, the choice of $\CX$ specifies a splitting of the charge lattice $\Gamma\to \CB$ by requiring that $\CX$ be in the fundamental Weyl chamber of $\ft$. Again, with respect to this splitting, a generic element $\gamma\in \Gamma$ decomposes  as $\gamma=\gamma_{\rm m}\oplus \gamma^{\rm e}$.

The electric charge is given by the eigenvalue of the semiclassical electric charge operator,
\begin{equation}
\hat{\gamma}^{\rm e} := \I \sum_{I=1}^{r} \alpha_{I} \Lie_{{\rm G}(h^I)}~.
\end{equation}
This operator commutes with the Dirac operator \eqref{DiracfMM} and hence the kernel can be graded by the eigenvalues $\gamma^{\rm e}$.  In the asymptotic region of singular monopole moduli space we have the analogous identifications to \eqref{exactTKVlimit} which results in 
\be
\hat\gamma^e=\I \sum_{\hat{i}=1}^{N}H^\ast_{I(\hat{i})}\pounds_{\partial_{\xi_{\hat{i}}}}~.
\ee
Under the asymptotic two-galaxy decomposition $\fMM\to \fMM_{\rm c}\times \CM_{{\rm h},0}\times \mathbb{R}_{\rm rel}^4$, this operator can be written as
\be
\hat\gamma^{\rm e}=\hat\gamma^{\rm e}_{\rm c}+ \hat{\gamma}_{\rm h}^{\rm e}~,
\ee
where 
\begin{align}\begin{split}
\hat\gamma^{\rm e}_{\rm c}=\I \sum_{\hat{a}=1}^{N_{\rm core}}H^\ast_{I(\hat{a})}\pounds_{\partial_{\xi_{\hat{a}}}} ~, \qquad
\hat\gamma^e_{\rm h}=  \I \gamma^\ast_{\rm h,m} \pounds_{\partial_{\uppsi}} + \I \sum_{\mathclap{p=N_{\rm core}+1}}^{N-1} \beta_{p} \pounds_{\partial_{\psi_{p}}} ~.
\end{split}\end{align}
The electric charge of the core and halo, $\gamma_{\rm c}^{\rm e}$ and $\gamma_{\rm h}^{\rm e}$, are given by the eigenvalues of $\hat\gamma^{\rm e}_{\rm c}$ and $\hat\gamma^{\rm e}_{\rm h}$ respectively.

Our goal is now to describe the explicit form of \eqref{DiracfMM} in the two-galaxy region of singular monopole moduli space.  As before we introduce the indices $\mu,\nu=1,... 4$ for the coordinates $y^{\mu i}=\{\vec{y}^{\,i},\psi_i\}$ and $R^\mu=\{\vec{R},\uppsi\}$ and the corresponding gamma matrices \eqref{gammamatrices}.  Since the metric \eqref{2singmetric} is identical in form to \eqref{hatmetric}, the result for large $R$ expansion of \eqref{DiracfMM} takes the form
\be\label{Dsingexp}
\Dsl^{\CY} = \Lambda\left( \Dsl_{{\rm c-h}}^{\CY} + \Dsl_{\rm rel}^{\CY} + O(1/R^2) \right)\Lambda^{-1}~.
\ee
This asymptotic Dirac operator commutes with the Lie derivatives $\{\Lie_{\pd_{\psi_i}}, \Lie_{\pd_{\uppsi}}\}$.  Let us denote the eigenvalues of these Lie derivatives by $\{\nu^i,\upnu\}$ as before.  Then $\slashed{D}_{\rm c-h}^{\YY}$ and $\slashed{D}_{\rm rel}^{\YY}$ restricted to a given $\{\nu^i,\upnu\}$ eigenspace are
\begin{align}
\left. \slashed{D}_{\rm c-h}^{\YY} \right|_{\{\nu^i,\upnu\}} =&~ \breve{\slashed{D}}_{\rm c-h}^{\YY} \otimes \mathbbm{1}_4 := \left( \slashed{D}_{\fMM_{\rm c} \times \MM_{{\rm h},0}}^{\YY} + (\breve{\slashed{D}}_{\rm c-h}^{\YY})^{(1)} \right) \otimes \mathbbm{1}_4~, \cr
\left. \slashed{D}_{\rm rel}^{\YY} \right|_{\{\nu^i,\upnu\}} =&~ \bar{\gamma}_{\rm c-h} \otimes \breve{\slashed{D}}_{\rm rel}^{\YY} ~,
\end{align}
where we've taken a basis of gamma matrices adapted to the asymptotic splitting $\SS(\fMM) \to \SS(\fMM_{\rm c} \times \MM_{{\rm h},0}) \times \SS(\mathbb{R}_{\rm rel}^4)$ of the spin bundle, with $\bar{\gamma}_{\rm c-h}$ the chirality operator on the first factor.  

Here, $\slashed{D}_{\fMM_{\rm c} \times \MM_{{\rm h},0}}^{\YY}$ is the Dirac operator on the product manifold, constructed using the Dirac operators on $\fMM_{\rm c}$ and $\MM_{{\rm h},0}$ in isolation.  $(\breve{\slashed{D}}_{\rm c-h}^{\YY})^{(1)}$ is the first order correction to this in the large $R$ expansion, which will have a form analogous to \eqref{D12correction}.  Finally, $\slashed{D}_{\rm rel}^{\YY}$, takes the same form as before, 
\begin{align}\label{DrelYsing}
\breve{\slashed{D}}_{\textrm{{\rm rel}}}^{\CY} :=&~ \frac{1}{\sqrt{m_{\rm halo}}} \left( 1 + \frac{ (\gamma_{{\rm c},{\rm m}}, \gamma_{{\rm h},{\rm m}})}{2 m_{\rm halo} R} \right) \bigg\{ \gamma^{\ualpha} \delta_{\ualpha}^{\phantom{\alpha}\alpha} \left[ \pd_{R^\alpha} - i p w_{\alpha} \right] - \I\gamma^{\underline{4}} \left( x - \frac{p}{2 R} \right) \bigg\} ~,
\end{align} 
but with $(\gamma_{1,{\rm m}},\gamma_{2,{\rm m}}) \to (\gamma_{\rm c,m},\gamma_{\rm h,m})$, $\mu \to m_{\rm halo}$,  and the parameters $x,p$ now given by
\begin{equation}\begin{split}
 x &:=  (\gamma_{\rm h,m}, \YY) - m_{\rm halo} \upnu = (\gamma_{\rm h,m}, \YY) + \langle \gamma_{\rm h}^{\rm e}, \CX \rangle~,\\
p &:= - \left[ (\gamma_{\rm c,m},\gamma_{\rm h,m}) \upnu - \overline{L}_i \nu^i  \right] =\llangle \gamma_{\rm c},\gamma_{\rm h}\rrangle ~.
\end{split}\end{equation}

The wavefunction in the asymptotic region of moduli space for a framed BPS state of definite electric charge can be written 
\be
\overline{\underline{\Psi}}^{(\gamma^{\rm e})}= \Lambda e^{\I \nu^i \psi_i + \I \upnu \uppsi }\overline{\underline{\Psi}}_{\nu^i,\upnu}(\vec{y}^i,\vec{R})~,
\ee
with
\begin{equation}
\gamma^{\rm e} = \gamma_{\rm c}^{\rm e} + \gamma_{\rm h}^{\rm e}~, \qquad \textrm{where} \quad \left\{ \begin{array}{l} \gamma_{\rm c}^{\rm e} =  -\sum_{\hat{a}=1}^{N_{\rm core}} \nu^{\hat{a}} H_{I(\hat{a})}^\ast~, \\[1ex]
\gamma_{\rm h}^{\rm e} = - \sum_{p=N_{\rm core}+1}^{N-1} \nu^p \beta_{p} - \gamma_{\rm h,m}^\ast \upnu~. \end{array} \right.
\end{equation}
Plugging this ansatz into $\slashed{D}^{\YY} \overline{\underline{\Psi}} = 0$, we again find normalizable solutions of the form 
\begin{align}\begin{split}\label{singsolutions}
\overline{\underline{\Psi}}_{\nu^i,\upnu}& = \left( \overline{\underline{\Psi}}^{(0)}_{{\rm c-h},\nu^i}(\vec{y}^{\,i}) + \frac{1}{R}\overline{\underline{\Psi}}^{(1)}_{{\rm c-h},\nu^i}(\vec{y}^{\,i},\uptheta,\upphi) + O(1/R^2)\right)
 \otimes \Psi_{\rm rel}^{(p)} (\vec{R})~,\\
\Psi_{\rm rel}^{(p)}& = R^j e^{-{\rm sgn}(p) x R}
\Psi_{j,m}^{(p)}(\uptheta,\upphi)~.
\end{split}\end{align}
In the first factor,
\begin{equation}
\overline{\underline{\Psi}}^{(0)}_{{\rm c-h},\nu^i}(\vec{y}^{\,i}) = \overline{\underline{\Psi}}_{\rm c}^{(\nu^{\hat{a}})}(\vec{y}^{\,\hat{a}}) \otimes \Psi_{{\rm h},0}^{(\nu^p)}(\vec{y}^{\,p})~,
\end{equation}
where $\overline{\underline{\Psi}}_{\rm c}^{(\nu^{\hat{a}})}$ is an $L^2$ zero mode of $\slashed{D}_{\fMM_{\rm c}}^{\YY}$ and $\Psi_{{\rm h},0}^{(\nu^p)}$ is an $L^2$ zero mode of $\slashed{D}_{\MM_{{\rm h},0}}^{\YY}$, where these operators are restricted to the corresponding $\nu$-eigenspaces.  These wavefunctions must be invariant under the exchange symmetry $S_{\rm c} \times S_{\rm h}$.  

Then $\overline{\underline{\Psi}}^{(1)}_{{\rm c-h},\nu^i}$ can be determined uniquely in terms of $\overline{\underline{\Psi}}^{(0)}$ in the same way as described under \eqref{firstperturbation}.  Finally, $\Psi_{\rm rel}^{(p)}$ is a zero mode of $\breve{\slashed{D}}_{\rm rel}^{\YY}$ just as before, where the angular momentum quantum number $j$ is constrained to $j = \half (|p|-1)$.

The same comments apply about the imposition of boundary conditions in the interior of the moduli space.  In arriving at the wavefunctions on $\mathbb{R}_{\rm rel}^4$ given in \eqref{singsolutions} we assumed that 
\be\label{framedell}
\ell_{\rm TN} = -\frac{(\gamma_{\rm c,m},\gamma_{\rm h,m})}{m_{\rm halo}}~,
\ee
which appears in the harmonic function via $\overline{H} = 1 + \ell_{\rm TN}/R$, is positive, and we imposed a regularity condition on the wavefunction at $R = 0$.  The model operator that we work with is therefore well-defined:  It is given by taking \eqref{Dsingexp}, dropping the $O(1/R^2)$ corrections, and replacing $\breve{\slashed{D}}_{\rm rel}^{\YY}$ by the Taub-NUT Dirac operator studied in Appendix \ref{app:AppD}.  It is then a valid question to ask whether or not this operator is a Fredholm deformation of the exact operator, \eqref{DiracfMM}.  As we will see momentarily, it apparently is when the core and halo electromagnetic charges are primitive, but not otherwise.  If, however, $\ell_{\rm TN} < 0$, then we would need to specify boundary conditions at $R = -\ell_{\rm TN}$ before we can even begin to address these points.  We leave this for future investigation.

The location of the walls of marginal stability and the change in the framed BPS spectrum are controlled by $x$ and $p$ respectively.  From the form of $\Psi_{\rm rel}$ we see that wavefunctions fail to be normalizable when $x\to 0$.  Hence the walls are at
\be
(\gamma_{\rm h,m}, \YY) + \langle \gamma_{\rm h}^{\rm e}, \CX \rangle=0~,
\ee
which agrees precisely with the computation from \cite{Moore:2015szp}.  Note that in the case of ${\rm PSU}({\mathfrak{N}})$ 't Hooft defects arising from $\mathfrak{su}({\mathfrak{N}}+1)$ monopoles in the $\XX_{\mathfrak{N}} \to \infty$ limit, this formula follows from the vanilla case, taking into account the restriction $\langle \gamma^{\rm e}, h^{\mathfrak{N}} \rangle=0$ on the total electric charge.

The leading large $R$ behavior of the solutions we have found is
\begin{equation}\label{L2singzmleading}
\overline{\underline{\Psi}}_{\nu^i,\upnu} \to  R^{j-1/2} e^{-\sgn(p)x R} \, \overline{\underline{\Psi}}_{\rm c}^{(\nu^{\hat{a}})}(\vec{y}^{\,\hat{a}}) \otimes \Psi_{0,{\rm h}}^{(\nu^p)}(\vec{y}^{\,p}) \otimes \Psi_{j,m}(\uptheta,\upphi)~, 
\end{equation}
and hence the number of states lost as we cross the wall is the product of the degeneracies associated with each factor.  If there is only a single defect present in the core, then the full moduli space has an ${\rm SO}(3)$ isometry, originating from rotations about the defect, and we can discuss index characters rather than simple degeneracies.  This leads to
\be
\Delta\fro(\gamma_{\rm c}+\gamma_{\rm h}, u;y)= \fro(\gamma_{\rm c},u;y) \, \Omega(\gamma_{\rm h},u;y) \, \chi_{|\llangle \gamma_{\rm c},\gamma_{\rm h}\rrangle|}(y)~,
\ee
thus reproducing the primitive wall-crossing formula from \cite{Gaiotto:2010be}.

The discussion of non-BPS boundstates in subsection \ref{ssec:nonBPS} is equally relevant here, since $\breve{\slashed{D}}_{\rm rel}^{\YY}$ is of exactly the same form.  Hence the same conclusions concerning the existence of stable non-BPS boundstates apply.

\section{Future Work}

We mention two directions for future work:
\begin{itemize}
\item In this paper we restricted consideration to pure vector-multiplet theories in the presence of pure 't Hooft defects.  Including matter representations and general Wilson--'t Hooft defects is accomplished by coupling the Dirac operator to a hyperholomorphic connection on a certain vector bundle over (singular) monopole moduli space \cite{Brennan:2016znk,Tong:2014yla,Gauntlett:2000ks}.  Work on extending the analysis of this paper to these cases is underway.
\item  We have shown that the two-galaxy region of monopole moduli space is associated with primitive wall-crossing phenomena.  It seems physically reasonable that the corrections to the primitive wall-crossing formula, due  to the fractionalization of non-primitive constituent charges, are associated with other asymptotic regions of moduli space.  There is clearly much to be understood, and it would be fascinating to gain a complete picture of wall-crossing in the semiclassical regime by utilizing the new analytic results on the compactification of monopole moduli space as a manifold with corners \cite{Kottke:2015rwx,FKS}.  For example, why is it that the jump in the kernel should only receive contributions from those corners of the moduli space associated with sub-partitions of a two-partition, corresponding to each of the two constituent electromagnetic charges being multiples of primitive charges?
\end{itemize}

\section*{Acknowledgements}

We thank Sergey Cherkis, Michael Singer, and Edward Witten for enlightening discussions.  GM and TBD are supported by the U.S. Department of Energy under grant DOE-SC0010008, and ABR is supported by the Mitchell Family Foundation.  ABR thanks the NHETC at Rutgers University for hospitality during the completion of this work.  

\appendix

\section{The Two-galaxy Region for Smooth Monopoles}\label{app:A}

In describing the two-galaxy limit of the asymptotic region of monopole moduli space, we introduce center of mass and relative coordinates according to \eqref{cmrelcovg1}.  Note that the difference of any two $\vec{x}^{\,\ha}$'s can be expressed as a linear combination of $\vec{y}^{\,a}$'s only.  The inverse transformations to \eqref{cmrelcovg1} are given explicitly by
\begin{equation}\label{cmrelcov2}
\vec{x}^{\,\ha} = \vec{X}_1 + ({\bf j}_1)^{\ha}_{\phantom{a}b} \vec{y}^{\,b} ~, \qquad \vec{x}^{\,\hp} = \vec{X}_2 + ({\bf j}_2)^{\hp}_{\phantom{p}q} \vec{y}^{\,q}~.
\end{equation}
where
\begin{equation}
{\bf j}_1 = \left( \begin{array}{c c c c} a_1 & a_2 & \cdots & a_{N_1-1}  \\ b_1 & a_2 & \cdots & a_{N_1-1} \\ b_1 & b_2 & \cdots & a_{N_1-1} \\ \vdots & \vdots & \ddots & \vdots \\ b_1 & b_2 & \cdots & b_{N_1-1} \end{array} \right)~, \quad {\bf j}_2 = \left( \begin{array}{c c c c} a_{N_1} & a_{N_1+1} & \cdots & a_{N-2}  \\ b_{N_1} & a_{N_1+1} & \cdots & a_{N-2} \\ b_{N_1} & b_{N_1+1} & \cdots & a_{N-2} \\ \vdots & \vdots & \ddots & \vdots \\ b_{N_1} & b_{N_1+1} & \cdots & b_{N-2} \end{array} \right) ~,
\end{equation}
with
\begin{align}\label{abcoef}
& a_a = \frac{m_{a+1} + \cdots + m_{N_1}}{m_{\rm gal1}}~, \quad b_a = - \frac{(m_1 + \cdots + m_a)}{m_{\rm gal1}}~, \quad (a = 1,\ldots, N_1 - 1)~, \cr
& a_p = \frac{m_{p+2} + \cdots + m_{N}}{m_{\rm gal2}}~, \quad b_p = - \frac{(m_{N_1 + 1} + \cdots + m_{p+1})}{m_{\rm gal2}}~, \quad (p = N_1,\ldots, N - 2)~. \qquad
\end{align}
These fit into square Jacobian matrices that give the map $( \{ \vec{y} \}, \vec{X})^T \mapsto \{ \vec{x} \}$ as follows:
\begin{align}\label{cmrelcov3}
(\vec{x}^{\ha}) =&~ ( {\bf j}_1 ~|~ {\bf 1}_1) \left( \begin{array}{c} \vec{y}^a \\ \vec{X}_1 \end{array} \right) \equiv {\bf J}_1  \left( \begin{array}{c} \vec{y}^a \\ \vec{X}_1 \end{array} \right)\quad,\qquad
(\vec{x}^{\hp}) = ( {\bf j}_2 ~|~ {\bf 1}_2) \left( \begin{array}{c} \vec{y}^s \\ \vec{X}_2 \end{array} \right) \equiv {\bf J}_2  \left( \begin{array}{c} \vec{y}^p \\ \vec{X}_2 \end{array} \right)~.
\end{align}
Here ${\bf 1}_1,{\bf 1}_2$ denote a length $N_{1},N_2$ column vector with all entries equal to $1$ respectively.  Similarly the angular coordinates transform as \eqref{fibercov}, or equivalently,
\begin{equation}\label{TKVcov}
\left( \frac{\pd}{\pd \xi^{\hat{a}}} \right) = {\bf J}_1 \left( \begin{array}{c} \frac{\pd}{\pd \psi^a} \\ \frac{\pd}{\pd \chi_1} \end{array} \right) ~, \qquad \left(  \frac{\pd}{\pd \xi^{\hat{s}}} \right) = {\bf J}_2 \left( \begin{array}{c} \frac{\pd}{\pd \psi^s} \\ \frac{\pd}{\pd \chi_2} \end{array} \right)~.
\end{equation}
And finally we implement the global center of mass and relative coordinates via \eqref{globalcom}.

In the limit that $R$ is much greater than all of the $y$, the matrix $M_{\hi\hj}$ has the structure
\begin{equation}
M_{\hi\hj} = \left( \begin{array}{c c} M_{\ha\hb} & \frac{D_{\ha\hq}}{R} + O(\frac{y}{R^2}) \\ \frac{D_{\hp\hb}}{R} + O(\frac{y}{R^2}) & M_{\hp\hq} \end{array} \right)~,
\end{equation}
where
\begin{align}
M_{\ha\hb} =&~ (M_1)_{\ha\hb} - \frac{1}{R} \delta_{\ha\hb} (H_{I(\ha)}, \gmtwo) + O\left(\frac{y}{R^2}\right)~, \cr
M_{\hp\hq} = &~(M_2)_{\hp\hq} - \frac{1}{R} \delta_{\hp\hq} (\gmone, H_{I(\hp)}) + O\left(\frac{y}{R^2}\right)~,
\end{align}
with
\begin{align}
(M_1)_{\ha\hb} = \left\{ \begin{array}{l l} m_{\ha} - \sum_{\hc\neq \ha} \frac{D_{\ha\hc}}{r_{\ha\hc}} ~, & \ha = \hb~, \\ \frac{D_{\ha\hb}}{r_{\ha\hb}}~, & \ha \neq \hb~, \end{array} \right.\quad \qquad
(M_2)_{\hp\hq} = \left\{ \begin{array}{l l} m_{\hp} - \sum_{\hr\neq \hp} \frac{D_{\hp\hr}}{r_{\hp\hr}} ~, & \hp = \hq~, \\ \frac{D_{\hp\hq}}{r_{\hp\hq}}~, & \hp \neq \hq~. \end{array} \right.
\end{align}
The latter are the matrices that would appear in the GM/LWY metrics for galaxies one and two in isolation.  Thus in the two-galaxy limit, the ``mass matrix'' takes the form
\begin{align}
M_{\hi\hj} =&~ \left( \begin{array}{c c} (M_1)_{\ha\hb} & 0 \\ 0 & (M_2)_{\hp\hq} \end{array} \right) + \frac{1}{R} \left( \begin{array}{c c} -\delta_{\ha\hb} (H_{I(\ha)},\gmtwo) & D_{\ha\hq} \\ D_{\hb\hp} & -\delta_{\hp\hq} (\gmone, H_{I(\hp)}) \end{array} \right) + O\left( \frac{y}{R^2} \right)~,
\end{align}
which we will write as
\begin{equation}\label{Mmatrixexp}
{\rM} = \left( \begin{array}{c c} \rM_1 & 0 \\ 0 & \rM_2 \end{array} \right) - \frac{1}{R} \left( \begin{array}{c c} {\bf D}_{11} & -{\bf D}_{12} \\ -{\bf D}_{21} & {\bf D}_{22} \end{array} \right) + O\left( \frac{y}{R^2} \right)~,
\end{equation}
where ${\bf D}_{11}$ and ${\bf D}_{22}$ are diagonal matrices and ${\bf D}_{21} = ({\bf D}_{12})^T$.

Then within each galaxy we make the change of variables \eqref{cmrelcov3}.  The relevant quantities to be computed are, at leading order,
\begin{equation}\label{Mt1diag}
{\bf J}_1^T \rM_1 {\bf J}_1 = \left( \begin{array}{c c} {\bf C}_1 & 0 \\ 0 & m_{\rm gal1} \end{array} \right) ~,\qquad {\bf J}_2^T \rM_2 {\bf J}_2 = \left( \begin{array}{c c} {\bf C}_2 & 0 \\ 0 & m_{\rm gal2} \end{array} \right)~,
\end{equation}
where ${\bf C}_{1} = \bf{j}_{1}^T {\bf M}_{1} {\bf j}_{1}$, \etc.
The $O(1/R)$ terms are
\begin{align}\label{deltaCpieces}
{\bf J}_{1}^T {\bf D}_{11} {\bf J}_1 =&~  \left( \begin{array}{c c} {\bf j}_{1}^T {\bf D}_{11} {\bf j}_1 & \langle \beta_a,\gamma_{2,{\rm m}} \rangle \\ \langle \beta_b, \gamma_{2,{\rm m}}\rangle & (\gamma_{1,{\rm m}}, \gamma_{2,{\rm m}}) \end{array} \right)~,\quad
{\bf J}_{2}^T {\bf D}_{22} {\bf J}_2 =&~ \left( \begin{array}{c c} {\bf j}_{2}^T {\bf D}_{22} {\bf j}_2 & \langle \beta_p,\gamma_{1,{\rm m}}\rangle \\ \langle \beta_q, \gamma_{1,{\rm m}} \rangle & (\gamma_{1,{\rm m}}, \gamma_{2,{\rm m}}) \end{array} \right)~, \nonumber \\[1ex]
{\bf J}_{1}^T {\bf D}_{12} {\bf J}_2 =&~ \left( \begin{array}{c c} {\bf j}_{1}^T {\bf D}_{12} {\bf j}_2 & \langle \beta_a,\gamma_{2,{\rm m}}\rangle \\ \langle \beta_q , \gamma_{1,{\rm m}} \rangle & (\gamma_{1,{\rm m}}, \gamma_{2,{\rm m}}) \end{array} \right)~,
\end{align}
where
\begin{equation}\label{beta1def}
\beta_a := ({\bf J}_{1}^T)_{a}^{\phantom{a}\hat{b}} H_{I(\hat{b})}^\ast = a_a \sum_{\hat{c}=1}^{a}H_{I(\hat{c})}^\ast + b_a \sum_{\hat{c}=a+1}^{N_1}H_{I(\hat{c})}^\ast~, \qquad a = 1,\ldots,N_1 - 1~,
\end{equation}
and similarly
\begin{equation}\label{beta2def}
\beta_p := ({\bf J}_{2}^T)_{p}^{\phantom{p}\hat{q}} H_{I(\hat{q})}^\ast =  a_p \sum_{\hat{r}=N_1 + 1}^{p+1}H_{I(\hat{r})}^\ast + b_p \sum_{\hat{r}=p+2}^{N}H_{I(\hat{r})}^\ast~, \qquad p = N_1 , \ldots, N-2 ~,
\end{equation}
with the $a$ and $b$ coefficients given in \eqref{abcoef}.  The key property to note of the $\beta_{a,p}$ is that they have zero pairing with $\XX$; for example we have $\langle \beta_a, \XX\rangle = -a_a b_a + b_a a_a = 0$.

Thus far we have described the transformation of the quadratic form $M_{\hi\hj}$ from the basis of differentials $d\vec{x}^{\,\hi}$ to the basis $(d\vec{y}^{\,a},d\vec{X}_1,d\vec{y}^{\,p},d\vec{X}_2)^T$.  Next we implement the transformation $(\vec{X}_1,\vec{X}_2) \mapsto (\vec{X}, \vec{R})$ in \eqref{globalcom}.  One finds that the quadratic form diagonalizes with respect to the overall center of mass coordinate, with this piece accounting for the first term on the right-hand side of \eqref{metproduct}.  Then collecting the remaining differentials into the block structure $(d\vec{{\bf y}}, d\vec{R}) = (d\vec{y}^a,d\vec{y}^p, d\vec{R})$, one obtains the first line of \eqref{hatmetric} with
\begin{align}\label{relquantities}
&  \wtbC := \left( \begin{array}{c c} {\bf C}_1 & 0 \\ 0 & {\bf C}_2 \end{array} \right)~,  \quad \delta {\bf C} := \left( \begin{array}{c c} - {\bf j}_{1}^T {\bf D}_{11} {\bf j}_1 & {\bf j}_{1}^T {\bf D}_{12} {\bf j}_2 \\ {\bf j}_{2}^T {\bf D}_{21} {\bf j}_1 & - {\bf j}_{2}^T {\bf D}_{22} {\bf j}_2 \end{array} \right)~, \cr
& {\bf L} := \left( \begin{array}{c} -\langle \beta_a, \gamma_{2,{\rm m}}\rangle \\ \langle \beta_p,\gamma_{1,{\rm m}} \rangle \end{array} \right)~, \qquad H(R) := 1 - \frac{(\gamma_{1,{\rm m}}, \gamma_{2,{\rm m}})}{\mu R}~.
\end{align}
Note that $\delta{\bf C}$ and ${\bf L}$ are coordinate independent.

Now we turn to the connection one-forms on the $N$-torus, $\Theta_{\hi}$.  We change variables in the fiber coordinates according to \eqref{fibercov}.  Denoting ${\bf J} = \diag( {\bf J}_1, {\bf J}_2)$, the quantity we want to investigate therefore is ${\bf J}^T \vec{{\bf W}} {\bf J}$, where $\vec{{\bf W}}$ is the matrix with components $\vec{W}_{\hi\hj}$ given in \eqref{MW}.  The reason for the ${\bf J}^T$ on the left is that we want $\Theta_{\hi}$ to transform like the legs along the fiber directions, \eqref{fibercov}.  The overall factors of ${\bf J}^T$ will then be of the right form to transform the inverse quadratic form, $(M^{-1})^{\hi\hj}$, to the $y$-$X$ basis.  The reason for the ${\bf J}$ on the right of $\vec{{\bf W}}$ is that we will put a ${\bf J} {\bf J}^{-1}$ between $\vec{W}_{\hi\hj}$ and $d\vec{x}^{\,\hj}$, using the ${\bf J}^{-1}$ to map the $d\vec{x}^{\,\hj}$ to $(d\vec{y}^{\,b},d\vec{X}_1,d\vec{y}^{\,q},d\vec{X}_2)^T$.  

However we only need $\vec{W}_{\hi\hj}$ through $O(1/R)$, which takes the form
\begin{equation}
(\vec{W}_{\hi\hj}) = \left( \begin{array}{c c} (\vec{W}_1)_{\ha\hb} & 0 \\ 0 & (\vec{W}_2)_{\hp\hq} \end{array} \right) + \left( \begin{array}{c c} -\delta_{\ha\hb} (H_{I(\ha)},\gmtwo) & D_{\ha\hq} \\ D_{\hb\hp} & -\delta_{\hp\hq} (\gmone, H_{I(\hp)}) \end{array} \right) \vec{w}(\vec{R}) + O\left(\frac{y}{R^2}\right)~,
\end{equation}
where $\vec{W}_{1,2}$ are the corresponding $\vec{W}$'s for galaxies one and two in isolation.  Wrapping the ${\bf J}^T$-${\bf J}$ around the first term, we observe that
\begin{equation}
{\bf J}_{1}^T \vec{{\bf W}}_1 {\bf J}_1 = \left(\begin{array}{c c} {\bf j}_{1}^T \vec{{\bf W}}_1 {\bf j}_{1} & {\bf 0} \\ {\bf 0}^T & 0 \end{array} \right)~,
\end{equation}
and similarly for $1 \mapsto 2$.  In the text we denoted the upper-left $(N_{1,2} - 1) \times (N_{1,2} -1)$ corners of thes expressions by $\vec{\boldsymbol{\WW}}_{1,2}$ respectively:
\begin{equation}\label{curlyW}
\vec{\boldsymbol{\WW}}_1 := {\bf j}_{1}^T \vec{{\bf W}}_1 {\bf j}_{1} ~, \qquad  \vec{\boldsymbol{\WW}}_2 := {\bf j}_{2}^T \vec{{\bf W}}_2 {\bf j}_{2} ~.
\end{equation}

For the ${\bf J}$-transformation of the $O(1/R)$ terms we can make use of \eqref{deltaCpieces}.  We then make the final transformation, corresponding to \eqref{globalcom}, on the relevant two-by-two block of ${\bf J}^T \vec{{\bf W}} {\bf J}$ --- that is, the block whose rows correspond to $\chi_1,\chi_2$ and whose columns correspond to $\vec{X}_1, \vec{X}_2$.  Making use of the definitions \eqref{relquantities}, one eventually finds
\begin{align}
\Theta_i (M^{-1})^{ij} \Theta_j =&~ \frac{d\chi^2}{(m_{\rm gal1} +m_{\rm gal2})} +  \left( \Theta_0, \Theta_\uppsi \right) \left( \begin{array}{c c} \wtbC + \frac{1}{R} \delta\mathbf{ C} & \frac{1}{R} {\bf L} \\  \frac{1}{R} {\bf L}^T & \mu H(R) \end{array} \right)^{-1} \left( \begin{array}{c} \Theta_0 \\ \Theta_{\uppsi} \end{array} \right)  +O\left(\frac{y}{R^2}\right),
\end{align}
with $(\Theta_0,\Theta_{\uppsi})$ defined as in \eqref{Thetarels1}, \eqref{Thetarels2}.  The first term on the right here is the remaining contribution to the metric on the center of mass factor in \eqref{metproduct}, while the last term completes our derivation of \eqref{hatmetric}.

\subsection{Hyperk\"ahlerity of the metric}\label{app:hk}

Here we address the hyperk\"ahlerity of the asymptotic metric \eqref{hatmetric} in the two-galaxy region of the strongly centered moduli space.  We collect the position and phase coordinates using indices $\ti,\tj = 1,\ldots, N-1$ and writing
\begin{equation}\label{ypsicollection}
\vec{y}^{\,\ti} = (\vec{y}^{\,a}, \vec{y}^{\,p}, \vec{R}) = (\vec{{\bf y}}, \vec{R})~, \qquad \psi_{\ti} = (\psi_a, \psi_p, \uppsi)~,
\end{equation}
the metric has the form
\begin{equation}\label{GHmetric}
d \hat{s}_{0}^2 = G_{\ti \tj} d\vec{y}^{\,\ti} \cdot d\vec{y}^{\,\tj} + (G^{-1})^{\ti \tj} \left( d\psi_{\ti} + \vec{V}_{\ti \tk} \cdot d\vec{y}^{\,\tk} \right) \left( d\psi_{\tj} + \vec{V}_{\tj \tl} \cdot d\vec{y}^{\,\tl} \right)~,
\end{equation}
where the matrices ${\bf G},\vec{{\bf V}}$ are given by
\begin{align}
{\bf G} :=&~ \left( \begin{array}{c c} \wtbC + \frac{1}{R} \delta{\bf C} & \frac{1}{R} {\bf L} \\  \frac{1}{R} {\bf L}^T & \mu H(R) \end{array} \right)\quad,\qquad
\vec{{\bf V}} := \left( \begin{array}{c c} \vec{\boldsymbol{\WW}}  + \delta{\bf C} \otimes \vec{w}(\vec{R}) &  {\bf L} \otimes \vec{w}(\vec{R}) \\ {\bf L}^T \otimes \vec{w}(\vec{R}) & -(\gamma_{1,{\rm m}}, \gamma_{2,{\rm m}}) \vec{w}(\vec{R}) \end{array} \right)~,
\end{align}
where $\vec{\boldsymbol{\WW}} = \diag(\vec{\boldsymbol{\WW}}_1, \vec{\boldsymbol{\WW}}_2)$.  Note that ${\bf G}, \vec{{\bf V}}$ are of the form
\begin{equation}\label{ourGV}
{\bf G} = \widetilde{{\bf G}} + \frac{1}{R} {\bf A}~, \qquad \vec{{\bf V}} = \vec{\widetilde{{\bf V}}} + {\bf A} \otimes \vec{w}(\vec{R})~,
\end{equation}
where $\widetilde{{\bf G}} = \diag(\wtbC, \mu)$, $\vec{\widetilde{{\bf V}}} = \diag(\vec{\boldsymbol{\WW}},0)$, and ${\bf A}$ is a constant matrix.

A metric of the form \eqref{GHmetric} is hyperk\"ahler iff \cite{MR953820} (letting $\alpha,\beta,\gamma = 1,2,3$)
\begin{equation}\label{hkrelns}
\frac{\pd}{\pd y^{\alpha \ti}} V_{\beta \tj \tk} - \frac{\pd}{\pd y^{\beta \tj}} V_{\alpha \ti \tk} = \epsilon_{\alpha\beta}^{\phantom{\alpha\beta}\gamma} \frac{\pd}{\pd y^{\gamma \ti}} G_{\tj \tk} \qquad \& \qquad \frac{\pd}{\pd y^{\alpha \ti}} G_{\tj \tk} = \frac{\pd}{\pd y^{\alpha \tj}} G_{\ti \tk} ~.
\end{equation}
These equations are satisfied on the leading order pieces, $({\bf G},{\bf V}) \to (\widetilde{{\bf G}},\widetilde{{\bf V}})$ because this just gives a direct product metric on $\MM_{1,0} \times \MM_{2,0} \times \mathbb{R}_{\rm rel}^4$ with the corresponding GM/LWY hyperk\"ahler metrics on the first two factors, (the strongly centered moduli spaces for galaxies one and two in isolation), and a flat metric on the third.  Since ${\bf A}$ is constant, the only derivatives that do not annihilate the correction term are those involving derivatives with respect to the components of $\vec{R}$.  It follows that the equations are indeed satisfied to order $O(1/R^2)$, and hence the metric \eqref{hatmetric} is hyperk\"ahler to the relevant order.

\section{Strong Centering and Deck Transformations for GM/LWY}\label{app:deck}

The universal cover of monopole moduli space splits according to $\widetilde{\MM} = \mathbb{R}^4 \times \MM_0$ and is acted on by the discrete group of deck transformations $\mathbb{D} \cong \mathbb{Z}$.  Here we show how the deck transformations and the strongly centered space $\MM_0$ arise in the asymptotic description of Gibbons--Manton and Lee--Weinberg--Yi.  In order to minimize the amount of additional notation we'll need, we discuss this for the moduli space associated to galaxy 1 in isolation.  Equivalently, we consider the trivial split in \eqref{2gsplit} where ${\rm S}_2$ is empty, so that $N = N_1$ and $\hat{i},\hat{j} \to \hat{a},\hat{b}$ \etc.

We will obtain a description of $\mathbb{D}$ by first understanding a certain subgroup $\mathbb{D}_{\rm g} \subset \mathbb{D}$.  We recall from \cite{Moore:2015szp,Moore:2015qyu} the description of this subgroup.  The ${\rm G}$-map furnishes a hyperk\"ahler torus action on $\MM$, with the triholomorphic Killing fields $\rG(h^I)$ generating $2\pi$-periodic isometries.  This induces a homomorphism $\mu: \mathbb{Z}^r \cong \Lambda_{\rm mw} \to \mathbb{D} \cong \mathbb{Z}$, the image of which is the subgroup $\mathbb{D}_g$.  Using the rational map construction it was proven in \cite{Moore:2015szp} that $\mathbb{D}_g \cong \ell \cdot \mathbb{Z}$, where $\ell = \gcd_I \{ n_{\rm m}^I \cp^I \}$, and furthermore that if $\phi$ is the generator of deck transformations and $h \in \Lambda_{\rm mw}$, we have \eqref{muhomo} with $\mu(h) = (\gm,h)$.  

In order to understand the $\mathbb{D}_{\rm g}$ action in the GM/LWY context, we first separate constituent monopoles by their type, $I = 1,\dots,r$.  Then, within each type, we choose an ordering (in general, non-unique) of the coordinates so that the centers of constituents of a given type $I$ are consecutive.  Thus, we will denote such centers as 
$\vec x^{\, \ha_I}$ where $\hat{a}_I$ runs from $1$ to $n_m^I$.  Proceeding similarly for the phases we then have 
\begin{equation}
\{ \vec{x}^{\, \hat{a}} \}_{\ha=1}^N = \bigcup_{I=1}^r \left\{ \vec{x}^{\, \ha_I} ~|~ \ha_I=1,\ldots, n_{\rm m}^I \right\}~, \qquad \{ \xi_{\ha} \}_{\ha=1}^N = \bigcup_{I=1}^r \left\{ \xi_{\ha_I} ~|~ \ha_I =1,\ldots,n_{\rm m}^I \right\}~.
\end{equation}

Then, within each type, we pass to center of mass and relative coordinates.  For the positions we define
\begin{equation}
\vec{X}^{I} = \frac{1}{n_{\rm m}^I} \sum_{\ha_I=1}^{n_{\rm m}^I} \vec{x}^{\, \ha_I} ~, \qquad \vec{y}^{\, a_I} = \vec{x}^{\, a_I} - \vec{x}^{\, a_I+1} ~, \quad a_I =1,\ldots, n_{\rm m}^I -1~,
\end{equation}
and for the phases,
\begin{equation}\label{phasebytype}
\chi^I = \sum_{\ha_I=1}^{n_{\rm m}^I} \xi_{\ha_I} ~, \qquad \psi_{a_I} = \sum_{b_I=1}^{a_I} \xi_{b_I} - \frac{a_I}{n_{\rm m}^I} \chi^I~, \quad a_I=1,\ldots, n_{\rm m}^I -1~.
\end{equation}
These transformations are dual.  Following the pattern of the previous appendix we define ${\bf J}^I$ so that the linear transformation $\{ \xi_{\ha_I} \} \mapsto \{ \psi_{a_I}, \chi^I \}$ is given by $({\bf J}^I)^T$.  Then the transformation $\{ \vec{x}^{\, \ha_I} \} \mapsto \{ \vec{y}^{\, a_I}, \vec{X}^I \}$ corresponds to $({\bf J}^I)^{-1}$.  Furthermore, with regards to the exchange symmetry \eqref{exchangesym}, the original $\{ \vec{x}^{\, \ha_I} \}$ transform in the reducible $n_{\rm m}^I$-dimensional representation of the symmetric group $S_{n_{\rm m}^I}$, $\vec{X}^I$ in the singlet, and $\{ \vec{y}^{\, a_I} \}$ in the irreducible $(n_{\rm m}^I - 1)$-dimensional ``standard'' representation.  For the phases, $\chi^I$ is also a singlet while the $\{ \psi_{a_I} \}$ transform in the dual of the standard representation.

The $\chi^I$ provide an explicit parameterization the torus action of hyperholomorphic isometries, \eqref{Gmap}, in the asymptotic region of moduli space.  To see this we note that a global gauge transformation generated by an element in the Cartan proportional to the magnetic weight $h^I$ will simultaneously rotate the phases of all monopoles of type $I$ by the same amount, while leaving the phases of all other monopoles invariant.  This indeed translates $\chi^I$ while leaving the $\psi_{a_I}$ fixed. In order to determine the precise relationship between $\rG(h^I)$ and $\pd_{\chi^I}$ we must discuss periodicities.  Since $\xi_{\ha_I} \sim \xi_{\ha_I} + 2\pi \cp^I$, it follows from \eqref{phasebytype} that
\begin{equation}\label{psiperiods1}
\chi^{I} \sim \chi^I + 2\pi \cp^I n_{\rm m}^I \qquad \textrm{and} \qquad \psi_{a_I} \sim \psi_{a_I} + 2\pi \cp^I~.
\end{equation}
To see the latter, shift $\xi_{a_I}$ forward and $\xi_{a_I + 1}$ backward by $2\pi \cp^I$; for the former, shift all $\xi_{\ha_I}$ forward by $2\pi \cp^I$.  In particular, since $\rG(h^I)$ generates $2\pi$-periodic isometries, we must have the asymptotic identification
\begin{equation}
\rG(h^I) \longrightarrow  \cp^I n_{\rm m}^I \pd_{\chi^I} = \cp^I \sum_{k=1}^{n_{\rm m}^I} \pd_{\xi_{I_k}} ~.
\end{equation}
The last step follows from the fact that the linear transformation $\{ \pd_{\psi_{a_I}} , \pd_{\chi^I} \} \mapsto \{ \pd_{\xi_{\ha_I}} \}$ is given by $({\bf J}^I)^{-1}$, the same transformation that sends $\{ \vec{x}^{\, \ha_I} \}$ to $\{ \vec{y}^{\, a_I}, \vec{X}^I \}$.

In order to make contact with the split, $\widetilde{\MM} = \mathbb{R}^4 \times \MM_0$, of the universal cover, we must transform from the $\vec{X}^I$ and $\chi^I$ to overall center of mass coordinates and relative coordinates.  For the center of mass coordinates we take
\begin{align}\label{typecentertocenter}
\vec{X} =&~ \frac{1}{(\gm, \XX)} \sum_{I=1}^r n_{\rm m}^I (H_I,\XX) \vec{X}^I = \frac{1}{(\gm,\XX)} \sum_{\hat{a}=1}^N m_{\hat{a}} \vec{x}^{\, \hat{a}}~, \cr
\chi =&~ \sum_{I=1}^r \chi^I = \sum_{\ha =1}^N \xi_{\ha}~,
\end{align}
which agrees with previous definitions.  

There is some freedom in how we define the relative coordinates, which we will denote by $\vec{Y}^I$ and $\uppsi_I$, $I = 1,\ldots, r-1$.  It is convenient to define the phases so that $\pd_{\uppsi_I} = \rG(h_{0}^I)$, where $\{ h_{0}^I ~|~ I=1,\ldots,r-1 \}$ is an integral basis for $\ker{\mu} \subset \Lambda_{\rm mw}$.  On the one hand, $h_{0} \in \ker{\mu}$ means $(\gm, h_0) =0$, and therefore, by \eqref{killing2metric}, $\rG(h_0)$ will be metric orthogonal to $\rG(\XX)$ and hence a triholomorphic Killing field on the strongly centered space $\MM_0$.  On the other hand, $h_0 \in \ker{\mu} \subset \Lambda_{\rm mw}$ means that $\rG(h_0)$ generates a periodic isometry of $\MM_0$, and if $\{ h_{0}^I \}$ is an integral basis for $\ker{\mu}$ then the $\rG(h_{0}^I)$ will generate $2\pi$-periodic isometries.

In Appendix C.3 of \cite{Moore:2015szp} we constructed an explicit $V \in \mathrm{SL}(r,\mathbb{Z})$ that maps the fundamental magnetic weights to a basis for $\ker{\mu}$ together with an element $h_{\rm g} \in \Lambda_{\rm mw}$ such that $\mu(h_{\rm g}) = \ell$.  Let us denote $h_{\rm g} \equiv h_{0}^r$.  Then, following the conventions of \cite{Moore:2015szp},
\begin{equation}
h_{0}^I = \sum_{J} (V^T)^{I}_{\phantom{I}J} h^J~,
\end{equation}
where $V^T$ is the transpose of $V$.  We introduce $2\pi$-periodic phases $\widetilde{\uppsi}_I$ such that $\pd_{\widetilde{\uppsi}_I} = \rG(h_{0}^I)$.  In particular the generator of $\mathbb{D}_{\rm g}$ acts by sending $\widetilde{\uppsi}_r \mapsto \widetilde{\uppsi}_r + 2\pi$ with the other $\widetilde{\uppsi}_I$ fixed.  (The $\uppsi_I$ that we are after will be related to these by another transformation as described below.)  In terms of the $\widetilde{\uppsi}_I$ we then have
\begin{equation}
\pd_{\widetilde{\uppsi}_I} = \sum_{J=1}^r (V^T)^{IJ} \ell^J \pd_{\chi^J} \quad \Rightarrow \quad \chi^I = \ell^I \sum_{J=1}^r V^{IJ} \widetilde{\uppsi}_J ~,
\end{equation}
where $\ell^I \equiv n_{\rm m}^I \cp^I$.  

The transformation is given by $V = V_{(1)} \cdots V_{(r-1)}$, where $V_{(I)}$ has the form of the identity matrix except for a $2 \times 2$ block along the diagonal, spanning the $I^{\rm th}$ and $(I+1)^{\rm th}$ row and column, where it takes the form
\begin{equation}
\left(\begin{array}{c c} \frac{\ell^{I+1}}{\lambda_{1\cdots (I+1)}} & x_{1\cdots I,I+1} \\[1ex] -\frac{\lambda_{1 \cdots I}}{\lambda_{1 \cdots (I+1)}} & y_{1\cdots I,I+1} \end{array} \right)~, \qquad \textrm{with} \quad \lambda_{1 \cdots I} := \gcd(\ell^1,\cdots,\ell^I)~,
\end{equation}
and where $x_{1\cdots I,I+1},y_{1\cdots I,I+1}$ is any pair of integers satisfying B\'ezout's identity: 
\begin{equation}\label{Bezout}
x_{1 \cdots I,I+1} \lambda_{1 \cdots I} + y_{1\cdots I,I+1} \ell^{I+1} = \lambda_{1 \cdots (I+1)}~.
\end{equation}
Note that $\lambda_{1\cdots r} = \ell$.  Also, if one solution to this equation is denoted $\{ x_{1\cdots I,I+1}^\ast, y_{1\cdots I,I+1}^\ast \}$, then all solutions are of the form $\{ x_{1\cdots I,I+1}^\ast + n \ell^I/\lambda_{1 \cdots (I+1)}, y_{1\cdots I,I+1}^\ast - n \lambda_{1\cdots I}/\lambda_{1\cdots (I+1)} \}$ for $n\in \mathbb{Z}$.  We will make some use of this shift freedom below.

Using these expressions we can determine how the generator of $\mathbb{D}_{\rm g}$ acts on the original $\chi^I$.  Let us denote this isometry by $\phi_{\rm g}$.  Setting $\widetilde{\uppsi}_I \xmapsto{\phi_{\rm g}} \widetilde{\uppsi}_I$ for $I < r$ and $\widetilde{\uppsi}_r \xmapsto{\phi_{\rm g}} \widetilde{\uppsi}_r + 2\pi$, we find
\begin{align}\label{Dgaction1}
\chi^1 \xmapsto{\phi_{\rm g}}&~ \chi^1 +  2\pi \ell^1 x_{1 \cdots (r-1),r} \cdots x_{12,3} x_{1,2}~, \cr
\chi^2 \xmapsto{\phi_{\rm g}}&~ \chi^2 +  2\pi \ell^2 x_{1 \cdots (r-1),r} \cdots x_{12,3} y_{1,2} ~, ~ \cdots \cr
\chi^I \xmapsto{\phi_{\rm g}}&~ \chi^I + 2\pi \ell^I x_{1 \cdots (r-1),r} \cdots x_{1 \cdots I, I+1} y_{1 \cdots (I-1),I}~, \quad 1 < I < r ~, ~ \cdots \cr
\chi^r \xmapsto{\phi_{\rm g}}&~ \chi^r + 2\pi \ell^r y_{1\cdots (r-1),r} ~.
\end{align}
One can check directly from these that $\chi \xmapsto{\phi_{\rm g}} \chi + 2\pi \ell$.  We will see this emerge from a different computation shortly.  We can also invert $V$ to obtain the $\widetilde{\uppsi}_I$ in terms of the $\chi^I$.  The result is
\begin{align}\label{chistotildepsis}
\widetilde{\uppsi}_I =&~ \frac{y_{1\cdots I,I+1}}{\lambda_{1 \cdots I}} (\chi^1 + \cdots + \chi^I ) - \frac{x_{1\cdots I,I+1}}{\ell^I} \chi_{I+1}~, \qquad I = 1,\ldots, r-1~, \cr
\widetilde{\uppsi}_r =&~ \frac{1}{\ell} (\chi^1 + \cdots + \chi^r)~.
\end{align}

Now let's transform from the $\{ \widetilde{\uppsi}_I \}_{I=1}^r$ to $\{\uppsi_I\}_{I=1}^{r-1}$ and $\chi$.  We set $\pd_{\widetilde{\uppsi}_I} = \rG(h_{0}^I) = \pd_{\uppsi_I} $ for $I=1,\ldots,r-1$, so we just need to express $\pd_{\widetilde{\uppsi}_r}$ in terms of $\pd_{\chi}$ and $\pd_{\uppsi_I}$.  This was also worked out in Appendix C.3 of \cite{Moore:2015szp}.  The result is
\begin{align}
\pd_{\widetilde{\uppsi}_r} =&~ \ell \pd_{\chi} + \ell \sum_{I=1}^{r-1} c_I \pd_{\uppsi_I} ~, \qquad \textrm{with} \cr
c_I =&~ \frac{x_{1\cdots I,I+1}}{\ell^{I+1}} \cdot \frac{ n_{\rm m}^{I+1} (H_{I+1},\XX)}{(\gm,\XX)} - \frac{y_{1\cdots I,I+1}}{\lambda_{1\cdots I}} \cdot \frac{\sum_{J=1}^I n_{\rm m}^J (H_J, \XX)}{(\gm,\XX)} ~.
\end{align}
These formulae imply the relations
\begin{equation}\label{psitildetopsi}
\uppsi_I = \widetilde{\uppsi}_I + \ell c_I \widetilde{\uppsi}_r ~, \quad I = 1,\ldots,r-1~, \qquad \chi = \ell \widetilde{\uppsi}_r~.
\end{equation}
This shows explicitly how $\phi_{\rm g}$ acts on $\mathbb{R} \times \MM_0$, where the first factor is parameterized by $\chi$.  Recall that this isometry shifts $\widetilde{\uppsi}_r$ by $2\pi$ while holding $\widetilde{\uppsi}_I$ fixed.  Hence
\begin{equation}\label{Dgaction2}
\uppsi_I \xmapsto{\phi_{\rm g}} \uppsi_I + 2\pi \ell c_I ~, \qquad \chi \xmapsto{\phi_{\rm g}} \chi + 2\pi \ell~.
\end{equation}
Note that the translation along the hyperk\"ahler torus of $\MM_0$ does not close for generic Higgs vevs.    \eqref{psitildetopsi} combined with \eqref{chistotildepsis} gives the transformation of the phases from the $\{\chi^I\}_{I=1}^r$ to the $\{ \uppsi_I \}_{I=1}^{r-1}$ and $\chi$.  As a check, the expression for $\chi$ reproduces \eqref{typecentertocenter}.

The transformation of position coordinates from $\{ \vec{X}^I \}_{I=1}^r$ to $\{\vec{Y}^I\}_{I=1}^{r-1}$ and $\vec{X}$ should be chosen to be dual -- \ie\ the position coordinates should transform as the vector fields of the phase coordinates.  We find
\begin{equation}
\vec{Y}^I = \frac{\ell^{I+1}}{\lambda_{1\cdots(I+1)}} \left( \lambda_{1\cdots I} (\vec{X}^I - \vec{X}^{I+1} ) + \sum_{J=1}^{I-1} x_{1\cdots J,J+1} \cdots x_{1\cdots (I-1),I} \lambda_{1\cdots J} (\vec{X}^{J} - \vec{X}^{J+1}) \right)~,
\end{equation}
while $\vec{X}$ is given as in \eqref{typecentertocenter}.  Note in the special case that all $\ell^I = 1$, we can set all of the $x$'s to zero, the $y$'s to one, and the $\lambda$'s to one.  Then these transformations reduce to the form we had in the previous appendix: $\vec{Y}^I = \vec{X}^I - \vec{X}^{I+1}$. 

Let us summarize what we have learned so far.  The asymptotic region of $\MM_0$ can be parameterized by position coordinates $\{ \vec{y}^{\, a_I} ; \vec{Y}^I \}$ and phase coordinates $\{ \psi_{a_I} ; \uppsi_I \}$, up to identification by the action of the exchange symmetry $S$.  The $I^{\rm th}$ factor of $S$ is a permutation group acting on the $\vec{y}^{\, a_I}$ through the standard irreducible representation and on the $\psi_{a_I}$ through the dual representation.  The $\vec{Y}^I$ and $\uppsi_I$ are $S$-invariant.

The advantages of this coordinate chart on $\MM_0$ are that it is easy to $(1)$ describe the action of $S$ and $(2)$ identify those circle isometries associated with the $\rG$-map -- namely translations of the $\uppsi_I$.  These are the isometries of the asymptotic metric that will extend to isometries of the exact strongly centered moduli space.  However these details are not crucial in the main text and we chose to work instead with the coordinates $\{ \vec{y}^{\, a},\psi_a \}$ on $\MM_0$ that are easier to describe because they do not keep track of the decomposition by type.  It is straightforward to determine the linear change of coordinates between the two systems, but we will not need it here.

Finally, the action of $\mathbb{D}_{\rm g} \subset \mathbb{D}$ on $\mathbb{R} \times \MM_0$ is given in \eqref{Dgaction2}, and is expressed in terms of the $\chi^I$ in \eqref{Dgaction1}.  The generator $\phi$ of $\mathbb{D}$ is a hyperk\"ahler isometery of the universal cover that must satisfy $\phi^{\ell} = \phi_{\rm g}$.  Hence it should act on the $\chi^I$ via
\begin{align}\label{Donchi}
\chi^1 \xmapsto{\phi} &~ \chi^1 + \frac{2\pi \ell^1}{\ell} x_{1\cdots (r-1),r} \cdots x_{12,3} x_{1,2}~, \cr
\chi^I \xmapsto{\phi}&~ \chi^I + \frac{2\pi \ell^I}{\ell} x_{1 \cdots (r-1),r} \cdots x_{1 \cdots I, I+1} y_{1 \cdots (I-1),I}~, \quad 1 < I < r ~, ~ \cdots \cr
\chi^r \xmapsto{\phi}&~ \chi^r + \frac{2\pi \ell^r}{\ell} y_{1\cdots (r-1),r} ~.
\end{align}
Equivalently it acts on the $\uppsi^I$ and $\chi$ via
\begin{equation}
\uppsi_I \xmapsto{\phi} \uppsi_I + 2\pi c_I ~, \qquad \chi \xmapsto{\phi} \chi + 2\pi~.
\end{equation}

Now it is straightforward to see how the identification by $\phi$ is implemented in the GM/LWY asymptotic description of $\MM$.  Recall that $\chi^I$ is a sum of the $n_{\rm m}^I$ phases $\xi_{\ha_I} \sim \xi_{\ha_I} + 2\pi \cp^I$.  For each $I$ we pick one phase, say $\xi_{\ha_I = 1}$, and shift it by an appropriate amount to account for \eqref{Donchi}.  Thus in the GM/LWY description, the periodicity of the constituent phases is ultimately responsible for enacting the quotient by $\mathbb{D}$.

A few comments are in order.
\begin{itemize}
\item First note that, while $\ell$ divides $\ell^I$ by definition, $\ell^I/\ell$ need not be an integer multiple of $\cp^I$.  Here we can use the shift freedom for the pairs $\{x,y\}$ to ensure that $r-1$ of the translations in \eqref{Donchi} are integer multiples of $2\pi \cp^I$.  But then for the final translation we can always assume that the corresponding $\cp = 1$.  The reason is that we have assumed all $n_{\rm m}^I$ are strictly greater than zero, so it is impossible to have only short roots.  In other words, what matters for the construction of the moduli space is not the Lie algebra \emph{per se}, but the effective Lie algebra -- that is, the Lie algebra determined from the Dynkin sub-diagram corresponding to those nodes that have monopoles.  Thus, for example, if we were to consider a case where all monopoles are along the short root in a $\mathrm{G}_2$ gauge theory, the effective Lie algebra would be $\mathfrak{su}(2)$ and all of the $\cp$'s would equal 1, not 3.
\item Second, a shift of $\xi_{\ha_I = 1} \xmapsto{\phi} \xi_{\ha_I = 1} + 2\pi n \cp^I$ not only generates a shift of $\chi^I$.  Additionally it implements a shift of the relative phases \eqref{psiperiods1}: $\psi_{a_I} \xmapsto{\phi} \psi_{a_I} + 2\pi n \cp^I (1- a_I/n_{\rm m}^I)$.  Hence the $\psi_{I_a}$, in addition to the $\uppsi_I$, are translated by the $\phi$ action on $\MM_0$.  Unlike the $\uppsi_I$ shift, $\ell$ times the $\psi_{a_I}$ shift will lead to an integer multiple of its period.  Hence, while $\phi$ can act nontrivially on all of the phase coordinates of $\MM_0$, $\phi_{\rm g}$ only acts on the $\uppsi_I$.  Since the $\pd_{\psi_{a_I}}$ do not extend to triholomorphic Killing vectors on the exact $\MM_0$, the $\phi$ action on the exact $\MM_0$ is not generated by a triholomorphic Killing field.  Nevertheless it is a hyperk\"ahler isometry of $\MM_0$.  In contrast, the $\phi_{\rm g}$ action is generated by a triholomorphic Killing field.  
\item Finally, we did not have to identify the $\phi$ action with a translation of $\xi_{\ha_I = 1}$.  We could have chosen any of the $\xi_{\ha_I}$, and indeed the exchange symmetry $S$ makes all of these choices equivalent.  We can see this explicitly by noting that two different choices, a shift of $\xi_{\ha_I}$ by $2\pi \cp^I$ versus a shift of $\xi_{\hb_I}$ by $2\pi \cp^I$, are related by a shift of some of the $\psi_{a_I}$ by full periods.  This is a trivial isometry on $\MM_0$. 
\end{itemize}         

\subsection{Example: $\mathfrak{su}(2)$ monopoles}

In this case there is only one type of monopole.  Thus $\chi = \chi^1$ and there are no $\uppsi_I$.  A charge $n = n_{\rm m}^1$ monopole has $\ell^1 = \ell = n$.  The $\mathbb{D}_{\rm g}$ action implies that $\chi$ is in fact periodic in this case: $\chi \sim \chi + 2\pi n$.  Hence the quotient by $\mathbb{D}$ simplifies to
\begin{equation}
\MM = \mathbb{R}^3 \times \frac{ S^1 \times \MM_0}{\mathbb{Z}_n}~,
\end{equation}
where the $S^1$ circle is parameterized by $\chi$.  

The asymptotic region of the strongly centered moduli space is parameterized by coordinates $\{ \vec{y}^{\, a}, \psi_a \}_{a=1}^{n-1}$ with $\psi_a \sim \psi_a + 2\pi$.  We identify by the $S_n$ action, with respect to which the $\vec{y}^{\, a}$ transform in the $(n-1)$-dimensional standard representation and the $\psi_a$ transform in the dual.

The generator $\phi$ of $\mathbb{D}$ can be taken to be $\xi_{\ha} \xmapsto{\phi} \xi_{\ha} + 2\pi \delta_{\ha 1}$.  This implies
\begin{equation}\label{su2case}
\psi_{a} \xmapsto{\phi} \psi_a + 2\pi \left( 1 - \frac{a}{n} \right)~, \qquad \chi \xmapsto{\phi} \chi + 2\pi~.
\end{equation}
Any other choice of generator, $\xi_{\ha} \mapsto \xi_{\ha} + 2\pi \delta_{\ha \hb}$, is related by a trivial isometry of $\MM_0$.  

Consider the case $n=2$, for example.  Then the asymptotic model for $\MM_0$ is an $S_2$ quotient of Taub-NUT with negative mass parameter and parameterized by $\{\vec{y},\psi \sim \psi + 2\pi\}$.  The $S_2$ quotient identifies points under the reflection $\{ \vec{y}, \psi \} \mapsto \{ -\vec{y},-\psi \}$.  The $S^1$ in \eqref{su2case} is parameterized by $\chi \sim \chi + 4\pi$.  The $\mathbb{Z}_2$ quotient identifies points on $S^1 \times \MM_0$ under $(\chi,\psi) \sim (\chi + 2\pi, \psi + \pi)$.

\section{The Dirac Operator in the Two-galaxy Region}
\label{app:B}

Now we will construct the asymptotic Dirac operator based on the metric \eqref{hatmetric} to order $O(1/R)$.  We will employ the coordinates $(y^{\alpha\ti}, \psi_{\ti})$ as in \eqref{ypsicollection}, and we will sometimes combine these together into $y^{\mu \ti}$, introducing $\mu,\nu = 1,\ldots,4$ with $y^{4 \ti} \equiv \psi_{\ti}$.  We refer to the components of the hatted metric on $\MM_0$ with respect to these coordinates as $\hat{G}_{\mu \ti,\nu \tj}$ so that
\begin{equation}
d\hat{s}_{0}^2 = \hat{G}_{\mu \ti,\nu \tj} dy^{\mu \ti} dy^{\nu \tj}~.
\end{equation}
We will use underlined indices to refer to the corresponding tangent space directions.

\subsection{Vielbein}

The nonzero components of the vielbein, $\hat{e}^{\underline{\mu \ti}}_{\phantom{\mu \ti}\mu \ti}$, are taken to be
\begin{align}\label{thees}
\hat{e}^{\underline{\alpha \ti}}_{\phantom{\alpha\ti} \alpha\ti} = \delta^{\underline{\alpha}}_{\phantom{\alpha}\alpha} \hat{e}^{\underline{\ti}}_{\phantom{\ti}\ti} ~, \qquad \hat{e}^{\underline{4 \ti}}_{\phantom{4\ti}\alpha \ti} = (\hat{e}_\alpha)^{\underline{\ti}}_{\phantom{\ti}\ti}~, \qquad \hat{e}^{\underline{4\ti}}_{\phantom{4\ti} 4\ti} = (\hat{e}_4)^{\underline{\ti}}_{\phantom{\ti}\ti}~,
\end{align}
where the matrices are given by
\begin{align}\label{blockes}
\hat{{\bf e}} =&~ \left( \begin{array}{c c} \wtbC^{1/2} + \frac{1}{2R} \wtbC^{-1/2} \delta {\bf C} & \frac{1}{R} \wtbC^{-1/2} {\bf L} \\ 0 & \mu^{1/2} \left( 1 - \frac{1}{2\mu R} (\gamma_{{\rm m},1}, \gamma_{{\rm m},2}) \right) \end{array} \right)  , \cr
\hat{{\bf e}}_\alpha =&~ \left( \begin{array}{c c} \left( \wtbC^{-1/2} - \frac{1}{2 R} \wtbC^{-1/2} \delta {\bf C} \wtbC^{-1} \right) \boldsymbol{\WW}_\alpha + w_\alpha \wtbC^{-1/2} \delta {\bf C} & w_\alpha \wtbC^{-1/2} {\bf L} \\ \mu^{-1/2} w_\alpha {\bf L}^T & - \mu^{-1/2} (\gamma_{{\rm m},1},\gamma_{{\rm m},2}) w_\alpha \end{array} \right)  , \cr
\hat{{\bf e}}_4 =&~ \left( \begin{array}{c c} \wtbC^{-1/2} - \frac{1}{2R} \wtbC^{-1/2} \delta {\bf C} \wtbC^{-1} & -\frac{1}{\mu R} \wtbC^{-1/2} {\bf L} \\ 0 & \mu^{-1/2} \left( 1 + \frac{1}{2\mu R} (\gamma_{{\rm m},1}, \gamma_{{\rm m},2}) \right) \end{array} \right)  ~.
\end{align}
Similarly, the components of the inverse vielbein, $\hat{E}^{\mu \ti}_{\phantom{\mu i}\underline{\mu \ti}}$, are given by
\begin{equation}\label{theEs}
\hat{E}^{\alpha\ti}_{\phantom{\alpha\ti}\underline{\alpha\ti}} = \delta^{\alpha}_{\phantom{\alpha}\ualpha} \hat{E}^{\ti}_{\phantom{\ti}\underline{\ti}} ~, \qquad \hat{E}^{4 \ti}_{\phantom{4\ti}\underline{\alpha\ti}} = \delta_{\ualpha}^{\phantom{\alpha}\alpha} (\hat{E}_{\alpha})^{\ti}_{\phantom{\ti}\underline{\ti}}~, \qquad \hat{E}^{4\ti}_{\phantom{4\ti}\underline{4\ti}} = (\hat{E}_{\underline{4}})^{\ti}_{\phantom{\ti}\underline{\ti}}~,
\end{equation}
with
\begin{align}\label{blockEs}
\hat{{\bf E}} =& \left( \begin{array}{c c} \wtbC^{-1/2} - \frac{1}{2R} \wtbC^{-1} \delta{\bf C} \wtbC^{-1/2} & - \frac{1}{\mu^{1/2} R} \wtbC^{-1} {\bf L} \\ 0 & \mu^{-1/2} \left( 1 + \frac{1}{2\mu R} (\gamma_{{\rm m},1},\gamma_{{\rm m},2}) \right) \end{array} \right)  ~,\cr
\hat{{\bf E}}_{\alpha} =& \left( \begin{array}{c c} -\boldsymbol{\WW}_{\alpha} \left( \wtbC^{-1/2} - \frac{1}{2R} \wtbC^{-1} \delta {\bf C} \wtbC^{-1/2} \right) - w_{\alpha} \delta {\bf C} \wtbC^{-1/2} &  \frac{1}{\mu^{1/2} R} \boldsymbol{\WW}_{\alpha} \wtbC^{-1} {\bf L} - \mu^{-1/2} w_{\alpha} {\bf L} \\ - w_{\alpha} {\bf L}^T \wtbC^{-1/2} &  \mu^{-1/2} (\gamma_{{\rm m},1},\gamma_{{\rm m},2}) w_{\alpha} \end{array} \right) ~, \cr
\hat{{\bf E}}_{\underline{4}}  =& \left( \begin{array}{c c} \wtbC^{1/2} + \frac{1}{2 R} \delta {\bf C} \wtbC^{-1/2} & \frac{1}{\mu^{1/2}R} {\bf L} \\ 0 & \mu^{1/2} \left( 1 - \frac{1}{2\mu R}(\gamma_{{\rm m},1} , \gamma_{{\rm m},2}) \right) \end{array} \right) ~. 
\end{align}
These satisfy the necessary relations to the order we are working:
\begin{equation}
\hat{E}^{\mu \ti}_{\phantom{\mu \ti}\underline{\mu \ti}} \hat{e}^{\underline{\mu \ti}}_{\phantom{\mu \ti}\nu \tj} = {\delta^\mu}_\nu {\delta^{\ti}}_{\tj} + O(1/R^2)~, \qquad \delta_{\underline{\mu\nu}} \delta_{\underline{\ti\tj}} \hat{e}^{\underline{\mu \ti}}_{\phantom{\mu \ti}\mu \ti} \hat{e}^{\underline{\nu \tj}}_{\phantom{\nu \tj}\nu \tj} = \hat{G}_{\mu \ti,\nu \tj} + O(1/R^2)~.
\end{equation}
%

\subsection{Spin connection}

Using the vielbein above, one can compute the spin connection.  The discussion is organized according to how many of the $\ti,\tj,\tk$ indices take the last value, $N-1$, corresponding to $\vec{y}^{N-1} = \vec{R}$.  By a slight abuse of notation we will refer to this index value as ``$\, \ti = R \,$'' rather than $\ti = N-1$.  The remaining indices running over the $N-2$ relative positions $\vec{{\bf y}} = (\vec{y}^{\,a},\vec{y}^{\,p})^T$ are $i,j,k = 1,\ldots,N-2$.  We give the expressions below with all indices referred to the frame on the tangent space.

 When there are no $\underline{R}$ indices, there is a leading $O(1)$ piece and $O(1/R)$ corrections to it:
\begin{equation}
\omega_{\underline{\mu i \nu j}, \underline{\rho k}} = \widetilde{\omega}_{\underline{\mu i \nu j},\underline{\rho k}} + \delta \omega_{\underline{\mu i \nu j},\underline{\rho k}} + \cdots~,
\end{equation}
where $\delta \omega = O(1/R)$.  Explicitly, the nonzero leading-order spin connection is found to be 
\begin{align}\label{backgroundspincon}
\widetilde{\omega}_{\underline{\alpha i \beta j},\underline{\gamma k}} =&~ \half (\tC^{-1/2})_{\ui}^{\phantom{i}i} (\tC^{-1/2})_{\uj}^{\phantom{j}j}  (\tC^{-1/2})_{\uk}^{\phantom{k}k} \left( \delta_{\underline{\alpha\gamma}} \delta_{\ubeta}^{\phantom{\beta}\beta} \pd_{\beta j} \tC_{ik} - \delta_{\underline{\beta\gamma}} \delta_{\ualpha}^{\phantom{\alpha}\alpha} \pd_{\alpha i} \tC_{jk} \right) - \cr
&~  -\half \delta_{\underline{\alpha\beta}} \delta_{\ugamma}^{\phantom{\gamma}\gamma} (\tC^{-1/2})_{\underline{k}}^{\phantom{k}k} \left[ (\pd_{\gamma k} \tC^{1/2}) \tC^{-1/2} - \tC^{-1/2} (\pd_{\gamma k} \tC^{1/2}) \right]_{\underline{ij}} ~, \cr
\widetilde{\omega}_{\underline{4 i \beta j},\underline{\gamma k}} =&~  \half (\tC^{-1/2})_{\ui}^{\phantom{i}i} (\tC^{-1/2})_{\uj}^{\phantom{j}j}  (\tC^{-1/2})_{\uk}^{\phantom{k}k} \delta_{\ubeta}^{\phantom{\beta}\beta} \delta_{\ugamma}^{\phantom{\gamma}\gamma} \left( \pd_{\beta j} \WW_{\gamma ik} - \pd_{\gamma k} \WW_{\beta i j} \right)~, \cr
\widetilde{\omega}_{\underline{4i 4j},\underline{\gamma k}} =&~  -\half (\tC^{-1/2})_{\uk}^{\phantom{\uk}k} \delta_{\ugamma}^{\phantom{\gamma}\gamma} \left[ (\pd_{\gamma k} \tC^{1/2}) \tC^{-1/2} - \tC^{-1/2} (\pd_{\gamma k} \tC^{1/2}) \right]_{\underline{ij}} \cr
\widetilde{\omega}_{\underline{\alpha i \beta j},\underline{4 k}} =&~ - \half (\tC^{-1/2})_{\ui}^{\phantom{i}i} (\tC^{-1/2})_{\uj}^{\phantom{j}j}  (\tC^{-1/2})_{\uk}^{\phantom{k}k} \delta_{\ualpha}^{\phantom{\alpha}\alpha} \delta_{\ubeta}^{\phantom{\beta}\beta} \left( \pd_{\alpha i} \WW_{\beta kj} - \pd_{\beta j} \WW_{\alpha k i} \right) ~, \cr
\widetilde{\omega}_{\underline{\alpha i 4j},\underline{4k}} =&~ \half (\tC^{-1/2})_{\ui}^{\phantom{i}i} (\tC^{-1/2})_{\uj}^{\phantom{j}j}  (\tC^{-1/2})_{\uk}^{\phantom{k}k} \pd_{\ualpha i} \tC_{jk} ~.
\end{align}
This is the spin connection on $\MM_{1,0} \times \MM_{2,0} \times \mathbb{R}_{\rm rel}^4$, with the $\MM_0$ factors equipped with their respective GM/LWY metrics.  The $O(1/R)$ corrections to the above components are captured by the simple replacement rule
\begin{equation}\label{correctedC}
\wtbC \to {\bf C} = \wtbC + \frac{1}{R} \delta {\bf C}~,
\end{equation}
leaving $\boldsymbol{\WW}_\alpha$ unchanged, (and expanding the result to first order in $1/R$).  The reason this captures all $O(1/R)$ corrections to these components of the spin connection is that the contributions from the $w_\alpha$ terms in the vielbein cancel.

Note that the components involving the $\WW_{\alpha ij}$ can be simplified using relations \eqref{hkrelns}:
\begin{equation}
\widetilde{\omega}_{\underline{\alpha i 4 j},\underline{\beta k}} = \widetilde{\omega}_{\underline{\alpha i \beta j}, \underline{4 k}} = - \half (\tC^{-1/2})_{\ui}^{\phantom{i}i} (\tC^{-1/2})_{\uj}^{\phantom{j}j} (\tC^{-1/2})_{\uk}^{\phantom{k}k} \epsilon_{\underline{\alpha\beta}}^{\phantom{\alpha\beta}\ugamma} \delta_{\ugamma}^{\phantom{\gamma}\gamma} \pd_{\gamma i} \tC_{jk}~.
\end{equation}
Furthermore, this relation can be extended to the $O(1/R)$ corrections to these components using the replacement \eqref{correctedC}.  The reason is that \eqref{correctedC} is simply a shift of $\tC$ by a constant as far as the $y^{\alpha i}$ are concerned, and $\tC$ is always differentiated in the relations \eqref{hkrelns}.

Next we consider the components of the spin connection with one $\underline{R}$ index that have a non-vanishing $O(1/R)$ piece.  With some effort they can all be related to the \eqref{backgroundspincon} in a rather simple way:
\begin{align}\label{spin1Rsimp}
\omega_{\underline{\alpha i \beta j},\underline{\gamma R}} =&~ - \frac{1}{\sqrt{\mu} R} (\tC^{-1/2} L)^{\underline{k}} \, \widetilde{\omega}_{\underline{\alpha i \beta j},\underline{\gamma k}} + O(1/R^2) ~, \cr
\omega_{\underline{\alpha i \beta R},\underline{\gamma k}} =&~   - \frac{1}{\sqrt{\mu} R} (\tC^{-1/2} L)^{\uj} \, \widetilde{\omega}_{\underline{\alpha i \beta j},\underline{\gamma k}} -   \frac{1}{\sqrt{\mu} R} (\tC^{-1/2})_{\uk}^{\phantom{k}k} \delta_{\underline{\alpha\beta}} \delta_{\ugamma}^{\phantom{\gamma}\gamma} \pd_{\gamma k} (\tC^{-1/2} L)_{\ui} + O(1/R^2)~, \cr
\omega_{\underline{4i \beta j},\underline{\gamma R}} =&~  - \frac{1}{\sqrt{\mu} R} (\tC^{-1/2} L)^{\uk} \, \widetilde{\omega}_{\underline{4i \beta j},\underline{\gamma k}} + O(1/R^2) ~, \cr
\omega_{\underline{4i \beta R},\underline{\gamma k}} =&~ -  \frac{1}{\sqrt{\mu} R} (\tC^{-1/2} L)^{\uj} \, \widetilde{\omega}_{\underline{4i \beta j},\underline{\gamma k}} + O(1/R^2) ~, \cr
\omega_{\underline{\alpha i \beta R},\underline{4k}} =&~  - \frac{1}{\sqrt{\mu} R} (\tC^{-1/2} L)^{\uj} \, \widetilde{\omega}_{\underline{\alpha i \beta j},\underline{4k}} + O(1/R^2) ~, \cr
\omega_{\underline{4i \beta R},\underline{4k}} =&~ - \frac{1}{\sqrt{\mu} R} (\tC^{-1/2} L)^{\uj} \, \widetilde{\omega}_{\underline{4i \beta j},\underline{4k}} + O(1/R^2) ~, \cr
\omega_{\underline{4i 4j},\underline{\gamma R}} =&~ - \frac{1}{\sqrt{\mu}R} (\tC^{-1/2} L)^{\uk} \, \widetilde{\omega}_{\underline{4i 4j},\underline{\gamma k}} + O(1/R^2)~.
\end{align}

Finally, one can show that components of the spin connection with two or three $\underline{R}$ indices start at $O(1/R^2)$, and thus we can neglect them to the order we are working.

Now introduce gamma matrices $\Gamma^{\underline{\mu \ti}}$ satisfying the Clifford algebra
\begin{equation}
[ \Gamma^{\underline{\mu \ti}}, \Gamma^{\underline{\nu \tj}} ]_+ = 2 \delta^{\underline{\mu\nu}} \delta^{\underline{\ti \tj}}~,
\end{equation}
and define $\Gamma^{\underline{\mu \ti \nu \tj}} := \half [\Gamma^{\underline{\mu \ti}}, \Gamma^{\underline{\nu \tj}}]$ as usual.  When we contract the spin connection components with the gamma matrices to construct the Dirac operator, we can absorb almost all effects of the $\omega$ with one $\underline{R}$ index by introducing shifted gamma matrices:
\begin{align}\label{Gdotomegarel}
\Gamma^{\underline{\rho \tk}} \omega_{\underline{\mu \ti \nu \tj},\underline{\rho \tk}} \Gamma^{\underline{\mu \ti \nu \tj}} =&~  \hat{\Gamma}^{\underline{\rho k}} \omega_{\underline{\mu i \nu j},\underline{\rho k}} \hat{\Gamma}^{\underline{\mu i \nu j}} - 2 \Gamma^{\underline{\gamma k}} \Gamma^{\underline{\alpha i \beta R}} \frac{1}{\sqrt{\mu} R} \delta_{\underline{\alpha\beta}} (\tilde{C}^{-1/2})_{\uk}^{\phantom{k}k} \delta_{\ugamma}^{\phantom{\gamma}\gamma}\pd_{\gamma k} (\tilde{C}^{-1/2} L)_{\ui} + \cr
&~ +  O(1/R^2)~, \raisetag{20pt}
\end{align}
with
\begin{equation}\label{Gammahats}
\hat{\Gamma}^{\underline{\alpha i}} := \Gamma^{\underline{\alpha i}} - \frac{1}{\sqrt{\mu} R} (\tC^{-1/2} L)^{\ui} \Gamma^{\underline{\alpha R}} ~, \qquad \hat{\Gamma}^{\underline{4 i}} := \Gamma^{\underline{4 i}}~.
\end{equation}
We also account for the $O(1/R)$ corrections to the $\omega_{\underline{\mu i \nu j},\underline{\rho k}}$ by working with the corrected ${\bf C}$ as we discussed around \eqref{correctedC}.  So above we account for both types of $O(1/R)$ corrections to the spin connection by working with $\omega_{\underline{\alpha i \beta j},\underline{\gamma k}}$ which is expressed in terms of ${\bf C}$ rather than $\wtbC$, and working with the $\hat{\Gamma}^{\underline{\alpha i}}$.

These hatted gamma matrices can be realized by a frame rotation on the spin bundle.  To do so we first complete the definition of the $\hat{\Gamma}^{\underline{\mu \ti}}$ by setting
\begin{equation}
\hat{\Gamma}^{\underline{\alpha R}} := \Gamma^{\underline{\alpha R}} + \frac{1}{\sqrt{\mu} R}  (\tC^{-1/2} L)_{\underline{i}} \Gamma^{\underline{\alpha i}} ~, \qquad \hat{\Gamma}^{\underline{4 R}} := \Gamma^{\underline{4 R}}~.
\end{equation}
Then $\hat{\Gamma}^{\underline{\mu \ti}} = \RR^{\underline{\mu \ti}}_{\phantom{\mu\ti}\underline{\nu \tj}} \Gamma^{\underline{\nu \tj}}$, where the rotation is given by
\begin{equation}
\RR = \prod_{\alpha, i}  \RR_{\theta^{\ui}}^{(\alpha)} ~,
\end{equation}
where $\RR_{\theta^{\ui}}^{(\alpha)}$ is a local rotation in the $e^{\underline{\alpha i}}$ - $e^{\underline{\alpha R}}$ plane by angle
\begin{equation}
\theta^{\ui} := \frac{1}{\sqrt{\mu} R} (\tC^{-1/2} L)^{\ui} ~,
\end{equation}
and we work to linear order in the $\theta^{\ui}$.
This rotation on the frame index can in turn be implemented through an adjoint action on the spinor indices,
\begin{equation}\label{hatGadaction}
\hat{\Gamma}^{\underline{\mu \ti}} = \RR^{\underline{\mu \ti}}_{\phantom{\mu\ti}\underline{\nu \tj}} \Gamma^{\underline{\nu \tj}} = \Lambda \Gamma^{\underline{\mu \ti}} \Lambda^{-1}~.
\end{equation}
with
\begin{equation}
\Lambda = \prod_{\alpha,i} \exp \left( \half \theta^{\ui} \Gamma^{\underline{\alpha i}} \Gamma^{\underline{\alpha R}}\right) = \mathbbm{1} + \half \theta^{\ui} \Gamma_{\underline{\alpha i}} \Gamma^{\underline{\alpha R}} + O(\theta^2)~.
\end{equation}
%

\subsection{Dirac operator}\label{app:B3}

To construct the Dirac operator, $\slashed{D}^{\YY_0}$, on $\MM_0$ we will also need $\slashed{\rG}(\YY_0)$.  We first have  
\begin{align}\label{Gdecomp1}
\rG(\YY_0) =&~ \sum_{I=1}^{\rnk{\mathfrak{g}}} \langle \alpha_I, \YY_0 \rangle \rG(h^I) = \sum_{I} \langle \alpha_I, \YY_0 \rangle \cp^I \sum_{k_I = 1}^{n_{\rm m}^I} \frac{\pd}{\pd \xi_{k_I}^I} + \textrm{exp.~small} \cr
=&~ \sum_I (H_I, \YY_0) \sum_{k_I = 1}^{n_{\rm m}^I} \frac{\pd}{\pd \xi_{k_I}^I} + \textrm{exp.~small} \cr
=&~ \sum_{\hat{a} = 1}^{N_1} (H_{I(\hat{a})}, \YY_0) \frac{\pd}{\pd \xi^{\hat{a}}} +  \sum_{\hat{p} = N_1 + 1}^{N} (H_{I(\hat{p})}, \YY_0) \frac{\pd}{\pd \xi^{\hat{p}}} + \textrm{exp.~small} ~,
\end{align}
where we used that the exact triholomorphic Killing vectors $\rG(h^I)$ approach the linear combinations of those in the GM/LWY metric identified in \eqref{exactTKVlimit} exponentially fast (in the same sense that exact metric approaches the GM/LWY metric exponentially fast).  Then using \eqref{TKVcov}, \eqref{beta1def}, \eqref{beta2def}, followed by  \eqref{globalcom}, we find
\begin{align}\label{Gtwistpsis}
\rG(\YY_0) =&~ \sum_{a} \langle \beta_a, \YY_0 \rangle \frac{\pd}{\pd \psi^a} + \sum_{p} \langle \beta_p, \YY_0 \rangle \frac{\pd}{\pd \psi^p} + (\gmone,\YY_0) \frac{\pd}{\pd \chi_1} + (\gmtwo,\YY_0) \frac{\pd}{\pd \chi_2} + \textrm{exp.~small} \cr
=&~ \sum_i \langle \beta_i, \YY_0 \rangle \frac{\pd}{\pd \psi^i} + (\gmone, \YY_0) \frac{\pd}{\pd\uppsi} + \textrm{exp.~small} ~,\raisetag{20pt}
\end{align}
where in the last step we also used that $(\gm,\YY_0) = 0$, or equivalently $(\gmone,\YY_0) = -(\gmtwo,\YY_0)$.

Now we can compute
\begin{equation}
\slashed{D}^{\YY_0} = \Gamma^{\underline{\rho \tk}} \hat{E}_{\underline{\rho \tk}}^{\phantom{\rho k}\rho \tk}  \pd_{\rho \tk} + \frac{1}{4} \Gamma^{\underline{\rho \tk}} \omega_{\underline{\mu \ti \nu \tj},\underline{\rho \tk}} \Gamma^{\underline{\mu \ti \nu \tj}} -i  \Gamma_{\underline{\rho \tk}} \hat{e}^{\underline{\rho \tk}}_{\phantom{\rho k}\rho \tk} \rG(\YY_0)^{\rho \tk}~,
\end{equation}
to $O(1/R^2)$ using \eqref{blockes}, \eqref{blockEs}, \eqref{Gdotomegarel}, and \eqref{Gtwistpsis}.  Given the simplifications in the spin connection afforded by \eqref{Gdotomegarel}, the goal will be to express everything in terms of the rotated $\hat{\Gamma}$'s and then use \eqref{hatGadaction}.  The final result will be an expression for $\slashed{D}^{\YY_0}$, through to $O(1/R^2)$, in terms of the $\Lambda$-conjugation of another Dirac-type operator.  The advantage of this approach is that the $\Lambda$-conjugation, which implements the frame rotation on the spin bundle, is sufficient to block-diagonalize the Dirac operator with respect to the $\Gamma^{\underline{\mu i}}$ - $\Gamma^{\underline{\mu R}}$ decomposition of the Dirac spinor bundle.  A key point is that the `extra' term on the right-hand side of \eqref{Gdotomegarel}, which originates from the inhomogeneous term in the second line of \eqref{spin1Rsimp}, is exactly what is needed to account for the action of the derivative on $\Lambda$:
\begin{align}
\Lambda \left[ \Gamma^{\underline{\gamma k}} E_{\underline{\gamma k}}^{\phantom{\gamma k}\gamma k} \pd_{\gamma k} \right] \Lambda^{-1} =&~ \hat{\Gamma}^{\underline{\gamma k}} E_{\underline{\gamma k}}^{\phantom{\gamma k}\gamma k} \pd_{\gamma k} - \half \Gamma^{\underline{\gamma k}} \Gamma^{\underline{\alpha i \beta R}} \frac{1}{\sqrt{\mu} R} \delta_{\underline{\alpha \eta}} (\tC^{-1/2})_{\uk}^{\phantom{k}k} \delta_{\ugamma}^{\phantom{\gamma}\gamma} \pd_{\gamma k} (\tilde{C}^{-1/2} L)_{\ui} + \cr
&~ + O(1/R^2)~. \raisetag{20pt}
\end{align}

Thus, suppressing the details, we are able to bring the Dirac operator to the form quoted in the text:
\begin{equation}
\slashed{D}^{\YY_0} = \Lambda \left( \slashed{D}_{12}^{\YY_0} + \slashed{D}_{\rm rel}^{\YY_0} + O(1/R^2) \right) \Lambda^{-1}~,
\end{equation}
where $\slashed{D}_{12}^{\YY_0}$ consists of terms involving only $\underline{\mu i}$-type gamma matrices, and $\slashed{D}_{\rm rel}$ consists of terms involving only $\underline{\mu R}$-type gamma matrices.  For the first operator we have
\begin{equation}
\slashed{D}_{12} = \slashed{D}^{\YY_0}_{\MM_{1,0} \times \MM_{2,0}} + (\slashed{D}_{12}^{\YY_0})^{(1)} ~,
\end{equation}
where the first term is precisely the $\rG(\YY_0)$-twisted Dirac operator on $\MM_{1,0} \times \MM_{2,0}$, where each factor equipped with the GM/LWY metric, and the second term contains the $O(1/R)$ corrections.  Explicitly,
\begin{align}
\slashed{D}_{\MM_{1,0} \times \MM_{2,0}}^{\YY_0} :=&~  \Gamma^{\underline{\alpha i}} \delta_{\ualpha}^{\phantom{\alpha}\alpha} \left[ (\tC^{-1/2})^{i}_{\phantom{i}\ui} \pd_{\alpha i} - (\WW_{\alpha} \tC^{-1/2})^{i}_{\phantom{i}\ui} \pd_{4i} \right] + \Gamma^{\underline{4i}} (\tC^{1/2})^{i}_{\phantom{i}\ui} \pd_{4i} + \cr
&~ + \frac{1}{4} \Gamma^{\underline{\rho k}} \widetilde{\omega}_{\underline{\mu i \nu j},\underline{\rho k}}\Gamma^{\underline{\mu i \nu j}}  -i \Gamma^{\underline{4i}} (\tC^{-1/2})^{i}_{\phantom{i}\ui} (\beta_{i}, \YY_0)~,
\end{align}
and
\begin{align}\label{D12correction}
(\slashed{D}_{12}^{\YY_0})^{(1)} :=&~  \Gamma^{\underline{\alpha i}} \delta_{\ualpha}^{\phantom{\alpha}\alpha} \bigg\{ -\frac{1}{2R} (\tC^{-1} \delta C \tC^{-1/2})^{i}_{\phantom{i}\ui} \pd_{\alpha i} + ( (\tfrac{1}{2R} \WW_{\alpha} \tC^{-1} - w_{\alpha} \mathbbm{1} ) \delta C \tC^{-1/2} )^{i}_{\phantom{i}\ui} \pd_{4i} + \cr
&~ \qquad \quad - w_{\alpha} (L^T \tC^{-1/2})_{\ui} \pd_{4R} - \frac{1}{\mu R} (\tC^{-1/2} L)_{\ui} \pd_{\alpha R} \bigg\} + \cr
&~ + \Gamma^{\underline{4i}} \bigg\{ \frac{1}{2R} (\tC^{-1/2} \delta C)^{i}_{\phantom{i}\ui} \pd_{4i} - \frac{1}{2R} (\tC^{-1/2} \delta C \tC^{-1})^{i}_{\phantom{i}\ui} (\beta_i, \YY_0) + \cr
&~ \qquad  \quad + i \frac{(\gamma_{{\rm m},1} , \YY_0)}{\mu R} (\tC^{-1/2} L)_{\ui} \bigg\} +  \frac{1}{4} \Gamma^{\underline{\rho k}} \delta \omega_{\underline{\mu i \nu j},\underline{\rho k}}\Gamma^{\underline{\mu i \nu j}}~.  \raisetag{20pt}
\end{align}
Meanwhile the second operator takes the form
\begin{align}
\slashed{D}_{\rm rel}^{\YY_0} :=&~ \frac{1}{\sqrt{\mu}} \left( 1 + \frac{ (\gamma_{{\rm m},1}, \gamma_{{\rm m},2})}{2\mu R} \right) \bigg\{ \Gamma^{\underline{\alpha R}} \delta_{\ualpha}^{\phantom{\alpha}\alpha} \left[ \pd_{\alpha R} + (\gamma_{{\rm m},1},\gamma_{{\rm m},2}) w_{\alpha} \pd_{4R} - w_{\alpha} L^i \pd_{4i} \right] + \cr
&~ \qquad \qquad + \Gamma^{\underline{4 R}} \left[ \left( \mu - \frac{(\gamma_{{\rm m},1}, \gamma_{{\rm m},2})}{R} \right) \pd_{4 R} + \frac{L^i}{R} \pd_{4i} -i  (\gamma_{{\rm m},1},\YY_0) \right] \bigg\}~.
\end{align}
(Strictly speaking, this expression contains some $O(1/R^2)$ terms when the $R^{-1}$ multiplies $w_{\alpha}$ which should be dropped.)  This result leads to \eqref{DrelY0}, \eqref{pxparameters} when acting on spinors of the form \eqref{LWYspinor}.

\section{Singular Monopole Moduli Space}
\label{app:C}

In the case of singular monopoles, we have a core-halo system. Here we can choose our origin to be anywhere in the core, but to be explicit we choose it to be centered on one of the singular monopoles.  In this case we need only go to center of mass and relative coordinates in the halo galaxy.  We let indices $\hat{a},\hat{b} = 1,\ldots,N_{\rm core}$ run over fundamental (mobile) constituents in the core and indices $\hat{p},\hat{q} = N_{\rm core}+1, \ldots, N_{\rm core} + N_{\rm halo} = N$, run over fundamental constituents in the halo galaxy.  We set 
\begin{align}\begin{split}
& \vec{R} = \frac{\sum_{\hat{p}} m_{\hat{p}} \vec{x}^{\,\hat{p}}}{m_{\rm halo}}~,\qquad \vec{y}^{\,p} = \vec{x}^{\,p} - \vec{x}^{\,p+1}~,
\end{split}\end{align}
where $m_{\rm halo} = \sum_{\hat{p}} m_{\hat{p}}$ and we have introduced indices $p,q = N_{\rm core} + 1,\ldots, N-1$ that run over the relative positions of halo constituents.  The inverse is given by
\begin{equation}\label{cmrelcov2sing}
\vec{x}^{\,\hat{p}} = \vec{R} + ({\bf j}_{\rm h})^{\hat{p}}_{~q} \vec{y}^{\,q} ~,
\end{equation}
where ${\bf j}_{\rm h}$ has an identical form to ${\bf j}_2$ with the galaxy-two constituent masses replaced by halo constituent masses.  Constructing ${\bf J}_{\rm h}$ by appending a column of 1's in the same way, we introduce the halo fiber coordinates
\begin{equation}
\left( \begin{array}{c} \psi_p \\ \uppsi \end{array} \right) = {\bf J}_{\rm h}^T ( \xi_{\hat{p}} )~.
\end{equation}
Note that $(\vec{R},\uppsi)$ play the role here that was previously played by $(\vec{X}_2,\chi_2)$.  They parameterize the position of the center of mass of the halo galaxy relative to the \emph{fixed} core.

The large $R$ expansion of the quadratic form $\overline{M}_{\hat{i}\hat{j}}$, \eqref{singmetb}, can be written in block form
\be
(\overline{M})_{\hat{i}\hat{j}}=\left(\begin{array}{cc}
(\overline{M})_{\hat{a}\hat{b}}& \frac{D_{\hat{a}\hat{q}}}{R}+O\left(\frac{1}{R^2}\right)\\
\frac{D_{\hat{p}\hat{b}}}{R}+O\left(\frac{1}{R^2}\right)&(M)_{\hat{p}\hat{q}}~,
\end{array}\right)
\ee
with
\be
(\overline{M})_{\hat{a}\hat{b}}=(\overline{M}_{\rm c})_{\hat{a}\hat{b}}-\delta_{\hat{a}\hat{b}} \frac{(H_{I(\hat{a})},\gamma_{\rm h,m})}{R}+O\left(\frac{1}{R^2}\right)~,
\ee
where
\be
(\overline{M}_{\rm c})_{\hat{a}\hat{b}}=\begin{cases}
m_{\hat{a}}-\sum_{\hat{c}\neq \hat{a}}\frac{D_{\hat{a}\hat{c}}}{r_{\hat{a}\hat{c}}}-\sum_{n=1}^{N_{\rm def}}\frac{(P_n,H_I(\hat{a}))}{r_{n\hat{a}}}&\hat{a}=\hat{b}\\
\frac{D_{\hat{a}\hat{b}}}{r_{\hat{a}\hat{b}}}&\hat{a}\neq \hat{b}
\end{cases}~,
\ee
and similarly
\be
(M)_{\hat{p}\hat{q}}=(M_{\rm h})_{\hat{p}\hat{q}}-\delta_{\hat{p}\hat{q}} \frac{(H_{I(\hat{p})},\gamma_{\rm c,m})}{R}+O\left(\frac{1}{R^2}\right)~,
\ee
where 
\be
(M_{\rm h})_{\hat{p}\hat{q}}=\begin{cases}
m_{\hat{p}}-\sum_{\hat{u}\neq \hat{p}}\frac{D_{\hat{p}\hat{u}}}{r_{\hat{p}\hat{u}}}&\hat{p}=\hat{u}\\
\frac{D_{\hat{p}\hat{q}}}{r_{\hat{p}\hat{q}}}&\hat{p}\neq \hat{q}
\end{cases}~.
\ee
In these expressions the core and halo magnetic charges are given by
\begin{equation}
\gamma_{{\rm c},{\rm m}} = \sum_{\hat{a}=1}^{N_{\rm core}} H_{I(\hat{a})} + \sum_{n=1}^{\rm N_{\rm def}} P_n ~, \qquad \gamma_{{\rm h},{\rm m}} = \sum_{\hat{p}=N_{\rm core}+1}^{N} H_{I(\hat{p})} ~.
\end{equation}
Therefore in the limit of large $R$ we can write $(\overline{M})_{\hat{i}\hat{j}}$ as
\be
\overline{M}_{\hat{i}\hat{j}}=\left(\begin{array}{cc}
(\overline{M}_{\rm c})_{\hat{a}\hat{b}}&0\\
0&(M_{\rm h})_{\hat{p}\hat{q}}
\end{array}\right)+\frac{1}{R}\left(\begin{array}{cc}
-\delta_{\hat{a}\hat{b}}(\gamma_{h,m},H_{I(\hat{a})})&D_{\hat{a}\hat{q}}\\
D_{\hat{b}\hat{p}}&-\delta_{\hat{p}\hat{q}}(\gamma_{\rm c,m},H_{I(\hat{p})})
\end{array}\right)+O\left(1/R^2\right)~.
\ee
As before, we will write this as
\be
\overline{\bf M}
=\left(\begin{array}{cc}
\overline{\bf M}_{\rm c}
&0\\
0&{\bf M}_{\rm h}
\end{array}\right)-\frac{1}{R}\left(\begin{array}{cc}
\mathbf{D}_{{\rm cc}}&-\mathbf{D}_{{\rm ch}}\\
-\mathbf{D}_{{\rm hc}}&\mathbf{D}_{{\rm hh}}
\end{array}\right)+O\left(1/R^2\right)
\ee

Now we make the similarity transformation to the center of mass and relative coordinates in the halo, defining $\mathbf{C}_h$ with components $(C_{\rm h})_{pq}$ such that
\be
\mathbf{J}_{\rm h}^T \mathbf{M}_{\rm h} \mathbf{J}_{\rm h}= \left(\begin{array}{cc}
\mathbf{j}_{\rm h}^T \mathbf{M}_{\rm h} \mathbf{j}_{\rm h}&0\\
0^T&m_{\rm halo}
\end{array}\right) =  \left(\begin{array}{cc}
\mathbf{C}_{\rm h}&0\\
0^T&m_{\rm halo}
\end{array}\right)~.
\ee
We combine ${\bf C}_{\rm h}$ with the leading order core matrix, writing
\be
\overline{\mathbf{C}}=\left(\begin{array}{cc} \overline{\mathbf{M}}_{\rm c}&0\\0&\mathbf{C}_{\rm h} \end{array}\right)~.
\ee
The first order corrections to this are captured by
\be
\delta \mathbf{\overline{C}}=\left(\begin{array}{cc}
 -\mathbf{D}_{\rm cc}&\mathbf{D}_{\rm ch}\mathbf{j}_{\rm h}\\
\mathbf{j}_{\rm h}^T\mathbf{D}_{\rm hc}&- \mathbf{j}_{\rm h}^T\mathbf{D}_{\rm hh}\mathbf{j}_{\rm h}
\end{array}\right)~.
\ee
Finally the vector $\overline{{\bf L}}$ and the harmonic function $\overline{H}(R)$ appearing in \eqref{2singmetric} are
\be
\overline{\mathbf{L}}=\left(\begin{array}{c} -(H_{I(\hat{a})},\gamma_{\rm h,m}) \\  \langle \beta_p, \gamma_{\rm c,m} \rangle \end{array} \right)~, \qquad  \overline{H}(R)=\left(1-\frac{(\gamma_{\rm h,m},\gamma_{\rm c,m})}{m_{\rm halo} R}\right)~.
\ee
Here $\beta_p$ is defined as in \eqref{beta2def} but with ${\bf J}_2$ replaced by ${\bf J}_{\rm h}$.

The terms in \eqref{2singmetric} involving the connection one-forms $(\overline{\Theta}_0,\overline{\Theta}_{\uppsi})$ can be obtained by following analogous steps to those in Appendix \ref{app:A}.

\section{Discrete Spectrum of the Twisted Dirac Operator on Taub-NUT}\label{app:AppD}

In this appendix we compute the spectrum of the Dirac operator coupled to a triholomorphic vector field on Taub-NUT. That is we want to solve the equation\footnote{See \cite{Jante:2015xra} for an earlier derivation.  The presentation here has been included for convenience of the reader.}
\be\label{eq:TNDIRAC}
\I \slashed{D}^{\CY_0}_{\rm TN}\Psi=\lambda \Psi\quad,\qquad \slashed{D}^{\CY_0}_{\rm TN}=\slashed{D}_{\rm TN}-\I \slashed{\rG}(\CY_0)~,
\ee
on Taub-NUT which has the metric
\begin{align}\label{TNmetric}
&ds^2= \mu \left[ H(R)d\vec{R}\cdot d\vec{R}+ H^{-1}(R)(\ell_{\rm TN} dx^4+\Omega)^2 \right]~,\cr
&H=1+\frac{\ell_{\rm TN}}{R}~,\qquad \Omega = \ell_{\rm TN} (\epsilon - \cos{\uptheta})d\upphi~, \qquad x^4 \sim x^4+4\pi ~.
\end{align}
Our coordinates are $x^\mu = (\vec{R},x^4) = (R^\alpha,x^4)$.  We also employ spherical coordinates $(R,\uptheta,\upphi)$ on the $\mathbb{R}^3$ base while $x^4$ parameterizes the circle fiber with asymptotic radius $2|\ell_{\rm TN}| \sqrt{\mu}$.  The parameter $\ell_{\rm TN}$ should be taken positive for the standard, complete Taub-NUT manifold.  However the asymptotic (large $R$) region of this space can occur in the description of the asymptotic monopole moduli space for either sign of $\ell_{\rm TN}$.  Therefore we will, for a time, allow for both possibilities.  When $\ell_{\rm TN}$ is negative we restrict ourselves to the region $R > -\ell_{\rm TN}$.  Finally $\epsilon = \pm 1$, where the plus sign should be chosen in the patch $0 \leq \uptheta < \pi$ and the minus sign in the patch $0 < \uptheta \leq \pi$.  So, $x^4$ depends on the patch and $x^4_- = x^4_+ + 2\phi$.  

The triholomorphic Killing field $\rG(\CY_0)$ must be proportional to $\pd_{4}$ and we parameterize the proportionality constant as in Appendix G of \cite{Moore:2015szp}\footnote{The mass parameter denoted $m$ in \cite{Moore:2015szp} has been set to $\mu$ here, the reduced mass of the two-galaxy system, as this is what appears in the asymptotic form of the metric \eqref{hatmetric}.  We have also chosen the opposite sign for $\Omega$ relative to \cite{Moore:2015szp} so as to better match both the conventions in the literature for the GM/LWY metric and the form of the resulting Dirac operator with \cite{Moore:2014jfa}.  This is equivalent to flipping the orientation of Taub-NUT and is why the BPS spinors we find below have the opposite chirality to those in Addendix G of \cite{Moore:2015szp}.}:
\be
\rG(\CY_0)=\frac{C}{
\mu \ell_{\rm TN}^2} \pd_{4}~.
\ee

Following the computations in \cite{Moore:2015szp}, one finds that \eqref{eq:TNDIRAC} takes the form
\begin{align}\label{Dirac2}
& \frac{1}{\sqrt{H}} \displaystyle\biggl\{ \gamma^{\underline{\alpha}} \pd_\alpha + \frac{1}{4 H} \gamma^{\underline{\alpha}} \pd_\alpha H \left( \mathbf{1} + \bar{\gamma} \right) - \frac{\I C}{\ell_{\rm TN}} \gamma^{\underline{4}} +  \cr
&  \qquad \qquad + \left[ \frac{ (\epsilon - \cos{\uptheta})}{r \sin{\uptheta}} \left( \sin{\upphi} \gamma^{\underline{1}} - \cos{\upphi} \gamma^{\underline{2}} \right) + \frac{H}{\ell_{\rm TN}} \gamma^{\underline{4}} \right] \pd_4 \displaystyle\biggr\} \Psi = -\I \sqrt{\mu} \lambda \Psi~.
\end{align}
Now let us expand in eigenmodes of $\pd_4$, taking
\begin{equation}
\Psi = e^{\I p x^4/2} \Psi^{(p)}(\vec{R})~,
\end{equation}
where $p \in \mathbb{Z}$.  Doing so, we find that the resulting operator acting on $\Psi^{(p)}$ from the left-hand side of \eqref{Dirac2} is consistent in form with $\breve{\slashed{D}}_{\rm rel}^{\YY_0}$, \eqref{DrelY0}, through $O(1/R)$, provided we identify the $p$ here with the one there and
\begin{equation}\label{xellparameters}
\ell_{\rm TN} = - \frac{(\gmone,\gmtwo)}{\mu} ~, \qquad \frac{1}{\ell_{\rm TN}} (C - p/2) = x~.
\end{equation}
To see this one notes that the $\gamma^{\ualpha} \pd_\alpha H$ term in \eqref{Dirac2} is $O(1/R^2)$ and can be dropped for the purposes of the comparison.  In the special case of an ${\rm SU}(3)$ monopole with $\gmone = H_1$ and $\gmtwo = H_2$, the relative moduli space \emph{is} Taub-NUT.  We see that $\ell_{\rm TN} \to \mu^{-1}$, we can identify the $2\pi$-periodic coordinate $\uppsi$ with $x^4/2$, and $p \to \upnu$ consistently with \eqref{pxparameters}.

Returning to the exact solution of \eqref{Dirac2} we introduce a chiral basis for the $\gamma$-matrices
\be
\gamma^{\underline{\mu}}=\left(\begin{array}{cc}0 &\tau^{\underline{\mu}}\\ \bar{\tau}^{\underline{\mu}}&0\end{array}\right)\quad,\qquad \tau^{\underline{\mu}}=(\vec\sigma, -\I \mathbbm{1})\quad,\qquad \bar{\tau}^{\underline{\mu}}=(\vec\sigma, \I \mathbbm{1})~,
\ee
such that $\bar{\gamma} := \gamma^{\underline{1}} \gamma^{\underline{2}} \gamma^{\underline{3}} \gamma^{\underline{4}} = \diag(-\mathbbm{1},\mathbbm{1})$, and we decompose $\Psi$ into chiral components
\begin{equation}
\Psi = \left( \begin{array}{c} \psi_- \\ H^{-1/2} \, \psi_+  \end{array} \right)~,
\end{equation}
with respect to an orthonormal basis for the spinor bundle, where the factor of $H^{-1/2}$ has been introduced for convenience.  Note that the $L^2$ innerproduct, $\langle \cdot|\cdot\rangle$, is then given in terms of the components by
\begin{equation}\label{ip1}
\langle \Psi | \Psi' \rangle := \int \sqrt{g} \Psi^\dag \Psi' = \int_{0}^{4\pi} \ell_{\rm TN} dx^4 \int_{\mathbb{R}^3} d^3 R\left( \psi_{+}^\dag \psi_{+}' + H(R) \psi_{-}^\dag \psi_{-}' \right)~.
\end{equation}
(This expression is appropriate for $\ell_{\rm TN} > 0$; otherwise the integral over the base should be restricted to $R > -\ell_{\rm TN}$.)

Then, introducing the frame rotation
\begin{equation}
U := e^{-\I \upphi \sigma^3/2} e^{-\I \uptheta \sigma^2/2}~,
\end{equation}
and setting
\begin{equation}\label{Weylrot}
\psi_{\pm} = e^{\I p x^4/2} e^{\I (\epsilon p/2 - m) \upphi} U \tilde{\psi}_{\pm}(r,\uptheta)~,
\end{equation}
one finds that \eqref{Dirac2} reduces to the pair 
\begin{align}\label{Weyl1}
& \left\{ \sigma^3 \left( \pd_R + \frac{1}{R} \right) + \frac{1}{\ell_{\rm TN}} \left(C - \frac{p}{2} H\right) + \frac{1}{R} K \right\} \tilde{\psi}_- = -\I  \sqrt{\mu} \lambda \tilde{\psi}_+ ~ , \cr
& \left\{ \sigma^3 \left( \pd_R + \frac{1}{R} \right) - \frac{1}{\ell_{\rm TN}} \left( C - \frac{p}{2} H \right) + \frac{1}{R} K \right\} \tilde{\psi}_+ = - \I \sqrt{\mu} \lambda H  \tilde{\psi}_- ~,
\end{align}
where
\begin{align}
K :=&~ \sigma^1 \left[ \pd_\uptheta + \frac{\sigma^3}{\sin{\uptheta}} \left( m + \half (\sigma^3-p) \cos{\uptheta} \right) \right]~.
\end{align} 

The final separation of variables proceeds as in Appendix C of \cite{Moore:2014jfa}.  We set\footnote{We follow the conventions of \cite{Sakurai} for Wigner $d$ functions and $SU(2)$ representation matrices.  The combination $e^{-i m \phi} d^{j}_{m,m'}(\theta)$ can also be expressed in terms of spin-weighted spherical harmonics, ${}_{m'}Y_{jm}$.}
\begin{equation}\label{rthetasep}
\tilde{\psi}_+ = \frac{1}{R}\left( \begin{array}{c} g_1(R) \, d^{j}_{m,\half(p-1)}(\uptheta) \\ g_2(R) \, d^{j}_{m,\half (p+1)}(\uptheta) \end{array} \right)~, \qquad \tilde{\psi}_-  = \frac{1}{R}\left( \begin{array}{c} f_1(R) \, d^{j}_{m,\half (p-1)}(\uptheta) \\ f_2(R)  \, d^{j}_{m,\half(p+1)}(\uptheta) \end{array} \right)~.
\end{equation}
Here $(j,m)$ are standard $\mathfrak{su}(2)$ quantum numbers, except that the minimal value of $j$ is $j_{\rm min} = \half (|p|-1)$ if $p \neq 0$, and $j_{\rm min} = \half$ if $p = 0$, and the allowed values of $j$ increase from this minimal value in integer steps.  Furthermore, when $j = j_{\rm min}$ and $p \neq 0$ we are required to set some components of $f,g$ to zero for regularity.  If $j = j_{\rm min}$ and $p > 0$ then we must set $f_2,g_2 = 0$, while if $j = j_{\rm min}$ and $p < 0$ we must set $f_1,g_1 = 0$.  

The $d^{j}_{m,m'}(\uptheta)$ are Wigner's small $d$-functions.  They can be expressed in terms of Jacobi polynomials as follows. Let
\begin{equation}
\cn = \min \{ j+m', j-m', j+m, j-m \}~.
\end{equation}
We introduce $a,\uplambda$ depending on the value of $\cn$:
\begin{equation}
\cn = \left\{ \begin{array}{r l l} j+m' ~: ~& a = m-m' ~, & \uplambda = m - m' ~, \\
j -m' ~:~ & a = m' - m~, & \uplambda = 0~, \\
j + m~: ~& a = m' - m~, & \uplambda = 0~, \\
j-m ~: ~& a = m-m' ~, & \uplambda = m - m'~. \end{array} \right.
\end{equation}
Then, with $b = 2j - 2\cn - a$, we have
\begin{equation}\label{wigner1}
d^{j}_{m,m'}(\uptheta) = (-1)^{\uplambda} \left( \begin{array}{c} 2j - \cn \\ \cn + a \end{array} \right)^{1/2} \left( \begin{array}{c} \cn + b \\ b \end{array} \right)^{-1/2} \left( \sin{\frac{\uptheta}{2}} \right)^{a} \left( \cos{\frac{\uptheta}{2}} \right)^{b} P_{\cn}^{(a,b)}(\cos{\uptheta})~,
\end{equation}
where the Jacobi polynomials are given by
\begin{equation}\label{wigner2}
P_{\cn}^{(a,b)}(z) = \frac{(-1)^{\cn}}{2^{\cn} \cn !} (1-z)^{-a} (1+z)^{-b} \frac{d^{\cn}}{dz^{\cn}} \left\{ (1-z)^a (1+z)^b (1-z^2)^{\cn} \right\} ~.
\end{equation}

Inserting \eqref{rthetasep} into \eqref{Weyl1}, we obtain the following pair of coupled equations:
\begin{align}\label{Weylfg}
& \left[ \sigma^3 \pd_R + x + \frac{1}{R} {\bf C}_- \right] f = -\I \sqrt{\mu} \lambda g \cr
& \left[ \sigma^3 \pd_R - x + \frac{1}{R} {\bf C}_+ \right] g = -\I \sqrt{\mu} \lambda H f ~,
\end{align}
for $f = (f_1,f_2)^T$ and $g = (g_1,g_2)^T$, where we used \eqref{xellparameters} and
\begin{equation}
{\bf C}_{\pm} := \pm \frac{p}{2} \mathbbm{1} - \I \sigma^2 k~, \qquad \textrm{with} \quad k :=\sqrt{ j(j+1)-\frac{1}{4}(p^2-1)}~.
\end{equation}
Note that $j=j_{\rm min}$ corresponds to $k=0$.  The innerproduct \eqref{ip1} can be expressed in terms of $f,g$ using \eqref{Weylrot} and \eqref{rthetasep} with the result
\begin{equation}\label{ip2}
\langle \Psi | \Psi' \rangle = \frac{16 \pi^2}{2j +1} \delta_{p p'} \delta_{j j'} \delta_{m m'} \int_{0}^{\infty} dR \left[ g^\dag g' + H(R) f^\dag f' \right]~.
\end{equation}
Again the integral should be restricted to $R > -\ell_{\rm TN}$ if $\ell_{\rm TN}$ is negative.

The equations \eqref{Weylfg} are nearly identical to the system (C.22) in \cite{Moore:2014jfa}, except for the factor of $H(R)$ on the right-hand side of the second equation.  This of course changes the detailed form of the solutions when $\lambda \neq 0$, but nevertheless the same techniques can be used to obtain explicit solutions.  In contrast, when $\lambda = 0$ then they are identical, and this fact was exploited in \cite{Moore:2015szp} to give explicit results for the BPS spinors on Taub-NUT.

From \eqref{Weylfg} one can derive the following second-order equation for $f$:
\begin{equation}\label{fWeyleqn}
\left\{ \pd_{R}^2 + \mu \lambda^2 - x^2 + \frac{(p x + \ell_{\rm TN} \mu \lambda^2)}{R} - \frac{ (j+\frac{1}{2})^2}{R^2} - \frac{1}{R^2} \sigma^3 {\bf C}_- \right\} f = 0~.
\end{equation}
The last term is diagonalized by setting
\begin{equation}\label{fSim}
f = {\bf O}_f \left( \begin{array}{c} \hat{f}_{1} \\ \hat{f}_{2} \end{array} \right)~, \qquad \textrm{with} \quad {\bf O}_f := \frac{1}{\sqrt{2j+1}} \left( \begin{array}{c c} a_- & a_+ \\ - a_+ & a_- \end{array} \right)~, \quad a_{\pm} = \sqrt{ (j+\half) \pm \frac{p}{2}} ~.
\end{equation}
This ${\bf O}_f$ diagonalizes $\sigma^3 {\bf C}_-$,
\begin{equation}\label{Ofsim}
{\bf O}_{f}^T (\sigma^3 {\bf C}_-) {\bf O}_f = \left( \begin{array}{c c} j+ \half & 0 \\ 0 & - (j + \half) \end{array} \right)~,
\end{equation}
leading to the following equations for the components $\hat{f}_{1,2}$:
\begin{equation}\label{fhateqn}
\left\{ \pd_{R}^2 + \mu \lambda^2 - x^2 + \frac{(p x + \ell_{\rm TN} \mu \lambda^2)}{R} - \frac{j_{1,2} (j_{1,2} + 1)}{R^2} \right\} \hat{f}_{1,2} = 0~,
\end{equation}
where $j_1 := j + 1/2$ and $j_2 := j - 1/2$.  Recall that in the extremal cases $j = j_{\rm min} = \half (|p|-1)$, we are to set $f_2 = 0$ when $p > 0$, and $f_1 = 0$ when $p < 0$.  Considering the form of the similarity transformation ${\bf O}_f$ in the extremal cases, one finds that \emph{both} situations correspond to setting $\hat{f}_1 = 0$.

Equation \eqref{fhateqn} has the form of the radial Schr\"odinger equation for the hydrogen atom.  From the large $R$ behavior of the effective potential, we see that there is a gap $0 \leq |\lambda| < |x|/\sqrt{\mu}$ in which boundstates might exist, while for $|\lambda| > |x|/\sqrt{\mu}$ we have a continuum of scattering states.  Whether or not boundstates actually exist depends on the boundary conditions in the interior.

The boundary conditions in the interior, in turn, depend on the sign of $\ell_{\rm TN}$.  If $\ell_{\rm TN} > 0$ then we should impose a regularity condition at $R = 0$.  If, however, $\ell_{\rm TN} < 0$ then we must impose some type of boundary condition at $R = -\ell_{\rm TN}$, the form of which requires a more careful analysis.  In the following we first restrict to the $\ell_{\rm TN} > 0$ case.  Then we will discuss two examples with $\ell_{\rm TN} < 0$ that demonstrate that the $\ell_{\rm TN} > 0$ results cannot be used to draw any conclusions about the $\ell_{\rm TN} < 0$ case.

Proceeding then with \eqref{fhateqn} for $\ell_{\rm TN} > 0$, we set
\begin{equation}
\rho := \kappa_\lambda R~, \qquad \rho_0 := \frac{p x + \ell_{\rm TN} \mu \lambda^2}{\kappa_\lambda}~, \qquad \textrm{with} \quad \kappa_\lambda := \sqrt{ x^2 - \mu \lambda^2}~,
\end{equation}
so that \eqref{fhateqn} takes the standard form
\begin{equation}
\left\{ \pd_{\rho}^2 -1 + \frac{\rho_0}{\rho} - \frac{j_{1,2} (j_{1,2} + 1)}{\rho^2} \right\} f_{1,2} = 0~.
\end{equation}
The solution that is regular at $\rho = 0 $ is
\begin{equation}\label{fhatsol}
\hat{f}_{1,2} = c_{1,2} \rho^{j_{1,2} + 1} e^{-\rho} \, {}_1F_1\left( j_{1,2} + 1 - \frac{\rho_0}{2} \, ; 2 j_{1,2} + 2 \, ; 2\rho \right)~.
\end{equation}
For $|\lambda| > |x|/\sqrt{\mu}$, $\rho$ is imaginary and these are scattering states.  See \cite{Jante:2015xra,MurthyPioline,Pioline:2015wza} for a detailed study of these scattering states and the role they play in accounting for the index of the Dirac operator associated to the quantum mechanics of two point dyons.  Reference \cite{Avery:2007xf} also considered both boundstate and scattering state wavefunctions for this problem.  

Focusing on the case $|\lambda| < |x|/\sqrt{\mu}$, the growth of the hypergeometric function will overwhelm the $e^{-\rho/2}$ prefactor at large $\rho$ unless the hypergeometric series truncates.  This will happen iff the first argument is a non-positive integer.  Hence we introduce the radial quantum number, $n$, in analogy with the hydrogen atom by setting
\begin{equation}
n = \frac{\rho_0}{2}~,
\end{equation}
and we will see that $n \in \half \mathbb{N}$.  The latter can be solved for the allowed boundstate eigenvalues:
\begin{align}\label{DTNspectrum}
\lambda_{p,n}^2 =&~ \frac{2}{\mu \ell_{\rm TN}^2} \left[ - (n^2 + \tfrac{p}{2} \ell_{\rm TN} x) + \sqrt{ (n^2 + \tfrac{p}{2} \ell_{\rm TN} x)^2 + \ell_{\rm TN}^2 x^2 (n^2 - \tfrac{p^2}{4}) } \, \right]  \cr
=&~ \frac{2}{\mu \ell_{\rm TN}^2} \left[ \tfrac{p}{2} (\tfrac{p}{2} - C) - n^2 + \sqrt{ ( n^2 + \tfrac{p}{2} (C-\tfrac{p}{2}))^2 + (C - \tfrac{p}{2})^2 (n^2 - \tfrac{p^2}{4}) } \, \right]~.
\end{align}
Note that the Dirac operator we are studying is a supercharge operator, so its eigenvalues, $\lambda$, should have units of energy${}^{1/2}$.  A choice of sign was made in \eqref{DTNspectrum} to ensure that the energies are real.  In the second line we expressed the result in terms of the coefficient $C$ of the triholomorphic Killing field.  When $C = 0$ this boundstate spectrum reduces to the one found for the ordinary Dirac operator on Taub-NUT in \cite{Gibbons:1986df}.

BPS states correspond to $n = |p|/2$, assuming $p \neq 0$.  The allowed $n$'s increase from this value in integer steps.  In order for the hypergeometric series to truncate we must impose that $j_2 + 1 - n$ is a non-positive integer.  If this integer is zero, then we must additionally set $c_1 = 0$ so that $\hat{f}_1 = 0$, or else the solution will not be normalizable.  Note that $n$ is integer or half-integer when $p$ is even or odd respectively, and the same is true for $j_2 = j - 1/2$.  Therefore $j_2 + 1 - n$ will indeed always be an integer.  Hence for a given $p \neq 0$ and $n$, the allowed values of $j$ are
\begin{equation}
j \in \left\{ j_{\rm min} = \half (|p|-1) ~, j_{\rm min} + 1~, \ldots~, n - \half \right\} ~,
\end{equation}
which collapses to the single value, $j = j_{\rm min}$, for BPS states.  At both the minimum and maximum values of $j$ we must set $c_1 = 0$, killing $\hat{f}_1$.  This is required for the minimal $j$ so that the wavefunction is regular at $\uptheta = 0,\pi$ and for the maximal $j$ so that it is exponentially decaying at large $R$.  For $j$ values strictly in between the minimum and maximum, we have two linearly independent solutions to \eqref{fWeyleqn} for $f$, for given quantum numbers $(p \neq 0,n,j,m)$, corresponding to the coefficients $c_1,c_2$.

Meanwhile if $p=0$ we know from regularity in $\uptheta$ that $j \in \{\half,\tfrac{3}{2},\ldots\}$.  Therefore from normalizability we conclude that the allowed values of $n$ are $n = \{1,2, \ldots\}$.  Then $j \in \{ \half,\ldots n - \half\}$ and $c_1$ should be set to zero when $j = n- \half$.  There are no BPS states when $p=0$.

To complete the discussion of solutions when $\ell_{\rm TN} > 0$, we should return to the first order equations \eqref{Weylfg} and determine the $g$ that goes with a given solution for $f$.  When $\lambda = 0$ the equations for $f,g$ decouple and one can show directly that there are no normalizable solutions for $g$ \cite{Moore:2014jfa}.  For $\lambda \neq 0$ we can use the first of \eqref{Weylfg} to read off the solution for $g$.  Note that in addition to the quantum numbers $(p,n,j,m)$, the solution for $g$ depends on the sign of $\lambda_{p,n}$.  We can also use this equation to express the innerproduct \eqref{ip2} in terms of $f$, or $\hat{f}$, only.  After some simplification one finds a nice result:
\begin{align}\label{ip3}
\langle \Psi | \Psi' \rangle =&~ \frac{16 \pi^2 (\lambda_{p,n} + \lambda_{p',n'})}{(2j +1) \lambda_{p',n'}} \delta_{p p'} \delta_{j j'} \delta_{m m'} \int_{0}^{\infty} dR H(R) \hat{f}^\dag \hat{f}' ~.
\end{align}
Hence it must be that the $\hat{f}_{1,2}$'s, \eqref{fhatsol}, for different choices of $n$'s, are orthogonal on $[0,\infty)$ with respect to the measure $H(R)$.  One can analytically check that this is indeed the case using techniques from \cite{MR1976614}.

In the case of BPS states an additional condition beyond $j = j_{\rm min} = n- \half$ is required for normalizability.  This comes from requiring that $\rho_0$ is positive when $\lambda =0$, which is equivalent to $p$ and $x$ having the same sign.  This condition can also be found from an analysis of the decoupled equations for $f_1$ and $f_2$ that arise from \eqref{Weylfg} on setting $\lambda = k = 0$ \cite{Moore:2014jfa}.  Assuming this is the case then $j_2 + 1 - \rho_0/2 = 0$ and the hypergeometric function is simply 1.  Thus $\hat{f}_2 \propto R^{|p|/2} e^{-|x| R}$, while we set $\hat{f}_1 = 0$.  For $p > 0$ \eqref{fSim} gives $f_1 = \hat{f}_2$ and $f_2 = 0$, while for $p<0$ we get $f_1 = 0$ and $f_2 = \hat{f}_2$.  Going back to \eqref{rthetasep} and \eqref{Weylrot}, we find that BPS spinors take the form $\Psi = (0,\psi_-)^T$ with
\begin{align}
\psi_- =&~ c \, e^{\I p x^4/2} R^{j -1/2} e^{-|x| R} \breve{\psi}(\uptheta,\upphi)~, \qquad \textrm{(BPS case)}
\end{align}
where $c$ is a normalization constant,
\begin{align}\label{BPSangular}
\breve{\psi}(\uptheta,\upphi) =&~ e^{\I (\epsilon p/2 - m)\upphi} \left\{ \begin{array}{l l} d^{j}_{m,j}(\uptheta) \left( \begin{array}{c} e^{-\I \upphi/2} \cos{\frac{\uptheta}{2}} \\ e^{\I \upphi/2} \sin{\frac{\uptheta}{2}} \end{array} \right) ~, & p > 0~, \\[4ex] d^{j}_{m,-j}(\uptheta) \left( \begin{array}{c} - e^{-\I \upphi/2} \sin{\frac{\uptheta}{2}} \\ e^{\I \upphi/2} \cos{\frac{\uptheta}{2}} \end{array} \right)~, & p < 0~, \end{array} \right.
\end{align}
and $j = \half (|p|-1)$.  Note from \eqref{wigner1} and \eqref{wigner2} that when $m' = \pm j$ the Wigner $d$-functions simplify considerably, as $\cn = 0$ and $P_{0}^{(a,b)}(\uptheta) = 1$.  Explicitly, both cases are captured by
\begin{equation}
d^{j}_{m,\pm j}(\uptheta) = (\pm 1)^{j+m} \left( \begin{array}{c} 2j \\ j \mp m \end{array} \right) \left( \sin{\frac{\uptheta}{2}} \right)^{j  \mp m} \left( \cos{\frac{\uptheta}{2}} \right)^{j \pm m} ~, \quad \textrm{(signs correlated).}
\end{equation}

Now let us comment on the $\ell_{\rm TN} < 0$ case.  The metric \eqref{TNmetric} is singular at $R = -\ell_{\rm TN}$, and a boundary condition is required there to specify the eigenvalue problem.  This boundary condition might be incompatible with some or all of the spectrum of boundstates we found above.  We give two examples, focusing on the BPS spectrum.

The first example is $\mathfrak{su}(2)$ gauge theory with $\gmone = \gmtwo = H$.  The relative moduli space is the Atiyah--Hitchin manifold and $C =0$ since there are no triholomorphic Killing vectors.  A naive application of the above analysis would suggest that BPS states might exist --- $x$ and $p$, \eqref{xellparameters}, do have the same sign when $C = 0$ and $\ell_{\rm TN} < 0$, for example.  However a simple argument using the Lichnerowicz--Weitzenb\"ock formula shows that the exact Dirac operator has no $L^2$ zeromodes.

The second example is a framed example in $\mathfrak{su}(2)$ gauge theory with two smooth monopoles in the presence of a minimal 't Hooft defect.  Consider the region of moduli space corresponding to configurations in which one monopole remains close to the defect while the other is far away.  The magnetic core charge is $\gamma_{\rm c} = H - \half H = \half H$ while the magnetic halo charge is $\gamma_{\rm h} = H$.  Hence $\ell_{\rm TN}$, \eqref{framedell}, is again negative.  Unlike the previous example, however, the Dirac operator on the exact strongly centered moduli space does have $L^2$ zeromodes.  (See section 7.4 of \cite{Moore:2015szp}.)



\begin{thebibliography}{10}

\bibitem{Andriyash:2010yf}
E.~Andriyash, F.~Denef, D.~L. Jafferis, and G.~W. Moore, ``{Bound state
  transformation walls},''
  \href{http://dx.doi.org/10.1007/JHEP03(2012)007}{{\em JHEP} {\bfseries 1203}
  (2012) 007},
\href{http://arxiv.org/abs/1008.3555}{{\ttfamily arXiv:1008.3555 [hep-th]}}.

\bibitem{Avery:2007xf}
S.~G. Avery and J.~Michelson, ``{Mechanics and Quantum Supermechanics of a
  Monopole Probe Including a Coulomb Potential},''
  \href{http://dx.doi.org/10.1103/PhysRevD.77.085001}{{\em Phys. Rev.}
  {\bfseries D77} (2008) 085001},
\href{http://arxiv.org/abs/0712.0341}{{\ttfamily arXiv:0712.0341 [hep-th]}}.

\bibitem{Bais:1979gv}
F.~A. Bais and W.~Troost, ``{Zero Modes and Bound States of the Supersymmetric
  Monopole},''
\href{http://dx.doi.org/10.1016/0550-3213(81)90499-5}{{\em Nucl. Phys.}
  {\bfseries B178} (1981) 125--140}.

\bibitem{Bielawski:1998hk}
R.~Bielawski, ``{Asymptotic metrics for SU(N) monopoles with maximal symmetry
  breaking},'' \href{http://dx.doi.org/10.1007/s002200050503}{{\em Commun.
  Math. Phys.} {\bfseries 199} (1998) 297--325},
\href{http://arxiv.org/abs/hep-th/9801092}{{\ttfamily arXiv:hep-th/9801092
  [hep-th]}}.

\bibitem{Bielawski:1998hj}
R.~Bielawski, ``{Monopoles and the Gibbons-Manton metric},''
  \href{http://dx.doi.org/10.1007/s002200050359}{{\em Commun. Math. Phys.}
  {\bfseries 194} (1998) 297--321},
\href{http://arxiv.org/abs/hep-th/9801091}{{\ttfamily arXiv:hep-th/9801091
  [hep-th]}}.

\bibitem{Blair:2010vh}
C.~D. Blair and S.~A. Cherkis, ``{Singular Monopoles from Cheshire Bows},''
  \href{http://dx.doi.org/10.1016/j.nuclphysb.2010.11.014}{{\em Nucl.Phys.}
  {\bfseries B845} (2011) 140--164},
\href{http://arxiv.org/abs/1010.0740}{{\ttfamily arXiv:1010.0740 [hep-th]}}.

\bibitem{Boulton:2017jls}
L.~Boulton, B.~J. Schroers, and K.~Smedley-Williams, ``{Quantum Bound States in
  Yang-Mills-Higgs Theory},''
\href{http://arxiv.org/abs/1708.08363}{{\ttfamily arXiv:1708.08363 [math-ph]}}.

\bibitem{Brennan:2016znk}
T.~D. Brennan and G.~W. Moore, ``{A note on the semiclassical formulation of
  BPS states in four-dimensional $N=$ 2 theories},''
  \href{http://dx.doi.org/10.1093/ptep/ptw159}{{\em PTEP} {\bfseries 2016}
  no.~12, (2016) 12C110},
\href{http://arxiv.org/abs/1610.00697}{{\ttfamily arXiv:1610.00697 [hep-th]}}.

\bibitem{Cherkis:2010bn}
S.~A. Cherkis, ``{Instantons on Gravitons},''
  \href{http://dx.doi.org/10.1007/s00220-011-1293-y}{{\em Commun.Math.Phys.}
  {\bfseries 306} (2011) 449--483},
\href{http://arxiv.org/abs/1007.0044}{{\ttfamily arXiv:1007.0044 [hep-th]}}.

\bibitem{Cherkis:1997aa}
S.~A. Cherkis and A.~Kapustin, ``{Singular monopoles and supersymmetric gauge
  theories in three-dimensions},''
  \href{http://dx.doi.org/10.1016/S0550-3213(98)00341-1}{{\em Nucl.Phys.}
  {\bfseries B525} (1998) 215--234},
\href{http://arxiv.org/abs/hep-th/9711145}{{\ttfamily arXiv:hep-th/9711145
  [hep-th]}}.

\bibitem{Cherkis:1998hi}
S.~A. Cherkis and A.~Kapustin, ``{Singular monopoles and gravitational
  instantons},'' \href{http://dx.doi.org/10.1007/s002200050632}{{\em
  Commun.Math.Phys.} {\bfseries 203} (1999) 713--728},
\href{http://arxiv.org/abs/hep-th/9803160}{{\ttfamily arXiv:hep-th/9803160
  [hep-th]}}.

\bibitem{Cherkis:2011ee}
S.~A. Cherkis, C.~O'Hara, and C.~Saemann, ``{Super Yang-Mills Theory with
  Impurity Walls and Instanton Moduli Spaces},''
  \href{http://dx.doi.org/10.1103/PhysRevD.83.126009}{{\em Phys.Rev.}
  {\bfseries D83} (2011) 126009},
\href{http://arxiv.org/abs/1103.0042}{{\ttfamily arXiv:1103.0042 [hep-th]}}.

\bibitem{Denef:2000nb}
F.~Denef, ``{Supergravity flows and D-brane stability},''
  \href{http://dx.doi.org/10.1088/1126-6708/2000/08/050}{{\em JHEP} {\bfseries
  0008} (2000) 050},
\href{http://arxiv.org/abs/hep-th/0005049}{{\ttfamily arXiv:hep-th/0005049
  [hep-th]}}.

\bibitem{Denef:2007vg}
F.~Denef and G.~W. Moore, ``{Split states, entropy enigmas, holes and halos},''
  \href{http://dx.doi.org/10.1007/JHEP11(2011)129}{{\em JHEP} {\bfseries 1111}
  (2011) 129},
\href{http://arxiv.org/abs/hep-th/0702146}{{\ttfamily arXiv:hep-th/0702146
  [HEP-TH]}}.

\bibitem{Diaconescu:1996rk}
D.-E. Diaconescu, ``{D-branes, monopoles and Nahm equations},''
  \href{http://dx.doi.org/10.1016/S0550-3213(97)00438-0}{{\em Nucl.Phys.}
  {\bfseries B503} (1997) 220--238},
\href{http://arxiv.org/abs/hep-th/9608163}{{\ttfamily arXiv:hep-th/9608163
  [hep-th]}}.

\bibitem{Diaconescu:2007bf}
E.~Diaconescu and G.~W. Moore, ``{Crossing the wall: Branes versus bundles},''
  {\em Adv. Theor. Math. Phys.} {\bfseries 14} no.~6, (2010) 1621--1650,
\href{http://arxiv.org/abs/0706.3193}{{\ttfamily arXiv:0706.3193 [hep-th]}}.

\bibitem{MR1976614}
C.~F. Dunkl, ``A {L}aguerre polynomial orthogonality and the hydrogen atom,''
  \href{http://dx.doi.org/10.1142/S0219530503000132}{{\em Anal. Appl.
  (Singap.)} {\bfseries 1} no.~2, (2003) 177--188},
  \href{http://arxiv.org/abs/math-ph/0011021}{{\ttfamily arXiv:math-ph/0011021
  [math-ph]}}.

\bibitem{FKS}
K.~Fritzsch, C.~Kottke, and M.~Singer, ``{Monopoles and the Sen Conjecture},''
\href{http://arxiv.org/abs/1811.00601}{{\ttfamily arXiv:1811.00601 [math.DG]}}.

\bibitem{Gaiotto:2008cd}
D.~Gaiotto, G.~W. Moore, and A.~Neitzke, ``{Four-dimensional wall-crossing via
  three-dimensional field theory},''
  \href{http://dx.doi.org/10.1007/s00220-010-1071-2}{{\em Commun. Math. Phys.}
  {\bfseries 299} (2010) 163--224},
\href{http://arxiv.org/abs/0807.4723}{{\ttfamily arXiv:0807.4723 [hep-th]}}.

\bibitem{Gaiotto:2010be}
D.~Gaiotto, G.~W. Moore, and A.~Neitzke, ``{Framed BPS States},''
  \href{http://dx.doi.org/10.4310/ATMP.2013.v17.n2.a1}{{\em Adv. Theor. Math.
  Phys.} {\bfseries 17} no.~2, (2013) 241--397},
\href{http://arxiv.org/abs/1006.0146}{{\ttfamily arXiv:1006.0146 [hep-th]}}.

\bibitem{Gauntlett:1993sh}
J.~P. Gauntlett, ``{Low-energy dynamics of N=2 supersymmetric monopoles},''
  \href{http://dx.doi.org/10.1016/0550-3213(94)90457-X}{{\em Nucl.Phys.}
  {\bfseries B411} (1994) 443--460},
\href{http://arxiv.org/abs/hep-th/9305068}{{\ttfamily arXiv:hep-th/9305068
  [hep-th]}}.

\bibitem{Gauntlett:1995fu}
J.~P. Gauntlett and J.~A. Harvey, ``{S duality and the dyon spectrum in N=2
  superYang-Mills theory},''
  \href{http://dx.doi.org/10.1016/0550-3213(96)00035-1}{{\em Nucl.Phys.}
  {\bfseries B463} (1996) 287--314},
\href{http://arxiv.org/abs/hep-th/9508156}{{\ttfamily arXiv:hep-th/9508156
  [hep-th]}}.

\bibitem{Gauntlett:2000ks}
J.~P. Gauntlett, C.-j. Kim, K.-M. Lee, and P.~Yi, ``{General low-energy
  dynamics of supersymmetric monopoles},''
  \href{http://dx.doi.org/10.1103/PhysRevD.63.065020}{{\em Phys.Rev.}
  {\bfseries D63} (2001) 065020},
\href{http://arxiv.org/abs/hep-th/0008031}{{\ttfamily arXiv:hep-th/0008031
  [hep-th]}}.

\bibitem{Gauntlett:1999vc}
J.~P. Gauntlett, N.~Kim, J.~Park, and P.~Yi, ``{Monopole dynamics and BPS dyons
  N=2 superYang-Mills theories},''
  \href{http://dx.doi.org/10.1103/PhysRevD.61.125012}{{\em Phys.Rev.}
  {\bfseries D61} (2000) 125012},
\href{http://arxiv.org/abs/hep-th/9912082}{{\ttfamily arXiv:hep-th/9912082
  [hep-th]}}.

\bibitem{Gauntlett:1996cw}
J.~P. Gauntlett and D.~A. Lowe, ``{Dyons and S duality in N=4 supersymmetric
  gauge theory},'' \href{http://dx.doi.org/10.1016/0550-3213(96)00218-0}{{\em
  Nucl.Phys.} {\bfseries B472} (1996) 194--206},
\href{http://arxiv.org/abs/hep-th/9601085}{{\ttfamily arXiv:hep-th/9601085
  [hep-th]}}.

\bibitem{Gibbons:1986df}
G.~Gibbons and N.~Manton, ``{Classical and Quantum Dynamics of BPS
  Monopoles},''
\href{http://dx.doi.org/10.1016/0550-3213(86)90624-3}{{\em Nucl.Phys.}
  {\bfseries B274} (1986) 183}.

\bibitem{Gibbons:1995yw}
G.~Gibbons and N.~Manton, ``{The Moduli space metric for well separated BPS
  monopoles},'' \href{http://dx.doi.org/10.1016/0370-2693(95)00813-Z}{{\em
  Phys.Lett.} {\bfseries B356} (1995) 32--38},
\href{http://arxiv.org/abs/hep-th/9506052}{{\ttfamily arXiv:hep-th/9506052
  [hep-th]}}.

\bibitem{Hanany:1996ie}
A.~Hanany and E.~Witten, ``{Type IIB superstrings, BPS monopoles, and
  three-dimensional gauge dynamics},''
  \href{http://dx.doi.org/10.1016/S0550-3213(97)00157-0}{{\em Nucl.Phys.}
  {\bfseries B492} (1997) 152--190},
\href{http://arxiv.org/abs/hep-th/9611230}{{\ttfamily arXiv:hep-th/9611230
  [hep-th]}}.

\bibitem{Houghton:1997ei}
C.~J. Houghton, ``{New hyperKahler manifolds by fixing monopoles},''
  \href{http://dx.doi.org/10.1103/PhysRevD.56.1220}{{\em Phys. Rev.} {\bfseries
  D56} (1997) 1220--1227},
\href{http://arxiv.org/abs/hep-th/9702161}{{\ttfamily arXiv:hep-th/9702161
  [hep-th]}}.

\bibitem{Jante:2013kha}
R.~Jante and B.~J. Schroers, ``{Dirac operators on the Taub-NUT space,
  monopoles and SU(2) representations},''
  \href{http://dx.doi.org/10.1007/JHEP01(2014)114}{{\em JHEP} {\bfseries 01}
  (2014) 114},
\href{http://arxiv.org/abs/1312.4879}{{\ttfamily arXiv:1312.4879 [hep-th]}}.

\bibitem{Jante:2015xra}
R.~Jante and B.~J. Schroers, ``{Taub-NUT dynamics with a magnetic field},''
  \href{http://dx.doi.org/10.1016/j.geomphys.2016.02.016}{{\em J. Geom. Phys.}
  {\bfseries 104} (2016) 305--328},
\href{http://arxiv.org/abs/1507.08165}{{\ttfamily arXiv:1507.08165 [hep-th]}}.

\bibitem{Joyce:2008pc}
D.~Joyce and Y.~Song, ``{A Theory of generalized Donaldson-Thomas
  invariants},''
\href{http://arxiv.org/abs/0810.5645}{{\ttfamily arXiv:0810.5645 [math.AG]}}.

\bibitem{Kontsevich:2008fj}
M.~Kontsevich and Y.~Soibelman, ``{Stability structures, motivic
  Donaldson-Thomas invariants and cluster transformations},''
\href{http://arxiv.org/abs/0811.2435}{{\ttfamily arXiv:0811.2435 [math.AG]}}.

\bibitem{MR3330788}
M.~Kontsevich and Y.~Soibelman, ``Wall-crossing structures in
  {D}onaldson-{T}homas invariants, integrable systems and mirror symmetry,'' in
  {\em Homological mirror symmetry and tropical geometry}, vol.~15 of {\em
  Lect. Notes Unione Mat. Ital.}, pp.~197--308.
\newblock Springer, Cham, 2014.
\newblock \href{http://arxiv.org/abs/1303.3253}{{\ttfamily arXiv:1303.3253
  [math.AG]}}.

\bibitem{Kottke:2015rwx}
C.~Kottke and M.~Singer, ``{Partial compactification of monopoles and metric
  asymptotics},''
\href{http://arxiv.org/abs/1512.02979}{{\ttfamily arXiv:1512.02979 [math.DG]}}.

\bibitem{Kronheimer}
P.~B. Kronheimer, ``{Monopoles and Taub-NUT Metrics},'' M.Sc. thesis, Oxford,
  1985.

\bibitem{Lee:1996if}
K.-M. Lee, E.~J. Weinberg, and P.~Yi, ``{Electromagnetic duality and SU(3)
  monopoles},'' \href{http://dx.doi.org/10.1016/0370-2693(96)00286-9}{{\em
  Phys.Lett.} {\bfseries B376} (1996) 97--102},
\href{http://arxiv.org/abs/hep-th/9601097}{{\ttfamily arXiv:hep-th/9601097
  [hep-th]}}.

\bibitem{Lee:1996kz}
K.-M. Lee, E.~J. Weinberg, and P.~Yi, ``{The Moduli space of many BPS monopoles
  for arbitrary gauge groups},''
  \href{http://dx.doi.org/10.1103/PhysRevD.54.1633}{{\em Phys.Rev.} {\bfseries
  D54} (1996) 1633--1643},
\href{http://arxiv.org/abs/hep-th/9602167}{{\ttfamily arXiv:hep-th/9602167
  [hep-th]}}.

\bibitem{Manschot:2010xp}
J.~Manschot, ``{Wall-crossing of D4-branes using flow trees},''
  \href{http://dx.doi.org/10.4310/ATMP.2011.v15.n1.a1}{{\em Adv. Theor. Math.
  Phys.} {\bfseries 15} no.~1, (2011) 1--42},
\href{http://arxiv.org/abs/1003.1570}{{\ttfamily arXiv:1003.1570 [hep-th]}}.

\bibitem{Manschot:2010qz}
J.~Manschot, B.~Pioline, and A.~Sen, ``{Wall Crossing from Boltzmann Black Hole
  Halos},'' \href{http://dx.doi.org/10.1007/JHEP07(2011)059}{{\em JHEP}
  {\bfseries 07} (2011) 059},
\href{http://arxiv.org/abs/1011.1258}{{\ttfamily arXiv:1011.1258 [hep-th]}}.

\bibitem{Manton:1981mp}
N.~Manton, ``{A Remark on the Scattering of BPS Monopoles},''
\href{http://dx.doi.org/10.1016/0370-2693(82)90950-9}{{\em Phys.Lett.}
  {\bfseries B110} (1982) 54--56}.

\bibitem{Moore:2014gua}
G.~W. Moore, A.~B. Royston, and D.~Van~den Bleeken, ``{Brane bending and
  monopole moduli},'' \href{http://dx.doi.org/10.1007/JHEP10(2014)157}{{\em
  JHEP} {\bfseries 10} (2014) 157},
\href{http://arxiv.org/abs/1404.7158}{{\ttfamily arXiv:1404.7158 [hep-th]}}.

\bibitem{Moore:2014jfa}
G.~W. Moore, A.~B. Royston, and D.~Van~den Bleeken, ``{Parameter counting for
  singular monopoles on $\mathbbm R^3$},''
  \href{http://dx.doi.org/10.1007/JHEP10(2014)142}{{\em JHEP} {\bfseries 10}
  (2014) 142},
\href{http://arxiv.org/abs/1404.5616}{{\ttfamily arXiv:1404.5616 [hep-th]}}.

\bibitem{Moore:2015qyu}
G.~W. Moore, A.~B. Royston, and D.~Van~den Bleeken, ``{$L^2$-Kernels Of
  Dirac-Type Operators On Monopole Moduli Spaces},''
\href{http://arxiv.org/abs/1512.08923}{{\ttfamily arXiv:1512.08923 [hep-th]}}.

\bibitem{Moore:2015szp}
G.~W. Moore, A.~B. Royston, and D.~Van~den Bleeken, ``{Semiclassical framed BPS
  states},'' \href{http://dx.doi.org/10.1007/JHEP07(2016)071}{{\em JHEP}
  {\bfseries 07} (2016) 071},
\href{http://arxiv.org/abs/1512.08924}{{\ttfamily arXiv:1512.08924 [hep-th]}}.

\bibitem{Murray:1996hi}
M.~K. Murray, ``{A Note on the (1, 1,..., 1) monopole metric},''
  \href{http://dx.doi.org/10.1016/S0393-0440(96)00044-7}{{\em J.Geom.Phys.}
  {\bfseries 23} (1997) 31--41},
\href{http://arxiv.org/abs/hep-th/9605054}{{\ttfamily arXiv:hep-th/9605054
  [hep-th]}}.

\bibitem{MurthyPioline}
S.~Murthy and B.~Pioline, ``{Mock modularity from black hole scattering
  states},'' \href{http://dx.doi.org/10.1007/JHEP12(2018)119}{{\em JHEP}
  {\bfseries 12} (2018) 119},
\href{http://arxiv.org/abs/1808.05606}{{\ttfamily arXiv:1808.05606 [hep-th]}}.

\bibitem{MR953820}
H.~Pedersen and Y.~S. Poon, ``Hyper-{K}\"ahler metrics and a generalization of
  the {B}ogomolny equations,'' {\em Comm. Math. Phys.} {\bfseries 117} no.~4,
  (1988) 569--580.

\bibitem{Pioline:2011gf}
B.~Pioline, ``{Four ways across the wall},''
  \href{http://dx.doi.org/10.1088/1742-6596/346/1/012017}{{\em J. Phys. Conf.
  Ser.} {\bfseries 346} (2012) 012017},
\href{http://arxiv.org/abs/1103.0261}{{\ttfamily arXiv:1103.0261 [hep-th]}}.

\bibitem{Pioline:2015wza}
B.~Pioline, ``{Wall-crossing made smooth},''
  \href{http://dx.doi.org/10.1007/JHEP04(2015)092}{{\em JHEP} {\bfseries 04}
  (2015) 092},
\href{http://arxiv.org/abs/1501.01643}{{\ttfamily arXiv:1501.01643 [hep-th]}}.

\bibitem{Pope:1978zx}
C.~Pope, ``{Axial Vector Anomalies and the Index Theorem In Charged
  Schwarzschild and Taub - Nut Spaces},''
\href{http://dx.doi.org/10.1016/0550-3213(78)90038-X}{{\em Nucl.Phys.}
  {\bfseries B141} (1978) 432}.

\bibitem{Ritz:2008jf}
A.~Ritz and A.~Vainshtein, ``{Dyon dynamics near marginal stability and non-BPS
  states},'' \href{http://dx.doi.org/10.1016/j.physletb.2008.08.016}{{\em Phys.
  Lett.} {\bfseries B668} (2008) 148--152},
\href{http://arxiv.org/abs/0807.2419}{{\ttfamily arXiv:0807.2419 [hep-th]}}.

\bibitem{Ritz:2001jk}
A.~Ritz and A.~I. Vainshtein, ``{Long range forces and supersymmetric bound
  states},'' \href{http://dx.doi.org/10.1016/S0550-3213(01)00483-7}{{\em Nucl.
  Phys.} {\bfseries B617} (2001) 43--70},
\href{http://arxiv.org/abs/hep-th/0102121}{{\ttfamily arXiv:hep-th/0102121
  [hep-th]}}.

\bibitem{Sakurai}
J.~J. Sakurai, {\em Modern Quantum Mechanics (Revised Edition)}.
\newblock Addison Wesley, 1~ed., 1993.

\bibitem{Seiberg:1994rs}
N.~Seiberg and E.~Witten, ``{Electric - magnetic duality, monopole
  condensation, and confinement in N=2 supersymmetric Yang-Mills theory},''
  \href{http://dx.doi.org/10.1016/0550-3213(94)90124-4}{{\em Nucl.Phys.}
  {\bfseries B426} (1994) 19--52},
\href{http://arxiv.org/abs/hep-th/9407087}{{\ttfamily arXiv:hep-th/9407087
  [hep-th]}}.

\bibitem{Seiberg:1994aj}
N.~Seiberg and E.~Witten, ``{Monopoles, duality and chiral symmetry breaking in
  N=2 supersymmetric QCD},''
  \href{http://dx.doi.org/10.1016/0550-3213(94)90214-3}{{\em Nucl.Phys.}
  {\bfseries B431} (1994) 484--550},
\href{http://arxiv.org/abs/hep-th/9408099}{{\ttfamily arXiv:hep-th/9408099
  [hep-th]}}.

\bibitem{Sethi:1995zm}
S.~Sethi, M.~Stern, and E.~Zaslow, ``{Monopole and Dyon bound states in N=2
  supersymmetric Yang-Mills theories},''
  \href{http://dx.doi.org/10.1016/0550-3213(95)00517-X}{{\em Nucl.Phys.}
  {\bfseries B457} (1995) 484--512},
\href{http://arxiv.org/abs/hep-th/9508117}{{\ttfamily arXiv:hep-th/9508117
  [hep-th]}}.

\bibitem{Stern:2000ie}
M.~Stern and P.~Yi, ``{Counting Yang-Mills dyons with index theorems},''
  \href{http://dx.doi.org/10.1103/PhysRevD.62.125006}{{\em Phys.Rev.}
  {\bfseries D62} (2000) 125006},
\href{http://arxiv.org/abs/hep-th/0005275}{{\ttfamily arXiv:hep-th/0005275
  [hep-th]}}.

\bibitem{Taubes:1981gw}
C.~H. Taubes, ``{The Existence of Multi - Monopole Solutions to the Nonabelian,
  Yang-Mills Higgs Equations for Arbitrary Simple Gauge Groups},''
\href{http://dx.doi.org/10.1007/BF01208275}{{\em Commun.Math.Phys.} {\bfseries
  80} (1981) 343}.

\bibitem{Tong:2014yla}
D.~Tong and K.~Wong, ``{Monopoles and Wilson Lines},''
  \href{http://dx.doi.org/10.1007/JHEP06(2014)048}{{\em JHEP} {\bfseries 1406}
  (2014) 048},
\href{http://arxiv.org/abs/1401.6167}{{\ttfamily arXiv:1401.6167 [hep-th]}}.

\bibitem{Weinberg:1979ma}
E.~J. Weinberg, ``{Parameter Counting for Multi-Monopole Solutions},''
\href{http://dx.doi.org/10.1103/PhysRevD.20.936}{{\em Phys.Rev.} {\bfseries
  D20} (1979) 936--944}.

\bibitem{Weinberg:1979zt}
E.~J. Weinberg, ``{Fundamental Monopoles and Multi-Monopole Solutions for
  Arbitrary Simple Gauge Groups},''
\href{http://dx.doi.org/10.1016/0550-3213(80)90245-X}{{\em Nucl.Phys.}
  {\bfseries B167} (1980) 500}.

\end{thebibliography}

\providecommand{\href}[2]{#2}\begingroup\raggedright\endgroup
\end{document}